\documentclass[titlepage, 12pt]{article}

\usepackage[english]{babel}

\usepackage[
    letterpaper,
    top=1in,
    bottom=1in,
    left=1in,
    right=1in,
    ]{geometry}

\usepackage{amsmath}
\usepackage{graphicx}
\graphicspath{{./images/}}
\usepackage[colorlinks=true, allcolors=blue]{hyperref}
\usepackage{titling}
\usepackage{lscape}
\usepackage{wasysym}
\usepackage{wrapfig}
\usepackage[title]{appendix}
\usepackage{caption}
\usepackage{subcaption}
\usepackage{booktabs}

\usepackage[authoryear,compress]{natbib}
\usepackage[nottoc]{tocbibind}
\usepackage{array}
\newcolumntype{L}[1]{>{\raggedright\let\newline\\\arraybackslash\hspace{0pt}}m{#1}}
\newcolumntype{C}[1]{>{\centering\let\newline\\\arraybackslash\hspace{0pt}}m{#1}}
\newcolumntype{R}[1]{>{\raggedleft\let\newline\\\arraybackslash\hspace{0pt}}m{#1}}

\predate{\centering\placetitlepicture\large}

\newcommand{\titlepicture}[2][]{%
  \renewcommand\placetitlepicture{%
    \includegraphics[#1]{#2}\par\medskip
  }%
}
\newcommand{\placetitlepicture}{} 
\date{February 21, 2024}
\title{Great Observatory for Long Wavelengths (GO-LoW) \\
NIAC Phase I Final Report}
\author{Mary Knapp (MIT Haystack Observatory, \url{mknapp@mit.edu}) \\ Lenny Paritsky (MIT Haystack Observatory) \\ Ekaterina Kononov (MIT Dept.\ of Aeronautics and Astronautics) \\ Melodie M. Kao (UC Santa Cruz, Lowell Observatory)}
\titlepicture[width=\linewidth]{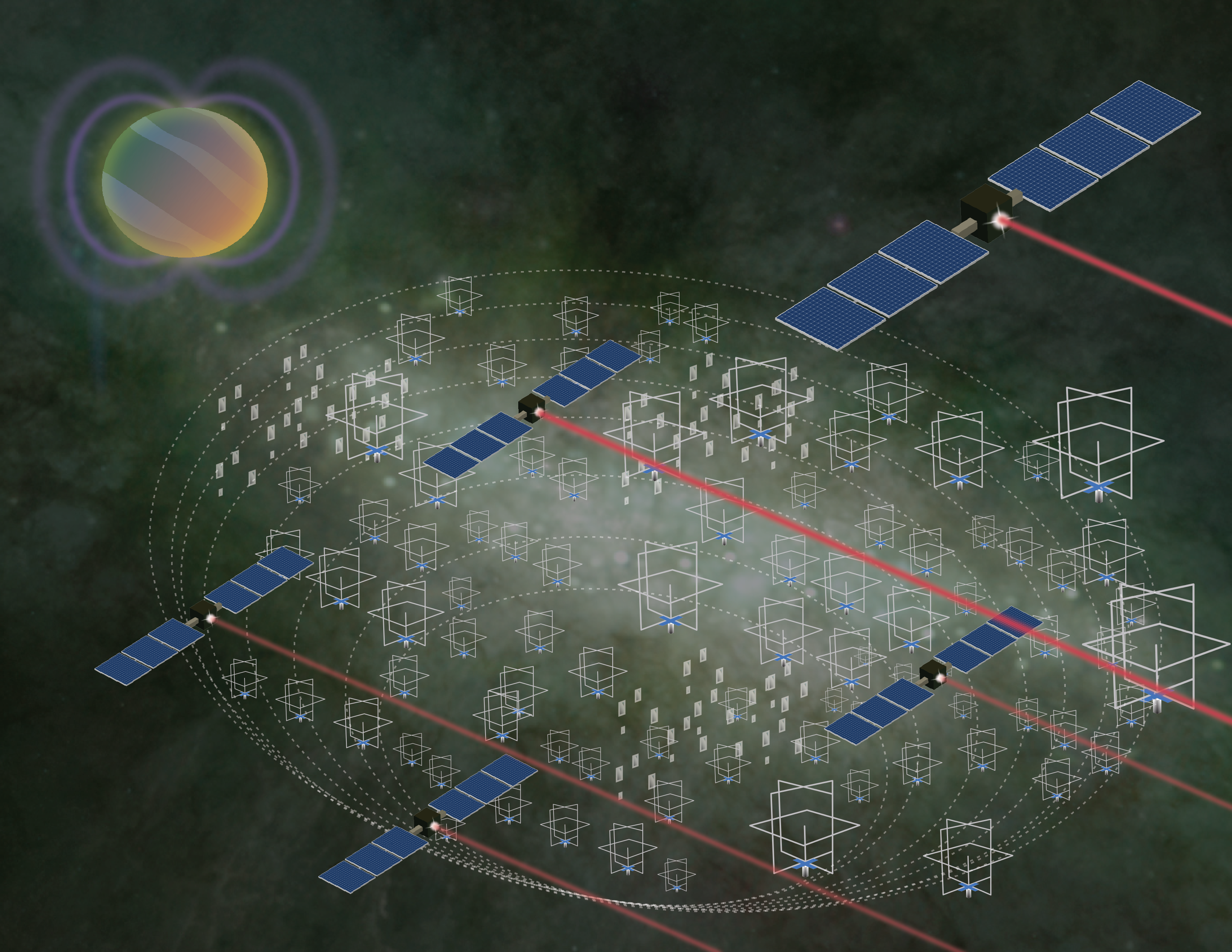}

\begin{document}

\begin{titlepage}
    \begin{center}
    \LARGE
        \textbf{Great Observatory for Long Wavelengths (GO-LoW) \\ \Large
NIAC Phase I Final Report}

\vspace{1.5cm}

\large
Mary Knapp (MIT Haystack Observatory, \url{mknapp@mit.edu} \\
Lenny Paritsky (MIT Haystack Observatory) \\
Ekaterina Kononov (MIT Dept.\ of Aeronautics and Astronautics) \\
Melodie M. Kao (UC Santa Cruz, Lowell Observatory) \\
\vspace{1.1cm}
\includegraphics[width=\linewidth]{images/NIAC-II_GO-LoW_final.png}
\vspace{1cm}
February 21, 2024

    \end{center}
\end{titlepage}



\tableofcontents

\clearpage

\section{Executive Summary and Key Findings} 
    \label{sec:exsum}
    \subsection{Motivation} \label{sec:exsum-motivation}
\begin{figure} [hbt!]
    \centering
    \includegraphics[width=1\linewidth]{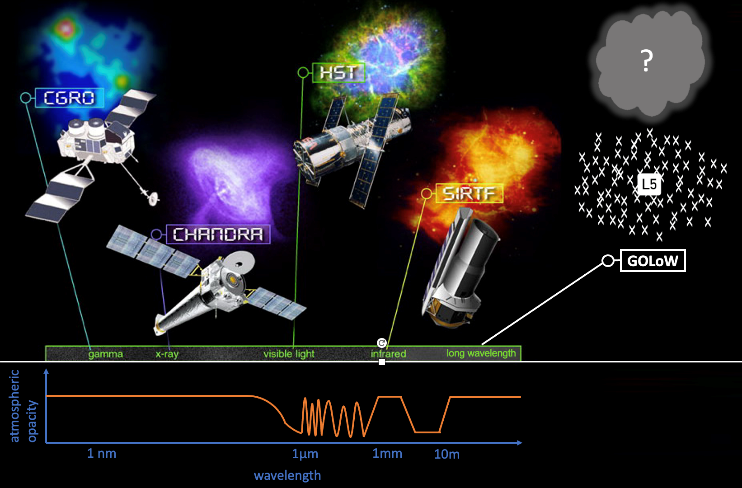}
    \caption{NASA's Great Observatories with GO-LoW.  Earth's atmospheric opacity is indicated in orange; GO-LoW accesses the last window of the EM spectrum  that cannot be observed from Earth’s surface. Great Observatories figure credit: NASA.}
    \label{fig:greatobs_wavelength}
\end{figure}

The Earth's neutral atmosphere and ionosphere block some wavelengths of light from reaching the ground (Figure \ref{fig:greatobs_wavelength}, bottom panel).  Humans have built space telescopes to access high-energy electromagnetic (EM) radiation from space (\emph{e.g.}, gamma rays with CGRO and later SWIFT, X-rays with Chandra, UV with the Hubble Space Telescope) as well as the parts of the infrared band that are blocked by the atmosphere (\emph{e.g.}, Spitzer, JWST).  With these Great Observatory telescopes, humanity has a resolved view of the universe at nearly all wavelength bands---except one.  

The low-frequency sky below $\sim$15 MHz (20 m) is obscured by the Earth's ionosphere, the layer of charged particles above the neutral atmosphere.  Single spacecraft have made measurements in this band, but cannot achieve high or even moderate angular resolution because a telescope's resolution ($\theta$) is set by $\theta = \lambda/D$, where $\lambda$ is the wavelength and $D$ is the telescope diameter.  For wavelengths that range from tens of meters to kilometers, a telescope must be hundreds of meters to many kilometers in diameter for even moderate resolution.

The \textbf{G}reat \textbf{O}bservatory for \textbf{Lo}ng \textbf{W}avelengths (\textbf{GO-LoW}) is an interferometric mega-constellation space telescope operating between 300 kHz and 15 MHz (see Table \ref{tab:keyparam}).  In a departure from the traditional approach of a single, large, expensive spacecraft (\emph{e.g.}, HST, Chandra, JWST), GO-LoW is an \textit{interferometric} Great Observatory comprising thousands of small, inexpensive, and reconfigurable nodes. Interferometry, a technique that combines signals from many spatially separated receivers to form a large ``virtual'' telescope, is ideally suited to long-wavelength astronomy. The individual antenna and receiver systems are simple and require no large structures, and the wide spacing between nodes provides high spatial resolution. 

\textbf{A distributed constellation of sensing elements provides (1) reliability and robustness to failures, (2) longevity by allowing for growth over time and infusion of new technology via staged replacement of nodes, (3) reduced costs through leveraging mass production, and (4) formation reconfigurability to optimize the observatory for diverse science cases.}

\begin{table}
\setlength{\arrayrulewidth}{0.5mm}
\renewcommand{\arraystretch}{1.5}
    \small
    \centering
    \begin{tabular}{|p{0.2\linewidth}|p{0.2\linewidth}|p{0.5\linewidth}|}
    \hline
    \textbf{Parameter} & \textbf{Value}  & \textbf{Rationale}  \\
    \hline
    \hline
    Frequency & 300 kHz (1 km)-- \par 15 MHz (20 m)  &  Largely inaccessible from Earth's surface, though some overlap between 10--15 MHz.  \\
    Location     & Earth-Sun L4 or L5  & Stable Lagrange point reduces stationkeeping needs and Earth-based radio frequency interference (RFI). Provides stable thermal environment. \\
    Constellation \par diameter     & maximum 6000 km & Scattering in interstellar/interplanetary media (ISM/IPM) limit resolution on long baselines. \\
    Constellation \par number     & 100,000 nodes & See Section \ref{sec:sensitivity_calc} for sensitivity discussion. \\
   Angular resolution     & \raggedright 34 arcsec at 300 kHz, \linebreak 0.7 arcsec at 15 MHz & Calculated, $\theta = \lambda/D$ \\
    \hline
    \end{tabular}
    \caption{GO-LoW key parameters}
    \label{tab:keyparam}
\end{table}

\subsection{Applications} \label{execsum-applications}
A low-frequency mega-constellation revolutionizes a number of compelling science cases: high-resolution all-sky mapping, Dark Ages/Epoch of Reionization cosmology, interstellar medium mapping, solar/planetary magnetic activity, and exoplanetary magnetospheric radio emission \citep{Jester2009}.  Simply mapping the sky at moderate spatial resolution ($<$1 arcminute) will be a huge leap forward in our multispectral view of the universe. No high-resolution sky maps exist below $\sim$15 MHz due to ionospheric shielding, yet each time humanity has mapped the sky in a new spectral band it has led to numerous discoveries such as radio and X-ray pulsars and gamma-ray bursts. We expect similar paradigm-changing discoveries at long wavelengths as well.  

Like all astrophysical observatories, GO-LoW will also be a time machine, peering back into the cosmological Dark Ages and Epoch of Reionization by observing highly redshifted 21-cm hydrogen emission.  This era is the infancy of the universe, from $\sim$10 million to 1 billion years after the Big Bang.  Observations of this era will test models of how dark matter interacts with baryonic matter.

Furthermore, GO-LoW will provide the capability of surveying exoplanet magnetic fields in the solar neighborhood.  This study has focused primarily on this application because it is most challenging in terms of sensitivity (and therefore constellation size).  It also fits squarely within the Astro2020 Decadal Survey \citep{astro2020} ``Worlds and Suns in Context'' science theme --- GO-LoW will allow us to learn about the magnetic fields of planets in neighboring stellar systems and add that context to atmospheric characterization and other measurements (see \S\ref{sec:estexoaurora} and \S\ref{sec:STM}).

\subsection{Phase I study goals}
The GO-LoW NIAC Phase I study had the following high-level goals:
\begin{enumerate}
    \item Determine the antenna design that optimizes per-element sensitivity between 100 kHz and 15 MHz.
    \item Develop the constellation architecture for GO-LoW, including a trade study of interferometric correlation strategies.
    \item Determine the optimal correlation architecture.
    \item Map out the technological development required to make GO-LoW feasible in the next 10--20 years.
\end{enumerate}

We address (1) in \S\ref{sec:antenna_options}, (2) in \S\ref{sec:ccc}, \S\ref{sec:lasercommarchitecture}, and \S\ref{sec:mission_architecture}, and (3) in \S\ref{sec:techmap}.  We also further develop the science case and the substantial, paradigm-shifting benefits to NASA in \S\ref{sec:sci_exo-stellar} and \S\ref{sec:resiliency}, \ref{sec:broaderimpacts}, respectively.  Key findings are summarized in the next section.

\subsection{Key findings from Phase I study}

Overview:
\begin{itemize}
    \item The Great Observatory for Long Wavelengths (GO-LoW) is a mega-constellation radio telescope composed of small satellites at an Earth-Sun Lagrange point.
    \item GO-LoW’s architecture is scalable, with important science possible with just 10 nodes. At full size, the constellation consists of 100,000 nodes. See \S\ref{sec:const-growth}.
    \item Phase I focused on defining the spacecraft, data, and communications architectures required to achieve GO-LoW’s ambitious science goals.
\end{itemize}

\noindent Science goals:
\begin{itemize}
    \item Representative mission: Survey exoplanet magnetic fields within 5 pc via auroral radio emissions.  See \S\ref{sec:sci_exo-stellar}, \S\ref{sec:estexoaurora}, \S\ref{sec:mission_architecture}.
    \item Other science goals include low-frequency sky mapping and 21-cm early universe cosmology.  See \S\ref{sec:cosmology}, \ref{sec:helioplanet}, \ref{sec:skymapping}.
\end{itemize}

\noindent Constellation architecture:
\begin{itemize}
    \item GO-LoW consists of two types of spacecraft: small, simple Listener Nodes (LNs) and larger, more capable Computation \& Communication Nodes (CCNs). This design provides an optimal balance between minimizing the number of launches required, delivering enough sensing nodes (LNs), and being able to downlink high data volumes via CCNs. (See Figures \ref{fig:hybrid-architecture}, \ref{fig:LN-stowed}, \ref{fig:CCN-diagram}.)
    \item LNs carry an optimized vector sensor antenna, which was determined to be the best compromise between sensitivity and ease of deployment. (See \S\ref{sec:antenna_options}.)
    \item CCNs are ESPA-class spacecraft, envisioned to support 100 to 1,000 LNs each.
    \item LNs are 3U spacecraft, optimized for packing within next-generation super-heavy lift launch vehicles (LVs, e.g. SpaceX Starship). Over 21,250 LNs can be launched from a single Starship, using reasonable assumptions about support structure. (See \S\ref{sec:launchpacking}.) 
    \item The full GO-LoW constellation can be deployed with 6--11 Starship launches. (See \S\ref{sec:mission_architecture}.)
    \item GO-LoW spans several thousand km but need not be bigger; scattering in the interplanetary/interstellar media limits higher resolution with a larger-diameter array \citep{Jester2009}.
\end{itemize}

\noindent Data \& communications architecture:
\begin{itemize}
    \item Science data is generated at ~1 Gbps at each LN. Grouping LNs into beamforming clusters, ranging from 100--1000 in number, is necessary to close the link budget to Earth. (See \S\ref{sec:corrstudy}.)
    \item For each beamforming cluster, a single CCN aggregates the LNs’ data (via RF link, \S\ref{sec:LN-CCN-link}) and transmits it to Earth via lasercom (\S\ref{sec:laser-link-L4}).
    \item Achieving ~1 Gbps via CCN-to-Earth lasercom is feasible based on the technology roadmap of optical communications. (See \S\ref{sec:lasercommarchitecture}, \S\ref{sec:techmap}.)
    \item Achieving ~1 Gbps via LN-to-CCN RF link is feasible based on existing technology because of the small ($<$100 km) distances involved. (See \S\ref{sec:rf_link_method}.)
    \item A single lasercom ground station could simultaneously downlink from all 100--1k CCNs.
    \item Interferometric correlation occurs on Earth, as opposed to in space. Total raw data scales as n, where n is the number of nodes, but correlated products scale as n$^2$. (See \S\ref{sec:corrstudy}.)
\end{itemize}

\noindent Key unknowns remaining (\S\ref{sec:conclusion}):
\begin{itemize}
    \item Autonomous operations framework required for a deep space mega-constellation. (See \S\ref{sec:autonomy}, \ref{sec:autonomy_future}.)
    \item Strategies for deploying GO-LoW at L4/5 and maintaining required distances between spacecraft (\emph{e.g.}, interferometric baselines). (See \S\ref{sec:orbit}.)
    \item Timescale of constellation’s orbital evolution, and degree of node “intermixing.” (See \S\ref{sec:orbit}.)
    \begin{itemize}
        \item Telemetry and update cadence needed for accurate in-space beamforming.
    \end{itemize}
    \item Effect of constellation motion, telecomm delays, clock offsets/drift, etc. on science data product quality (noise level, resolution, imaging artifacts). \S\ref{sec:timecal_future}
\end{itemize}
\clearpage

\section{Introduction}
    
Humankind has never before seen a detailed map of the low-frequency ($<$15 MHz) radio sky. Earth’s ionosphere hides it from ground-based telescopes, and traditional space missions struggle to access it because long meter- to kilometer-scale wavelengths require massive telescopes. Electromagnetic radiation at these low frequencies carries crucial information about key ingredients to planetary evolution and habitability, such as exoplanet magnetic fields and stellar space weather, as well as the interstellar and intergalactic medium, and the earliest stars and galaxies.

The Great Observatory for Long Wavelengths (GO-LoW) mission concept is an interferometric array composed of two types of small satellites to an Earth-Sun Lagrange point (\emph{e.g.}, L4).  GO-LoW’s representative mission takes on the challenge of detecting and characterizing exoplanetary magnetic fields within 5 parsecs of the solar system via auroral radio emission. Each spacecraft must be equipped with a cutting-edge antenna that maximizes sensitivity in order to detect the faint whispers of radio emission from exoplanets in our stellar neighborhood.  

For the first time in human history, the low-frequency ($<$15 MHz) radio sky may be within reach: GO-LoW leverages the unique capabilities of vector sensor antennas, private sector technology developments driving low Earth orbit mega-constellations, accelerating economic trends behind the SmallSat revolution, and a concept innovation to NASA's traditional mission design process (\S\ref{sec:resiliency}).

 \subsection{A new approach to risk management for NASA: Multi-Phase Great Observatory Arrays}       \label{sec:resiliency}  

GO-LoW applies NASA's existing heritage-focused risk management practices to the end-to-end design process for a Great Observatory. Namely, carefully implementing a Great Observatory in multiple phases reduces the instantaneous risk of an ambitious design that may initially appear to exceed NASA's risk and/or cost budget. Our representative mission science case demonstrates how this innovation to NASA's mission design approach offers a feasible and attractive pathway for NASA to extend its current mission development portfolio to invest in higher-risk missions with transformative potential. 

Large uncertainties in predicted sensitivities for exoplanet radio emissions ($\sim$1 mJy to $<$1 $\mu$Jy; \S\ref{sec:sensitivity_calc}) reflect the wide-ranging potential scientific impact of GO-LoW on multiple Divisions under NASA's Science Mission Directorate, including Planetary Science, Heliophysics, and Astrophysics. Detections of exoplanet radio emission will allow for comparative magnetospheric physics, tying together detailed models informed by planetary science missions (\emph{e.g.}, Cassini, JUNO, MAVEN, and Parker Solar Probe) to new discoveries from systems outside of the solar system. 

\textbf{However, traditional mission design approaches struggle to account for larger uncertainties inherent to fields with large discovery spaces}. If GO-LoW accommodates only the most optimistic flux predictions, it may be under-designed for real exoplanet systems. Yet taking the most conservative assumptions --- as is typical in space mission design --- can lead to prohibitively over-designed hardware requirements.  If cost and complexity appear too great, GO-LoW may never get built at all, and humankind will remain blind to the low-frequency radio sky. \textit{This all-or-nothing approach hinders NASA's ability to invest in fields with large discovery spaces}. 
 
\textbf{Resiliency to sensitivity requirements:} GO-LoW showcases a new approach to Great Observatory design: multi-phase evolution of science requirements and capabilities. By design, GO-LoW is resilient to uncertainties in predicted sensitivity; it is easily updated with additional nodes to meet each successive set of science requirements. This multi-phase approach offers additional benefits:
\begin{enumerate}
    \item \textbf{Eliminates catastrophic single-point hardware failures.} The loss of a few nodes does not meaningfully impact the overall sensitivity of the array (\S\ref{sec:sensitivity_calc}), allowing GO-LoW to dynamically compensate as nodes fail. When a critical number of nodes needs replacement, GO-LoW can integrate a new delivery of nodes. In effect, this evolves Hubble's successful telescope repair missions to an uncrewed approach with much lower risk to the observatory, to astronauts, and to NASA. 
    \item \textbf{Reduces resource waste} by recycling the previous array into the new array. Instead of retiring the nodes comprising obsolete interim phases, GO-LoW leverages them. 
    \item \textbf{Delivers science returns at regular intervals} while gradually building the array to its full capacity. This eliminates the science vacuum that currently takes place in the long decades between Great Observatory mission conception to first light.
     \item \textbf{Insulates NASA from budget bottlenecks}, enabling concurrent development of other missions in NASA's SMD portfolio. This reduces the winner-takes-all dynamic that existing too-big-to-fail missions can introduce within the scientific community. 
    \item \textbf{Adapts to future advances in scientific knowledge} that may call for updates in sensitivity requirements, ensuring long-term returns on GO-LoW's science impact.
    \item \textbf{Adapts to advances in technology} by incorporating components and subsystems that have matured along the technology readiness level (TRL) ladder into replenishment or upgrade launches to GO-LoW. 
\end{enumerate}

To implement a phased design approach, we propose modifying NASA's existing Science Traceability Matrix (STM) to incorporate interim phases.  NASA may then accordingly align the funding structure of a multi-phase Great Observatory to the specified interim phases. In Future Work (\S\ref{sec:conclusion}), we discuss a key next step of specifying requirements for a technology demonstration mission; this mission and its STM would be the first interim phase on the development path to GO-LoW.

\subsection{Representative mission architecture} \label{sec:rep-mission-architecture}
During this study, the GO-LoW team identified the optimal, hierarchical architecture (see \S\ref{sec:ccc} for justification) for the mission, which consists of two types of spacecraft: 

1. \textbf{Listener Nodes (LNs)}, shown in Figure \ref{fig:LN-stowed}. These $\sim$3U spacecraft, roughly half the size of a shoebox, carry GO-LoW's innovative science payload: the deployable Vector Sensor Antenna (VSA) (Figure \ref{fig:LN-deployed}, see also \S\ref{sec:antenna_options}) and its associated science radio.
 \begin{figure}
    \centering
    \includegraphics[width=1\linewidth]{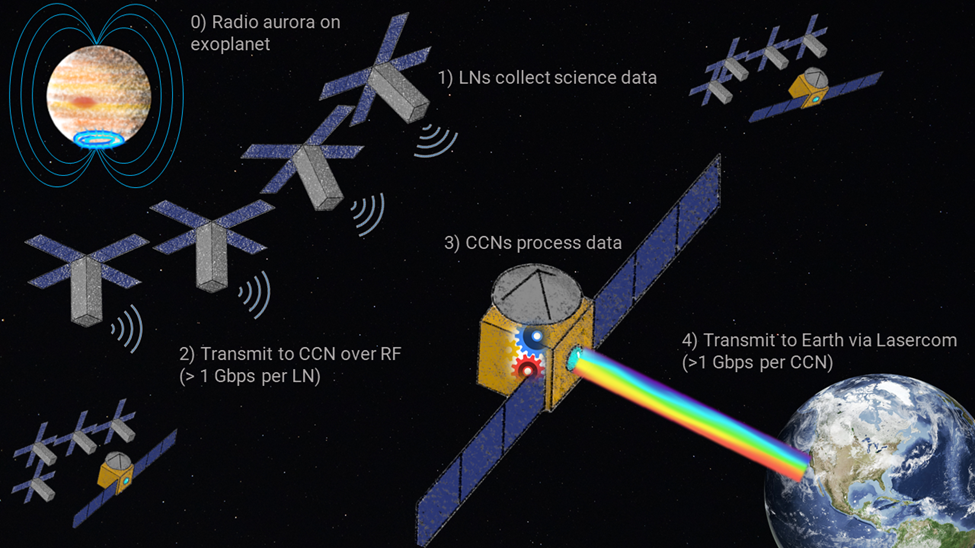}
    \caption{GO-LoW’s hybrid architecture. Small, easily packable Listener Nodes (LNs) form the bulk of the constellation, collecting radio frequency (RF) data from the environment, digitizing it and transmitting it to the more capable Communication \& Computation Nodes (CCNs). CCNs process data via beamforming and then transmit the reduced data to Earth using a lasercom link. The whole constellation will be within the field of view of a single lasercom system on Earth, allowing for multiplexed communications. The beamformed data is cross-correlated and further processed into higher-level science data products on the ground.}
    \label{fig:hybrid-architecture}
\end{figure}

2. \textbf{Computation \& Communication Nodes (CCNs)}, shown in Figure \ref{fig:CCN-diagram}. These ESPA-class spacecraft (about the size of a mini-fridge, $\sim$1 m$^3$) are tasked with \textbf{(a)} collecting and processing data (beamforming) from a local cluster of 100-1,000 LNs, \textbf{(b)} relaying commands back to the LNs, and \textbf{(c)} communicating with Earth, including transmitting science data and receiving high-level commands.

\begin{figure}
     \centering
     \includegraphics[width=1\linewidth]{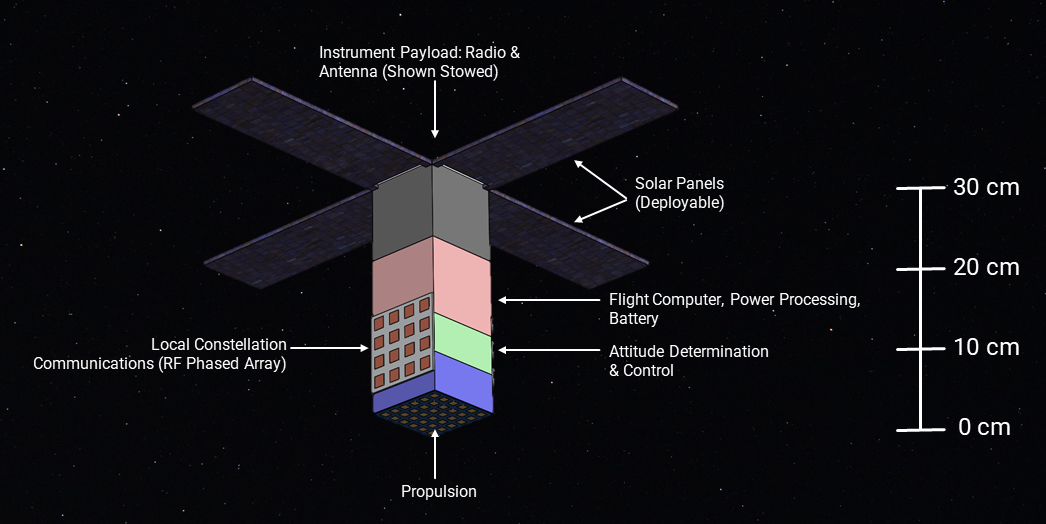}
     \caption{Listener Node (LN): a 3U CubeSat, shown with its science antenna stowed. The major components of the LN are highlighted, including its radio payload, propulsion system, and phased array for local RF communications.}
     \label{fig:LN-stowed}
 \end{figure}

  \begin{wrapfigure}{r}{0.6\textwidth}
      \includegraphics[width=1\linewidth]{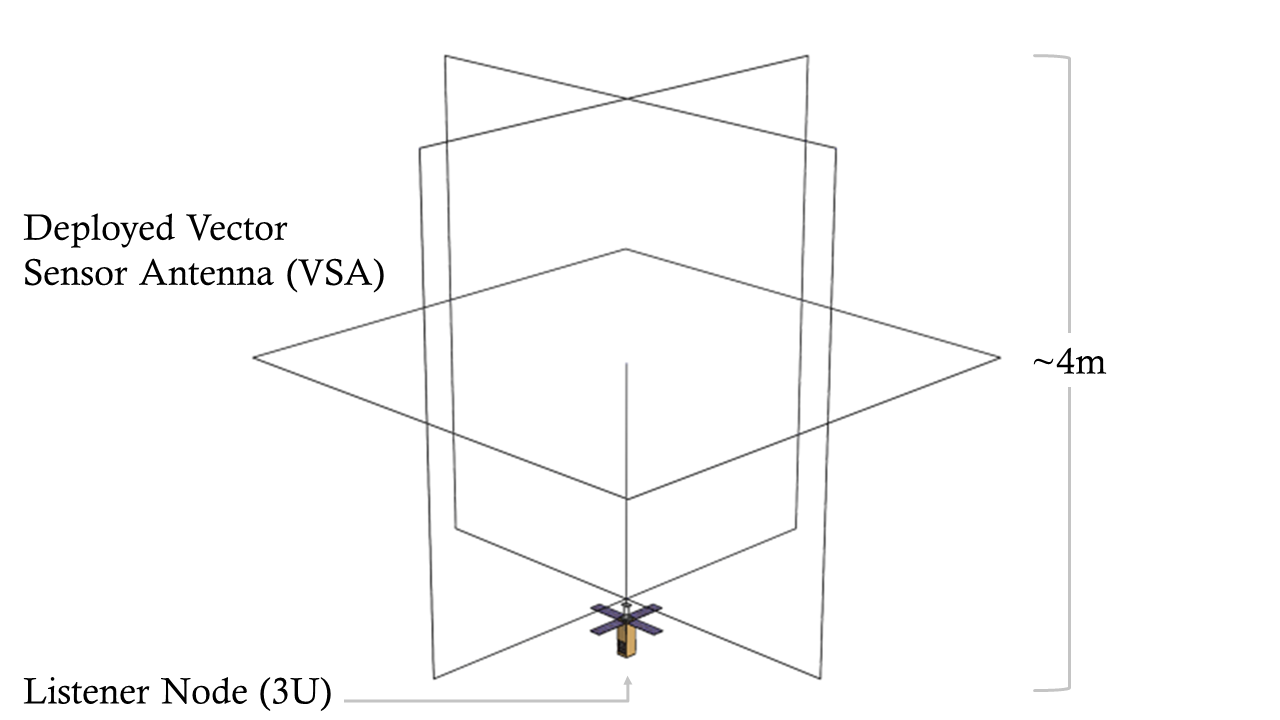}
      \caption{Listener Node (LN) shown with the Vector Sensor Antenna (VSA) fully deployed. The VSA is the primary science sensor of the GO-LoW mission.}
      \label{fig:LN-deployed}
  \end{wrapfigure}
  
\begin{figure}
    \centering
    \includegraphics[width=1\linewidth]{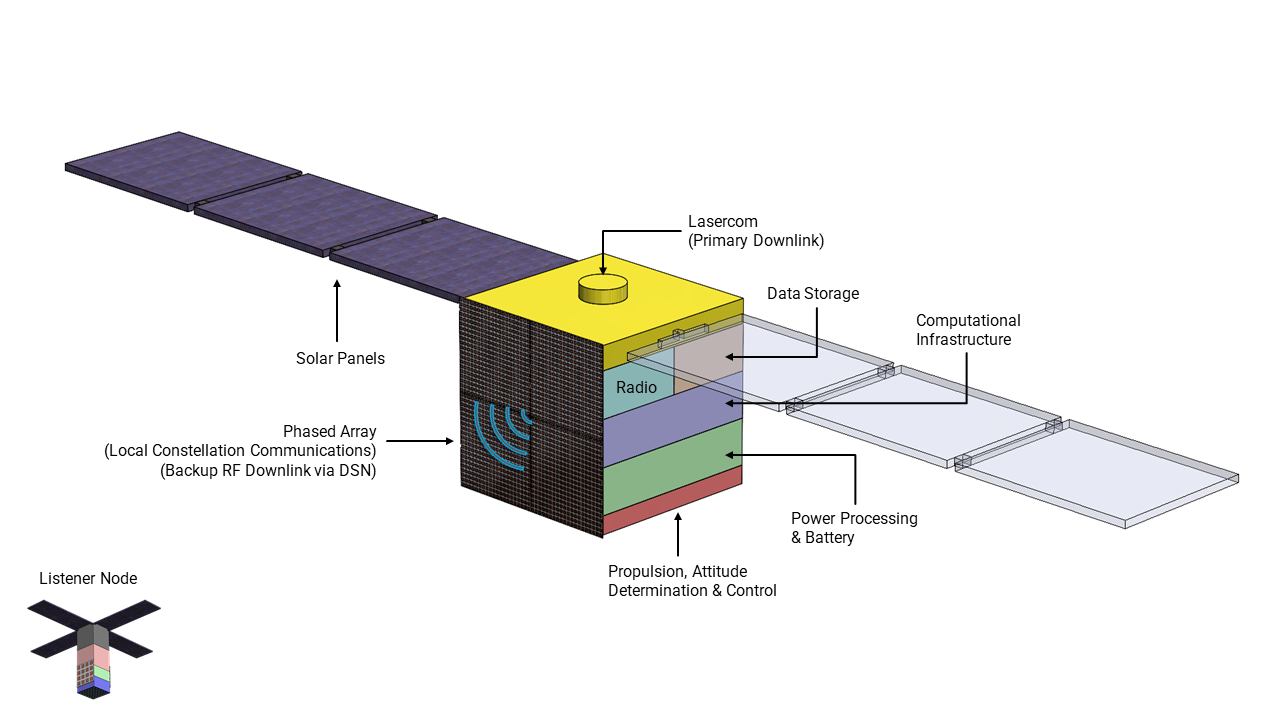}
    \caption{Communication \& Computation Node (CCN) with major subsystems labeled. The CCN is an ESPA-class spacecraft approximately 1m in scale, and is shown with Listener Node (LN) for comparison. }
    \label{fig:CCN-diagram}
\end{figure}
  
Figure \ref{fig:hybrid-architecture} illustrates the architecture outlined in the Phase I study. Science data is generated by the LNs at $\sim$1 Gbps. The raw or minimally processed data is then transmitted to the CCNs via radio-frequency (RF) link. The CCNs reduce the data volume by beamforming all the input LN data and then transmit that beamformed data to Earth via laser link. Commands from Earth may be transmitted by laser link and/or DSN RF communication.

In Section \ref{sec:science_overview}, we describe the science motivation for GO-LoW's representative mission as well as additional science investigations. Section \ref{sec:ccc} describes correlation of radio data and communication within the constellation and to Earth.  Section \ref{sec:mission_architecture} describes the launch and constellation mission architecture.  In Section \ref{sec:techmap}, we describe the technology developments needed to build out GO-LoW.  Finally, Section \ref{sec:conclusion} describes open questions and next steps for continuing to demonstrate the feasibility of GO-LoW.

\clearpage
\section{Science}
    \label{sec:science_overview}
   To showcase the versatility of GO-LoW's science impact, we examined a broad set of science cases. Exoplanet and stellar science (\S\ref{sec:sci_exo-stellar}) drive the science requirements for GO-LoW  because the faint signals from nearby exoplanet radio aurorae are the most challenging in terms of sensitivity.  Exact predictions of exoplanet radio fluxes are very challenging, however (\S\ref{sec:estexoaurora}), due to extremely limited data and wide-open discovery space. 
   
    \subsection{Exoplanet and stellar science} 
        \label{sec:sci_exo-stellar}
        


Magnetic fields are the last ingredient needed to understand the thousands of exoplanets that have been discovered in the last 20 years; magnetized planets are thought to be ubiquitous based on the solar system example (though this is yet to be confirmed). Planetary magnetic fields influence a variety of planetary processes including atmospheric escape \citep{Ramstad2021, Airapetian2020, Gronoff2020}, planet migration \citep{Strugarek2015}, and the photocatalysis and destruction of the chemical building blocks of life \citep{Griessmeier2005, Dartnell2011,Atri2020}. Exoplanet magnetic fields play central roles in extrasolar aurorae \citep{Zarka2001}, which can themselves trace moons \citep{Noyola2014} and their (cryo)volcanism. Finally exoplanet magnetic fields diagnose the fluid properties and structures of the deep interior dynamo regions that generate them \citep{Stevenson2010,Driscoll2011}, for instance by anchoring dynamo predictions \citep{Blaske2021, Driscoll2011, Stamenkovic2012}.  For a more thorough review of the rich science at hand, we refer the interested reader to \citet{Zarka2007, Lazio2018}. 

Studying the magnetic signatures of exoplanets also offer a means to assessing the space weather environments around their host stars.  Terrestrial planets are particularly well-suited to probing the stellar winds of M dwarfs \citep[e.g.,][]{Pineda2023, Kavanagh2023}, which are most likely to host close-in rocky planets and whose wind properties remain almost entirely unconstrained. 

Our own solar system showcases both the diversity of planetary magnetic fields and the only sure means for assessing them in non-transiting exoplanet systems. Better understanding the radio signatures of solar system planets, where we have in-situ measurements of the magnetic field and particle environments to complement radio observations, is an essential step in the process of characterizing exoplanet magnetic fields.  

Charged particles in planetary magnetospheres produce coherent auroral radio emissions \citep{Zarka1998c} at frequencies that are directly proportional to the local magnetic field strength \citep{Treumann2006}. Currently, only auroral radio emission can characterize planetary magnetic fields across the full range of planet mass and orbital distance. Adding planetary radio emissions to the existing arsenal of IR, optical, UV, and X-ray imaging and spectroscopy is essential. 

Direct measurements of exoplanet magnetic fields will drive major advancements in exo\-planet science by: 
\begin{enumerate}
    \item Enabling statistical studies of the relationship between planet properties and magnetic fields.
    \item Directly testing magnetic field predictions from exoplanet dynamo models, thereby driving their development with many more data points than solar system planets can provide. 
    \item Broadening and enhancing the study exo-auroral physics, a young field that currently relies on brown dwarfs as test beds \citep{Hallinan2015, Kao2016, Pineda2017}.
    \item Providing inputs into exoplanet atmospheric evolution modeling, informing habitability and astrobiology studies.
    \item Probing exoplanet interior conditions: all magnetized exoplanets must have a vigorously convecting conductive fluid layer in their interiors in order to sustain a dynamo field; measuring a magnetic field therefore constrains interior structure/composition and heat flow.
\end{enumerate}

\subsubsection{Estimating exoplanet auroral emission brightness} \label{sec:estexoaurora}
Properly accounting for the large uncertainty in expected auroral radio emissions from exoplanets is a major challenge in sizing GO-LoW to meet its primary science requirement of surveying nearby exoplanet systems. 

The wide range of predictions for nearby exoplanet systems depend on assumptions: planetary magnetic field strength, stellar magnetic field and spatially dependent wind parameters, space weather, and so on. Even for well-studied systems that have measured or modeled values for some of these parameters, the flux predictions remain model-dependent.  Without a confirmed direct detection of radio emission from an exoplanet, models used to make predictions are unvalidated outside of the solar system.  On the other hand, simply scaling solar system planet auroral radio fluxes out to 5 or 10 parsecs does not account for the diversity of stellar systems (age, stellar type, stellar activity, planet size, orbit, composition, formation, etc.).  In the age of large exoplanet surveys from Kepler, TESS, and ground-based instruments, the solar system is no longer the only or best model for extrapolations.  

Exoplanet radio flux predictions based on extreme stellar wind or space weather may overestimate fluxes, while scaling solar system planets may be overly conservative.  Instead, we take a middle road and specified a constellation with 100,000 nodes and sensitivity between 1 mJy and 100 $\mu$Jy in 24 hours (10--100 $\mu$Jy in 2500 hours, Figure \ref{fig:sensitivity}).  Such an array will detect many nearby planets given moderately enhanced stellar wind and space weather conditions.  GO-LoW requirements will continue to be refined as we learn more about exoplanetary magnetic fields and radio emission from ground-based measurements (LOFAR \citep{van2013lofar, Edler2021}, NenuFAR \citep{Zarka2015}, SKA \citep{Dewdney2009}, OVRO  LWA \citep{Hallinan2014}).

\pagebreak

\subsection{Science Traceability Matrix}
        \label{sec:STM}
        GO-LoW is a general purpose scientific observatory and therefore it serves multiple science cases.  The primary science case studied here is for stars and exoplanets (summarized in \S\ref{sec:sci_exo-stellar}), since these sensitivity requirements are expected to drive key design decisions.  In \S\ref{sec:othersci}, we summarize several other science cases and their key requirements.  See \emph{e.g.}, \citet{Cecconi2018} for additional details.  Much work remains to be done to refine the values in the STM; some of this work is ongoing in the scientific community as part of other efforts and some could be continued in a NIAC Phase II study, if awarded.

\clearpage
\begin{landscape}
\begin{figure}
    \centering
    \includegraphics[trim = 0.15in 0.85in 0.15in 0.75in, clip, width=1\linewidth]{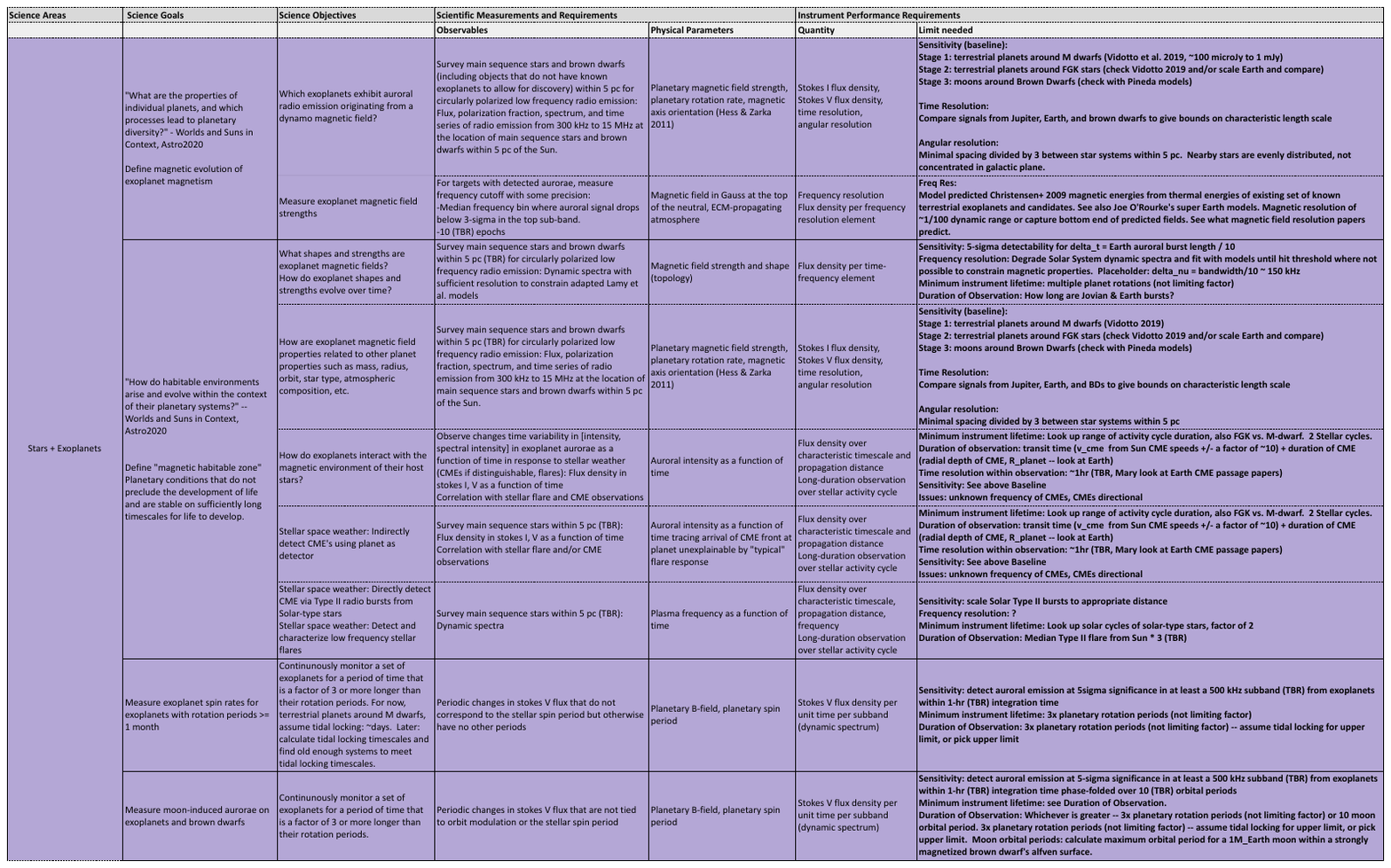}
    \caption{Science Traceability Matrix outline for Stars \& Exoplanets science case.  In this study, we laid out key surveys and targeted observations for stars an exoplanets.  The "Limits" column of the STM describes future investigations that are needed to set requirements based on each of these science objectives.}
    \label{fig:stm-exo}
\end{figure}
\end{landscape}

\subsection{Additional science domains} \label{sec:othersci}

\subsubsection{Heliophysics and solar system planets} \label{sec:helioplanet}
Exoplanet science starts close to home.  Developing our understanding of the solar weather environment and its effect on more easily studied solar system planets will inform exoplanet radio science in the next decades.  There are a plethora of intrinsically interesting science investigations that GO-LoW enables within our own solar system, showing that GO-LoW is a truly cross-disciplinary Great Observatory.

\paragraph{Heliophysics} As with cosmology (\S\ref{sec:cosmology}), the capabilities of GO-LoW scale as the sensitivity and resolution of the array increase.  Solar radio bursts are very bright and therefore observable with a small number of spacecraft ($<$ 10).  SunRISE \citep{Kasper2021}, currently readying for launch, is a constellation of 6 spacecraft at a near-geostationary orbit that will track solar radio bursts as they propagate away from the sun. GO-LoW would offer (1) a different vantage point at L4 or L5, viewing the sun from either 60 degrees ahead or behind the Earth; (2) significantly increased sensitivity to track weaker bursts (\emph{e.g.}, \citep{Sharma2022}) and perhaps even nanoflares \citep{Parker1988} and associated radio emission; and (3) much higher angular resolution (see Table \ref{tab:keyparam}), allowing for extremely precise localization and tracking of solar radio bursts as they propagate through the interplanetary medium.

\paragraph{Gas giant auroral emission} Gas giant (Jupiter, Saturn) auroral radio emission, as well as terrestrial radio emission, is similarly observable with a small number of spacecraft (fewer than 10, Figure \ref{fig:sensitivity}.  Longer baselines ($>$ 1000 km) would enable spot mapping, or localizing auroral emission to scales smaller than the diameter of these planets.  Observing the ice giants (Uranus, Neptune) requires a larger, but still modest, constellation of $\sim$100 spacecraft to obtain the required sensitivity at appropriately short integration times and bandwidth.  Solar system planetary aurorae are dynamic in both total and spectral intensity, so relatively short integration times and narrower bandwidths are needed to ensure that interesting features are not washed out by integration.  Monitoring the auroral spectra of all solar system planets would provide an unprecedented dataset that links planetary processes with heliophysics.  This dataset could be used to test and refine our understanding of how gas giant/ice giant magnetospheres respond to solar wind events such as CMEs and how CMEs propagate and evolve beyond 1 AU.  Studying the aurorae of our solar system's ice giants (Uranus, Neptune) will provide insight into similar exoplanets as well.

\paragraph{Planetary lightning} GO-LoW will also be capable of detecting lightning signatures from solar system planets.  \citet{Zarka2008b} summarizes current knowledge of lightning detections from solar system planets, noting that Venus is a particularly interesting target because the existence of Venusian lighting is disputed and could have important implications for atmospheric chemistry (\emph{e.g.}, \citet{Delitsky2015}). \citet{griessmeier_future_2018} notes that ``In the solar system, [planetary lightning] emission is considerably (several orders of magnitude) weaker than the auroral radio emission."  GO-LoW, at full scale, would be able to detect lightning emission from any planet within the solar system if it occurs within GO-LoW's frequency band.

\subsubsection{Sky mapping and interstellar medium} \label{sec:skymapping}
\citet{Cong2021} have produced simulated maps of the low frequency sky at 1, 3, and 10 MHz.  GO-LoW would produce a set of sky maps that could be compared to these predictions.  Deviations from predictions would highlight gaps in our knowledge about the structure of the interstellar medium within and above/below the galactic plane.  GO-LoW's mapping capability would substantially improve on RAE-2 \citep{Novaco1978, Alexander1974} well before the constellation reaches its full size of 100,000 nodes; with a 500 km diameter, GO-LoW could achieve resolution of 8.25 arcsec at 15 MHz, 2 arcmin at 1 MHz, and 5 arcmin at 500 kHz (not accounting for scattering).  As GO-LoW grows, sensitivity will increase both due to the addition of nodes and accumulated integration time, producing deeper maps of the low-frequency sky.  These maps will contribute to understanding and removing the foreground for 21-cm cosmology science (\ref{sec:cosmology}).  See \citet{Cecconi2016} for further discussion of all-sky mapping science goals.

\citet[sect.~4.2.5, fig.~5]{Jester2009} note that angular scattering and temporal broadening will limit the obtainable resolution for all-sky maps at GO-LoW frequencies.  All-sky maps will test these scattering predictions for both the interstellar and interplanetary media.  The scattering constraints, as we currently understand them, mean that GO-LoW has a natural bound on its longest useful baseline; maximum resolution will be set by nature rather than being a key requirement imposed by any science case.  As sky maps at sub 10 MHz frequencies are produced by the growing GO-LoW constellation, scattering effects will become better understood and the maximum practical baselines for GO-LoW will be refined.  Of course, a free-flying constellation can adjust the placement of nodes at any time to modify the distribution of baseline lengths.

\subsubsection{21-cm cosmology} \label{sec:cosmology}
The impact of a space-based low-frequency array for 21-cm early universe cosmology has been discussed in detail elsewhere (\emph{e.g.}, \citet{Cecconi2018, Burns2019, Jester2009, Pritchard2012}) and is therefore not a focus of this study.  

Studies of the early universe rely on the highly redshifted 21 cm neutral hydrogen ``spin-flip" line to track the distribution and state of neutral hydrogen as the universe evolved after the Big Bang.  There are two key goals for cosmic Dark Ages cosmology: (1) observe the global, or monopole, sky-averaged spectrum and (2) observe the spatial power spectrum of 21 cm hydrogen emission as a function of frequency (which corresponds to redshift).  The global signal measurement provides a test for predictions of the absorption spectrum as a function of time/redshift, while the spatial power spectrum provides a 3D reconstruction of how matter was distributed in the early universe.

As described in \citet{Burns2019}, the removal of foregrounds and the chromaticity of the antenna are key challenges for measuring the global 21-cm spectrum.  This calibration challenge will be investigated in the next phase of GO-LoW's concept development.

\subsubsection{Additional science applications}
There are additional science cases beyond those listed above, including transient events and pulsars.  See \emph{e.g.}, \citet{Jester2009, Cecconi2018, Burns2021lunarreview}.

\subsection{Science return from a growing constellation} 
\label{sec:const-growth}

\begin{figure}[hbt!]
    \centering
    \includegraphics[width=\textwidth]{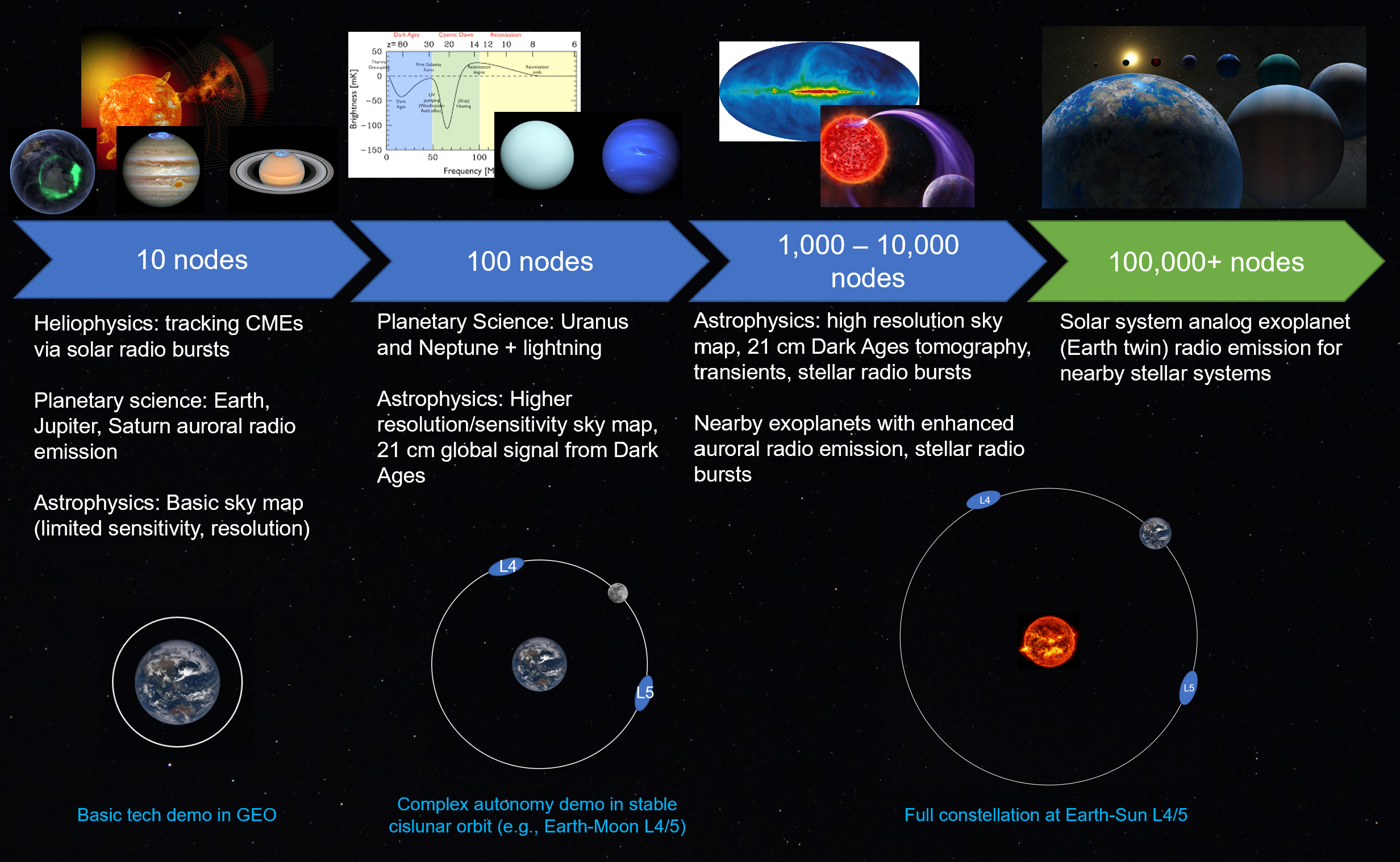}
    \caption{A low-frequency radio interferometric constellation is useful at a range of scales and can grow over time. The science cases listed are not exhaustive; for example, solar system science will continue to be relevant and improve in capability as constellation size grows. The bottom diagrams indicate potential locations for constellations as the number of nodes grows. Small constellations are already proposed for GEO; larger ($\sim$100 nodes) constellations could take advantage of stable cislunar geocentric orbits for a less crowded and radio noisy environment while remaining within relatively easy communications range. The full GO-LoW constellation is best suited to a more distant location (Earth-Sun L4/5) to avoid traffic management issues and to ensure a pristine radio-quiet environment.}
    \label{fig:ConstProg}
\end{figure}

GO-LoW will open a new electromagnetic spectral remote sensing window --- one of the last yet unexplored. There are applications for this capability across all NASA science directorates. Most importantly, GO-LoW will be scientifically useful at all scales: it does not need to reach the full strength envisioned in the representative mission to provide high-impact science return (\ref{fig:ConstProg}).  Missions already in development highlight the value of a tiny low-frequency constellation \citep{Kasper2021, Sundkvist2016, Erickson2018, Lind2019, SWIPE}. These small constellations (n$<$10) do not require the autonomy planned for GO-LoW, but they demonstrate the science benefits to heliophysics and planetary science of low-frequency sensing. A larger constellation of $\sim$100 nodes increases both sensitivity (see \ref{fig:sensitivity}) and angular resolution, opening up the potential for comprehensive solar system monitoring, significantly improved all-sky maps, and potentially early 21-cm Dark Ages cosmology results. At this scale, autonomy becomes useful for streamlining operations. Sensitivity and angular resolution continue to improve as the constellation grows beyond 1,000 nodes: in this range, exoplanetary radio emission from nearby systems with extreme forcing may be detectable. Autonomy is essential at this size. 

The full-strength constellation, GO-LoW’s representative mission, is designed to be capable of detecting radio emission from exoplanetary systems that are similar to the solar system. Exoplanets with magnetic field strength similar to Earth ($<$ 1 Gauss) emit at frequencies too low to detect on the ground since the cyclotron maser emission frequency decreases proportionally with magnetic field strength (\emph{e.g.}, \citet{Farrell1999, Zarka2008}). Assuming that at least some Earth-sized planets have magnetic field strengths similar to Earth, only a space-based high-sensitivity radio telescope can measure them. \textbf{GO-LoW complements other instruments like JWST in the campaign to characterize our neighboring worlds and understand their potential to support life}.  At every scale, GO-LoW will produce new views of the sky and of planets near and far that will engage the public in space exploration.

\subsection{Reconfigurability} \label{sec:reconfigure}
\begin{figure}
    \centering
    \begin{subfigure}[t]{1\textwidth}
    \centering
    \includegraphics[width=1\linewidth]{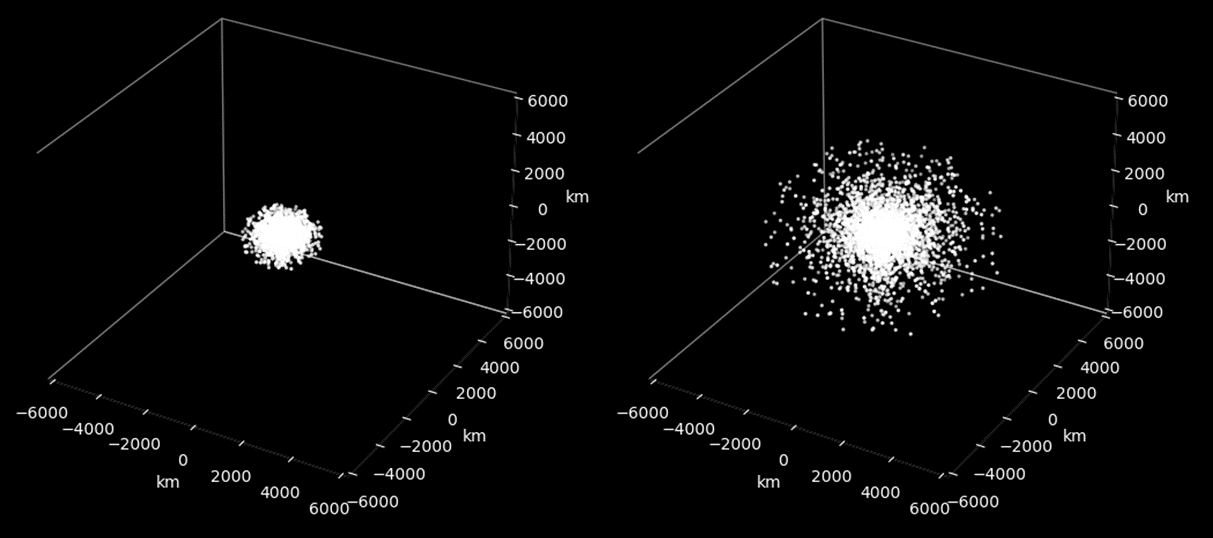}
    \caption{Steps 1 and 2 of array expansion}
    \label{fig:reconfig1}
    \end{subfigure}

    \begin{subfigure}[t]{1\textwidth}
        \centering
    \includegraphics[width=1\linewidth]{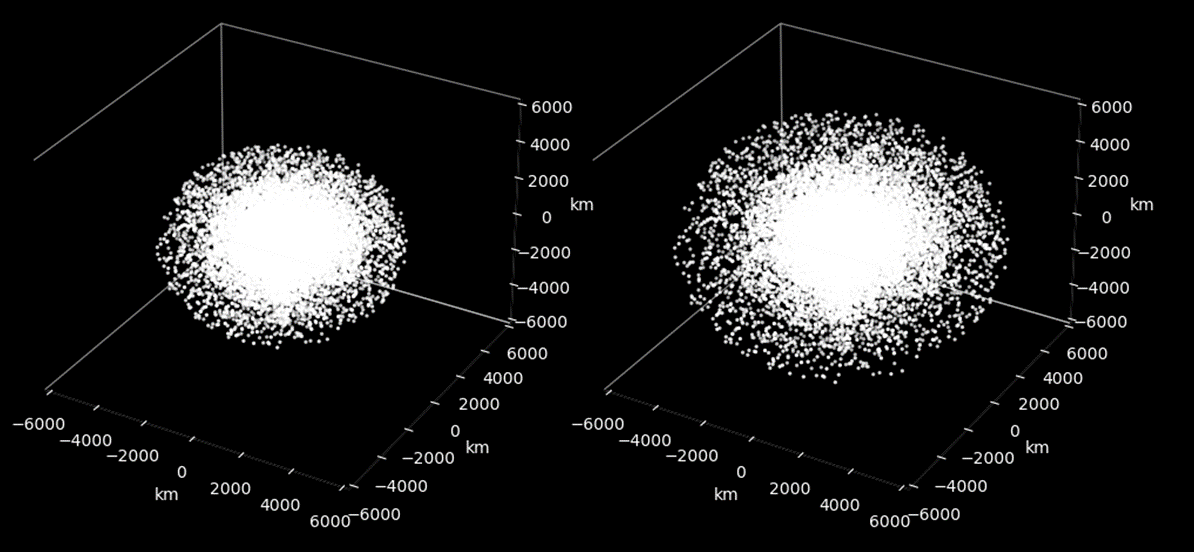}
    \caption{Steps 3 and 4 of array expansion}
    \label{fig:reconfig2}
    \end{subfigure}

\caption{Graduated expansion from a tightly packed core to a fully extended array.}\label{fig:reconfig_multi1}
\end{figure}

\begin{figure}
    \begin{subfigure}[t]{1\textwidth}
        \centering
    \includegraphics[width=1\linewidth]{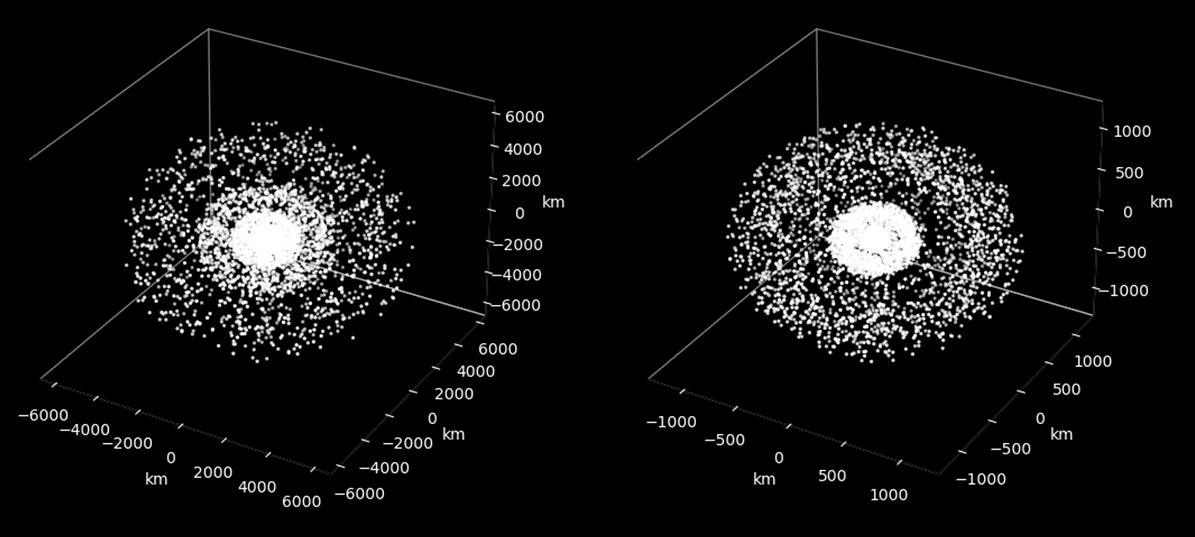}
    \caption{}
    \label{fig:reconfig3}
    \end{subfigure}

    \begin{subfigure}[t]{1\textwidth}
        \centering
    \includegraphics[width=1\linewidth]{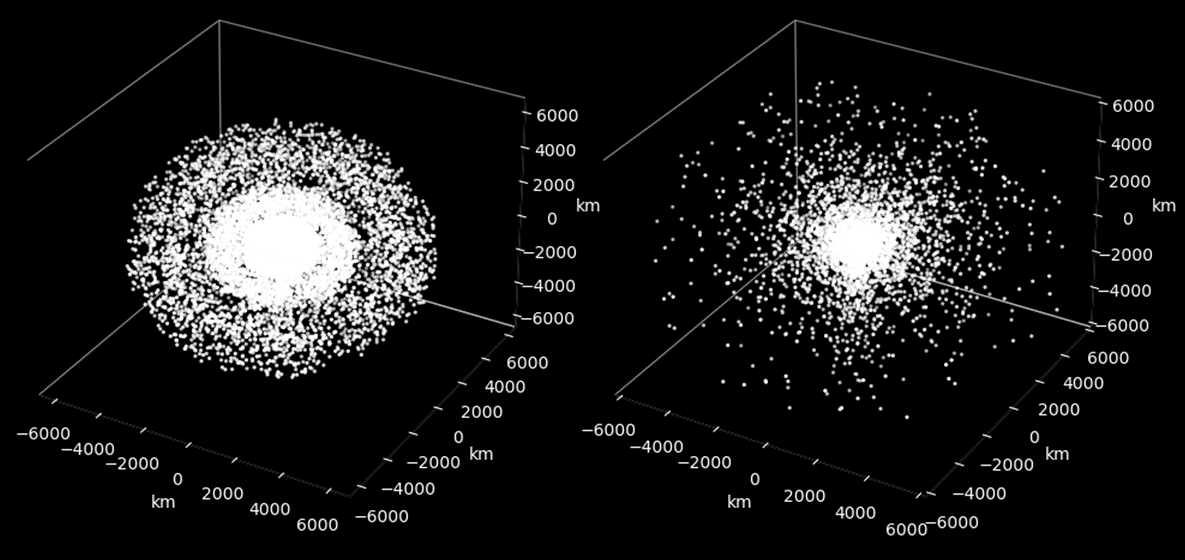}
    \caption{}
    \label{fig:reconfig4}
    \end{subfigure}

\caption{A range of configurations for GO-LoW, ranging from a stepped set of shells \ref{fig:reconfig3} to a dense core with outriggers \ref{fig:reconfig4}, right.}\label{fig:reconfig_multi2}
\end{figure}

A key advantage of GO-LoW compared to surface arrays (see \S\ref{sec:compare_lunar}) is that it can be reconfigured as needed for different science cases.  As with the VLA, science data collection can continue during reconfiguration.  Figures \ref{fig:reconfig_multi1}, \ref{fig:reconfig_multi2} shows a range of possible configurations, including growth from a small core (Figure \ref{fig:reconfig_multi1}) to an evenly filled configuration like DSA2000 radio camera \citep{hallinan2019dsa}, a dense core with outriggers like OVRO LWA \citep{Hallinan2014} or LOFAR \citep{van2013lofar} (Figure \ref{fig:reconfig4}, right).

The following figures were generated for a few thousand nodes.  A next step will be to model the full number of nodes and to simulate CCN beamforming and then cross correlation of the beamformed data.  When we simulate CCN beamforming clusters, we will also generate a baseline map between the chosen phase center of each cluster and and every other cluster.  This vector map can then be used to generate a 3D point spread function (PSF) that can be used for imaging [Kononov, 2024, in prep.\footnote{Thesis work, expected completion May 2024.}].

\clearpage
\section{Collecting Data: Sensitivity, Antenna Choice, and Calibration}
    \subsection{Sensitivity calculation}
        \label{sec:sensitivity_calc}
        
The sensitivity of an interferometric array depends on that of  an individual receiver element and the total number of receiver elements. After deriving the sensitivity of an individual element, scaling with the total number of elements is straightforward if all elements can be treated as identical. More detailed modeling to account for near field interactions between antennas is possible (e.g., \citet{Ellingson2011}), but beyond the scope of this study.

\cite{kononov_sensitivity_2024} derives the sensitivity for a single vector sensor. Briefly, a vector sensor's system equivalent flux density (SEFD) is found by beamforming the constituent elements and evaluating the improvement in signal-to-noise ratio at the beamformer output. \citet{kononov_sensitivity_2024} used AERO-VISTA's dual element vector sensor implementation as a case study, and we applied their method in our NIAC Phase I work to a re-designed version that is capable of higher sensitivity.

\begin{figure}[hbt!]
    \centering
    \includegraphics[width=\textwidth]{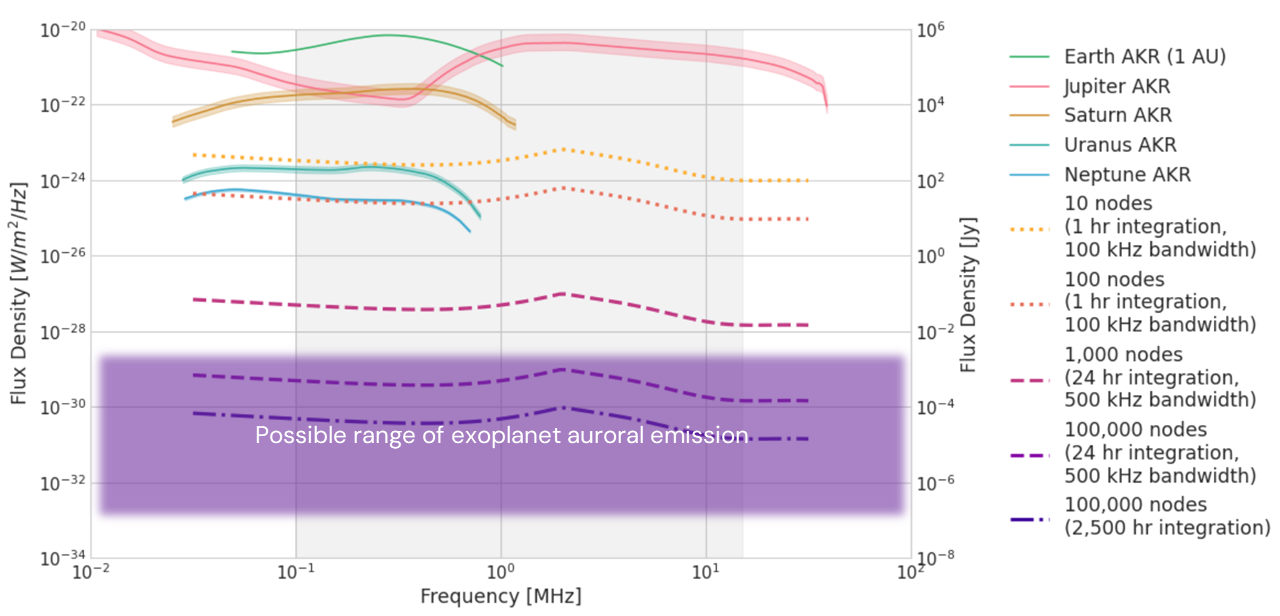}
    \caption{Sensitivity analysis with ideal electrically small vector sensor antennas. The SEFD for a single baseline is calculated then scaled by time/bandwidth (see legend) and then by number of nodes in the constellation; the shape of the dotted and dashed curves is set by the combination of antenna response and sky noise background. The dark purple dash-dot curve shows the sensitivity improvement for 2,500 hours of integration rather than 24 hours. Solar system planet auroral emissions with intensity ranges are shown as seen from Earth-Sun L4/5 \citep{Zarka2012}, data available \citep{ZarkadataSS}. The purple box represents the range of possible exoplanet auroral emission levels in the solar neighborhood. As described in \citet{Vidotto2019}, strong stellar winds and/or CMEs may significantly enhance radio emission, setting the upper edge of the box. Other scaling predictions span a wider range (\emph{e.g.}, \citet{griessmeier_future_2018}) while conservatively scaling solar system planet auroral emission to 5 pc sets the bottom of the box.}
    \label{fig:sensitivity}
\end{figure}

The sensitivity of an interferometer ($S_\text{rms}$) is determined by four factors: per-element SEFD, number of nodes ($N$), and integration in time ($t$) and bandwidth ($\Delta v$): 
\begin{equation}\label{eq:sensitivity}
    S_\text{rms} = \frac{\text{SEFD}}{\sqrt{N(N-1)/2t\Delta v}}
\end{equation}

Figure \ref{fig:sensitivity} shows the sensitivity of an array composed of 10, 100, 1000, and 100,000 vector sensor nodes; the last is GO-LoW’s representative mission. Smaller constellation sizes (10, 100) are shown with shorter integration times and smaller bandwidths more appropriate to heliophysics and solar system science. Larger array sizes (1000; 100,000) use 24 hours and 500 kHz bandwidth as the standard integrations to highlight the achievable sensitivity. A 2500 hour integration time is shown for direct comparison to FARSIDE \citep{Burns2019}, a proposed lunar surface low frequency interferometer.

In Figure \ref{fig:sensitivity}, the SEFD for a single baseline is calculated then scaled by time/bandwidth (see legend) and number of constellation nodes. The dark purple dash-dot curve shows the sensitivity improvement for 2500 hours of integration compared to 24 hours. Solar system planet auroral emissions with intensity ranges are shown as seen from Earth-Sun L4/5. The purple box represents the range of possible exoplanet auroral emission levels in the solar neighborhood. As described in \citet{Vidotto2019}, strong stellar winds and/or CMEs may significantly enhance radio emission, setting the upper edge of the box. Other scaling predictions span a wider range (\emph{e.g.}, \citet{griessmeier_future_2018}) while conservatively scaling solar system planet auroral emission to 5 pc sets the bottom edge.  Particularly extreme systems may show even stronger flux than indicated by the purple region on this plot.  \citet{Turner2021} reported a tentative detection of $>$400 mJy flux from tau Boo (15 pc); \citet{Vidotto2017} predict flux of 6--24 mJy from a young system much farther away (V830 Tau, 130 pc).

    \subsection{Antenna options trade study}
        \label{sec:antenna_options}
        We considered several antenna design parameters and evaluated several antenna types against those parameters. We scored each antenna type as good, medium, or poor in each of the parameters, and weighted the parameters according to their importance to the mission success. This method yielded a stoplight chart (Figure \ref{fig:antenna_stoplight}), and the best total score identified the type of antenna most suitable for this mission.

\subsubsection{Parameters}
\paragraph*{Directionality} Narrower beam is better for interference tolerance (interfering sources not in the beam are de-emphasized) and flat sky approximation in imaging. Good:$<10^\circ$, medium: $10^\circ$ - $90^\circ$, poor: $>90^\circ$.

\paragraph*{Sensitivity} Related to radiation efficiency and resonance. Calculated in terms of System Equivalent Flux Density (SEFD); low SEFD means lower noise and is therefore better. This parameter is weighted highly because it determines how many spacecraft would be needed to fulfill the mission.

\paragraph*{Bandwidth} Wide bandwidth is desired because signals of interest can occur over wide range of frequencies, 0.1 – 15 MHz. Good: $<20\%$, medium: $20-100\%$ 20-100\%, poor: $>$100\%.

\paragraph*{Polarimetric capabilities} Polarimetric measurements are desired for key science cases.

\paragraph*{Technology readiness level} Higher TRL, especially technologies with flight heritage, pose a lower risk and lower development cost. Good: $<$3, medium: 4-6, poor: $>$7.

\paragraph*{Size weight and power (SWaP)} should be minimized – relates to feasibility and launch cost. This parameter is weighted highly because the weight impacts the total launch mass and feasibility greatly due to the large number of spacecraft needed. Good: $<$8 kg, medium: 8-50 kg, poor: $>$50 kg.

\paragraph*{Cost} Production and raw materials cost is considered. The ability to make it automated versus needing technician labor. Non-recurring engineering cost is less important in the context of producing thousands of units.

\subsubsection{Antenna types}
\paragraph*{Electrically small crossed dipoles} Considering the already high TRL for dipoles and tripoles, there is little development work left to do.  Crossed dipoles had to be low cost to be deployed by the thousands in surface arrays (e.g., LOFAR \cite{van2013lofar}, LWA \citep{ellingson2009long}). An active impedance analog front end circuit can improve the sensitivity of electrically short antennas but it is difficult to design. Electrically small dipoles have a directivity of 1.7 dBi and a toroidal beam pattern \citep{balanis_antenna_2016}. Electrically short antennas are inefficient because their radiation resistance is often much lower than their ohmic loss. Electrically short dipoles and loops have been shown to have stable reception patterns for broad frequency ranges over several octaves \citep{robey_high_2016}. Front/back ambiguity assuming a ground plane is not used. Polarization aberration at off-boresight angles.  Monopole/dipole/tripole type antennas are a low deployment risk because they are mechanically simple and have significant heritage, with TRL 9. 

\paragraph*{Electrically small tripoles} There is more cost than crossed dipoles associated with the 3D deployment mechanism and 3-channel receiver. Front/back ambiguity but no polarization abberation. Based on photos of the Longjiang \citep{yan2023ultra} and NCLE \citep{arts2019design} payloads, they look small and flight heritage gives it TRL of 8 or 9.

\paragraph*{Electrically small vector sensor} The AERO-VISTA prototype demonstrates progress but work remains in de-risking and manufacturability. There is more cost than tripoles associated with the loop deployment mechanism and 6-channel receiver. There is no polarization aberration or direction of arrival ambiguity. A conventionally beamformed vector sensor will produce a 135 degree HPBW \citep{nehorai_minimumnoisevariance_1999}. In certain situations optimal beamforming can produce a much narrower beam, but in general, a beamformed vector sensor has a nearly hemispherical beam. The loops add complexity and risk of entanglement to the deployment. There is no space heritage, but the AERO-VISTA prototype places it at TRL 6. The loops, additional support structure, and actuators add mass above that of a tripole.

\paragraph*{Resonant crossed dipole and tripole} These are wavelength-scale antennas. A key development challenge would be miniaturization for a small platform. Large dipoles have at best 5 dBi directive gain, and toroidal beam pattern \cite{balanis_antenna_2016}.  The high efficiency in the resonant region is achieved through impedance matching and is only available in a narrow range about the resonant frequency.  Long dipoles/triples have the same polarization and directional degeneracy issues as short ones.  Flight heritage on RAE-1 and -2 \cite{weber_radio_1971, alexander_scientific_1975}, Voyager-1 and -2 \cite{warwick1977planetary} gives it TRL 9.  Mass associated with about 300 meters of antenna wire would be significant.

\paragraph*{Dish reflector, folding or inflatable} The parabolic dish antenna would need to be many wavelengths in diameter to achieve directive gain. The bandwidth would be limited by the feed, which are usually horn-like and serve up to an octave of bandwidth. The dish would need $O(\lambda^2)$ material, possibly $O(\lambda^3)$ inflation gas.   But this could be a coarse wire mesh due to low frequency.  It would be several kilometers in diameter, which would amount to a large quantity of material.  Missions: folding: RadioAstron \cite{kardashev_radioastron_2013}, or NISAR \cite{kellogg2020nasa}.  inflatable: STS-77 \cite{freeland1997large}.

\paragraph*{Patch antenna}  While patch antennas are popular for satcom in S and X bands, it would need significant development to create a deployment method for scaling it to low frequency.  Typical 3 dB beam width is 60 degrees with directivity 8-9 dBi. 
The basic patch has a bandwidth of 1-3\%. It can be increased up to 50\% using multiple layers or complex shapes like spirals and fractals, but we assume basic variant for evaluating cost and complexity.  A patch can receive both polarizations and they can be separated at the feed.  The size of the patch would be $O(\lambda^2)$, and it would need a ground plane just as large with substrate between them - see studies on solar sail propulsion for state of the art thin/lightweight materials.

\paragraph*{Horn and corrugated horn variants}  Horns can receive both polarizations, which are separated by an orthomode transducer coupled to the feed.  The horn is a standard component used in many communication and radar systems and is typically non-deployable.  The size of the horn's opening would be $O(\lambda^2)$ for the lowest frequency because of the sharp cutoff on the low frequency end. Smaller than a dish but still impractical at long wavelength scale. 

\paragraph*{Quadrifilar helix (QFH} These can be constructed cheaply (hobbyist versions from PVC pipe and copper wire exist), but the scaled up deployable space version will need significant technician assembly, which is prohibitive considering that up to 100,000 units will be needed. A QFH senses polarization via RHCP and LHCP beamformers: a right-wound QFH senses RHCP, but phasing the feeds in reverse allows it to sense LHCP.  Long helix: The HPBW is about 30 degrees and directivity about 15 dB. Compact helix has a hemispherical beam.  Bandwidth can range 10-70\% depending on design parameters, but is at tradeoff with directionality and efficiency. Helical antennas are available CoTS for UHF and above, and are very popular for satcom. GATOSS and OreSat are cubesats with deployable helical antennas. However, it would need to be scaled to low frequencies. The helix dimensions have to be $O(\lambda)$. The quadrifilar can be fed differentially, eliminating the requirement for a ground plane. Mass of the 4 conductors plus support structure will be no less than that of the resonant tripoles.

\paragraph*{Log-periodic}  Available CoTS for terrestrial systems at frequencies UFH and above, but significant work would be needed to scale it and make it deployable for low frequencies. Like the helical, raw materials may be cheap, but it will need significant assembly. 50-90 degree beam with 7-11 dBi directivity. Very wide bandwidth is a prominent benefit. Polarimetry with log-periodic would have the same ambiguities as crossed dipoles, but mitigated by the directional beampattern. The FreeFall Aerospace prototype plus L3 CoTS deployable put this at TRL 6, but cautiously because their frequency is no lower than VHF/UHF.  Since log-periodic is an array of half-wavelength dipoles fed in parallel, the mass would be several times that of the resonant dipoles, plus necessary support structure.

\subsubsection{Results}
\begin{figure}
    \centering
    \includegraphics[width=0.8\textwidth]{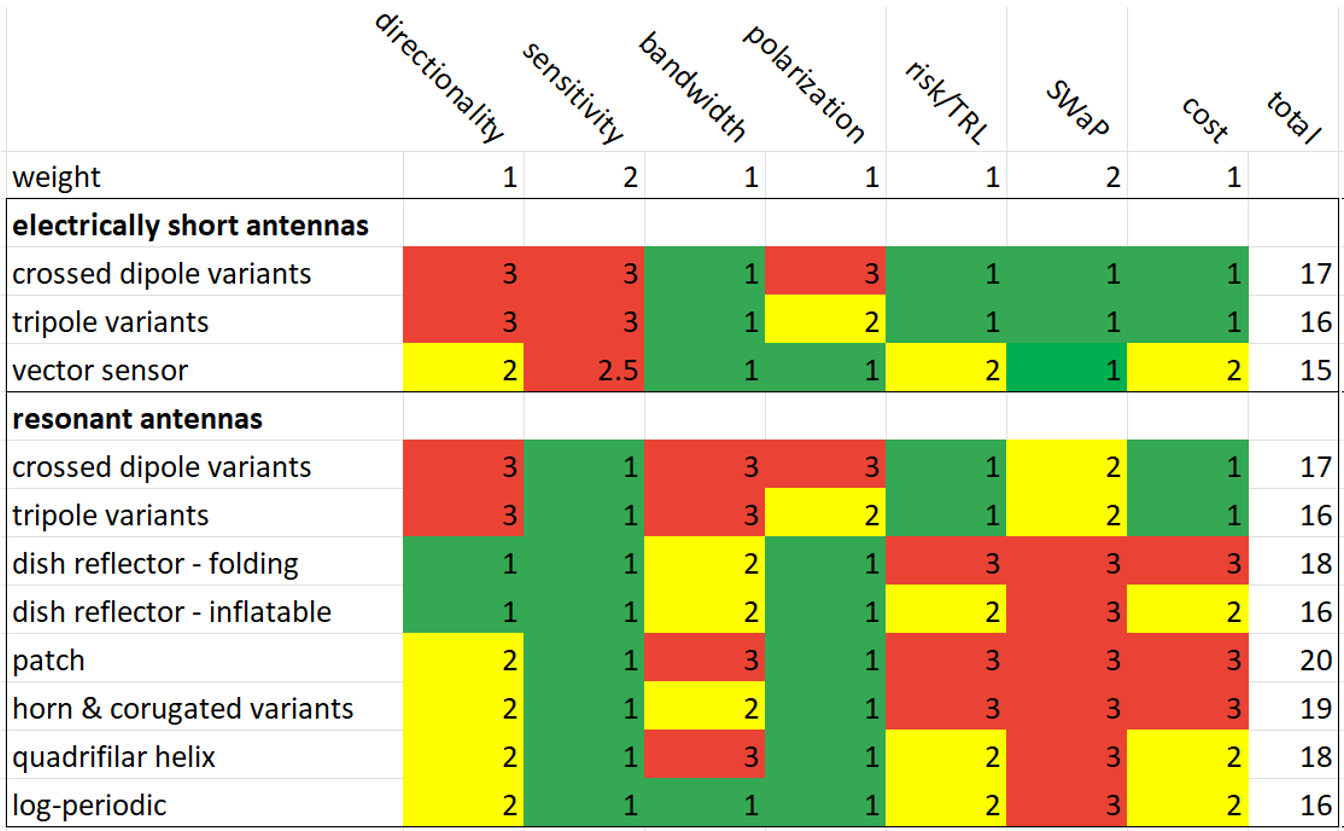}
    \caption{Stoplight chart of antenna options.  The vector sensor antenna had the best (lowest) score.}
    \label{fig:antenna_stoplight}
\end{figure}

The table in Figure \ref{fig:antenna_stoplight} summarizes the scores for each antenna in each parameter. \textbf{The analysis determined that vector sensor is the best antenna for the mission}. It is electrically small so it has a good chance to fit within the SWap budget, and its six channels improve the directionality, sensitivity, and polarization properties over its tripole counterparts. 

    \subsection{Ranging and calibration}
        \label{sec:rangecal}
        An interferometer functions by measuring the position-dependent phase differences in plane waves propagating across the array.  In order to do this, the vector between each node in the array must be known with precision of $\sim$1/16th of a wavelength \citep{TMS}.  With that vector, one can calculate the geometric phase delay along a baseline for any direction and frequency.  In order to turn the geometric phase delay information into scientifically useful data, other phase delays and amplitude differences within the system must be estimated and removed from the data.  \citep[Ch.~10]{TMS} describes this process for ground-based arrays.  Some adjustments are required for a free-flying space-based array.

\subsubsection{Ranging for baseline measurement} \label{sec:ranging}
At 15 MHz, 1/16th of a 20 m wavelength is 1.25 m.  This is the allowable error in the measurement of each baseline vector in the GO-LoW constellation.  Traditionally, the position of deep-space spacecraft is delta-differential one-way ranging (delta-DOR, \citep{book2011delta}).  Given the high demands on DSN \citep{DSNoverburdened} and the frequent updates GO-LoW would need to account for moving spacecraft, we chose to pursue an internal ranging scheme for GO-LoW.  

Constellations in cislunar space have access to the GNSS constellation, either directly if in low Earth orbit or via the backlobes (\emph{e.g.}, SunRISE, \citet{Alibay2017, Hegedus2019}).  At 1 AU, GO-LoW will be too far away to take advantage of GNSS signals.  An alternative solution is cooperative two-way ranging.  \citet{Rajan2011jointsync}, \citet{Rajan2015jointranging}, and \citet{Bentum2020} describe a cooperative ranging scheme, conducted over RF links, appropriate for an interferometric constellation.  Given that plausible schemes for measuring the positions of a cloud of nodes exist in the literature and the precision measurements are not particularly stringent, we chose to focus on other aspects of the study and leave a detailed error budget for ranging as future work.  The feasibility of conducting meter-class ranging over a few thousand kilometers is not in doubt.

The 1.25 m baseline knowledge requirement translates to a 4.17 ns error in time of flight measurement.  Baseline measurement precision will ultimately be a function of clock accuracy and our ability to synchronize clocks across the array.  The LNs will carry Chip-Scale Atomic Clocks (CSACs, \citet[refs. therein]{Kitching2018}), a flight-proven technology \citep{Ritz2023} that is planned for use on interferometric missions like AERO-VISTA \citep{Belsten2020}.  The CCNs will be responsible for providing a time synchronization signal, as GNSS does for Earth-based systems.  The CCNs will therefore need stable clocks and periodic synchronization amongst themselves based on a reference from Earth.  Accurate and synchronized timestamping of the raw data is required for CCN beamforming and for cross correlation on Earth.  Clock errors and drift that remain can be solved for as part of the calibration process.

\subsubsection{Calibration} \label{sec:cal}
Calibration of ground-based interferometric data is a multi-step process that solves for phase and amplitude errors in visibility data.  The radio interferometric measurement equation formalism \citep{Hamaker1996a, Smirnov2011} parameterizes error sources from environmental and instrument effects as complex Jones matrices.  This formalism is well developed and available in community software tools like CASA \citep{CASA}.  Space-based interferometry need not account for phase and/or amplitude errors imposed by the Earth's neutral atmosphere or ionosphere, which simplifies calibration somewhat.

A typical calibration procedure for a ground-based telescope is to observe one or more calibration sources which are well-characterized in position and flux density as a function of frequency.  These well-known sources, mostly distant quasars, do not change with time and are therefore a truth source for correct phase and amplitude.  The best set of low frequency calibration sources is provided by \citet{Scaife2012}.  Some of these sources may be appropriate calibration sources for GO-LoW, while others will become too faint at low frequencies (see \citet[fig.~2]{Scaife2012}).  The sources that are visible near the top of the GO-LoW band may be useful for bootstrapping other calibration sources.  Identifying and measuring low frequency calibration sources will be one of the first tasks for GO-LoW as it grows - a few 10s to 100s of nodes would be sufficient to identify strong sources and precisely measure their spectral flux density.  Our lack of knowledge of the low frequency sky is a challenge for calibration strategy, but it is one GO-LoW itself can solve as it grows.

At the lower end of the GO-LoW band, however, scattering and increasing plasma optical depth may render extragalactic calibration sources useless.  In this case, artificial calibration sources may be required.  CCNs could carry small coded transmitters within the science band for periodic calibration.  These near-field calibrators would be helpful in mapping out the LN beam patterns which will be imperfect due to manufacturing differences. A CCN could also be left in Earth orbit to serve as a far-field calibration source.  Finally, internal calibration sources will be built into the science receiver systems \citep{Lind2022} to account for local instrumental effects.  Calibrating LN data for such effects could take place on the CCNs as part of the automated data aggregation and beamforming process.

\clearpage
\section{Moving Data: Correlation and Communications}
    \label{sec:ccc}
    An interferometer is a data telescope.  The information collected at each node has little value until combined and correlated with data from every other node.  This is because interferometers measure phase difference over a baseline (the projected distance between two nodes).  Phase measurements over many baselines are combined across time and frequency into correlated data products called visibilities.  These visibilities are then turned into scientifically useful images through an inversion and deconvolution process (see e.g., \cite{TMS}).  This section discusses four key areas of the data collection, transmission, and processing: where and how the data from the telescope will be correlated (\S\ref{sec:corrstudy}), how the data will move within the constellation and to/from Earth (\S\ref{sec:dte-vs-hybrid}), laser communication (\S\ref{sec:lasercommarchitecture}), and radio frequency (RF) communication (\S\ref{sec:rf-comm-arch}).  

These study areas are tightly linked.  For example, correlation in space imposes very different requirements on both onboard computation and communications architecture than correlation on the ground.  In order to make progress, we examined two extreme cases.  First, we considered direct raw data transfer to Earth so that correlation of every node in the constellation could take place on Earth where computation is cheaper and easier.  Second, we considered the case where correlation is done entirely in space and only correlated products are sent to Earth, where they can be turned into images.  Neither of these architectures were found to be optimal, so we also considered a hybrid model where some data combining is done in space before sending reduced data products to Earth for correlation and imaging.  The physical size of the required spacecraft and how it packs into a launch vehicle is another key consideration discussed in \S\ref{sec:launchpacking}; an architecture that cannot be fielded due to cost and complexity is not useful.

    \subsection{In-space vs. ground correlation trade study}
        \label{sec:corrstudy}
In the context of interferometry, correlation is the process of ``measur[ing] the complex cross-correlation function" of an interferometer \cite{TMS}.  There are two main steps in the correlation process: a) transforming the data from the time domain to the frequency domain, usually through a Fourier transform and b) cross multiplying data from different telescopes.  The two most commonly implemented correlator types get their names from the order in which these steps are performed.  In FX correlators, the frequency domain transformation happens first (F for Fourier transform) followed by the cross multiplication step.  In XF correlators, those steps are reversed.  See \cite{} for details on the performance differences between these architectures and \cite{} for additional information on other correlation architectures.  The FX architecture is most widely used, so we assume this architecture for GO-LoW.  The F and X steps are separable; they do not need to be performed on the same hardware or proximate in time.  This feature has been leveraged to develop distributed correlation concepts. 

\paragraph{Centralized correlation}
The most common type of correlation architecture is centralized.  Correlators may operate on analog or digital data (digital is now more common due to the falling cost and increasing capabilities of digital electronics) and that data may be at the observation frequency or an intermediate frequency from which the observation frequency has been mixed down.  The analog or digital signals in the \textit{time domain} are transmitted from all of the antennas to a central location for correlation.  Centralized correlators are used by e.g. the VLA \citep{perley2009expanded}.  Transmission to the central correlator may be in real time or via recorded data, as is the case for the Event Horizon Telescope (EHT) \citep{wardle2019first}.

\paragraph{Hybrid or station-based correlation}
Station-based architectures are a variant on centralized correlation, but they reduce the total number of data streams coming into the central correlator by beamforming a group of antennas into a single signal before transmission.  In telescopes that use this model, a group of antennas that are physically close together are beamformed, or phased up, to form a virtual single antenna with a single output signal in the time domain.  The beamformed data rate is nearly the same as the data rate from a single antenna in the station, so the data is reduced by a factor of n where n is the number of antennas in a station.  The beamformed signal from each station is then transmitted to a central correlator.  This central correlator functions in the same way as a correlator in a centralized system as described above.  This station-based architecture is used by e.g. LOFAR \citep{van2013lofar}, MWA \citep{Lonsdale2009}, and the SKA \citep{Dewdney2009}.  This architecture reduces the number of cross correlation calculations that must be performed while allowing for high sensitivity through the use of many antennas.

\paragraph{Fully distributed correlation}
A distributed correlation architecture splits the computational tasks of correlation across physically separate processing nodes \cite{gunst2003correlator, Gunst2008}.  This is accomplished by splitting up the data collected at each node and transmitting a subset to other nodes for processing.  The parallelization may be achieved by splitting the data into antenna subsets, time chunks, or frequency bands.  Following \cite{Gunst2008}, we focus on parallelizing via frequency bands as this is most efficient and conceptually simple.  RF data is digitized and then converted from the time domain to the frequency domain via fast Fourier transform (FFT) or polyphase filter bank (PFB).  In the architecture proposed by \cite{Gunst2008}, each node in the array is responsible for correlating data in a particular frequency band.  The F step is done locally, data is transmitted around the array, and then each node performs the X step on its assigned frequency sub-band.  See also \citet{Rajan2013corr, moonen2014implementation, Hegedus2019}.

\subsubsection{Data volume (raw products)} \label{sec:datvol-raw}
For this and the following section we use the following definitions:
\begin{itemize}
    \item Raw data: digitized timeseries data from an antenna, whether in the time domain as voltages or in the frequency domain as spectral intensity as a function of time.  
    \item Correlated products: output of the correlation process - a timeseries of complex visibilities for each baseline. 
\end{itemize}

Table \ref{tab:datvol} shows the calculation of raw data rate for a single LN.  The calculation is based on the Aurora radio \citep{Lind2022} from the AERO-VISTA mission, which receives, digitizes, and stores data from a 6-channel vector sensor.  The six channels are the six components of the vector sensor (three dipoles and three loops).  The radio directly samples the RF at a rate of 33 million samples per second (Msamp/sec), slightly higher than Nyquist sampling for 15 MHz, GO-LoW's highest frequency.  16 bits per channel are used to preserve dynamic range.  The data are processed down from 6 channels to two orthogonal polarization channels to reduce bandwidth before transmitting the data to a CCN.  All of these values are subject to further refinement as engineering design progresses, but they are representative of a real system that could perform GO-LoW observations.  \textbf{Every LN in the constellation produces data at this rate when observing.}  For 100,000 nodes, the total raw data production rate is \textbf{105.6 Tbps}.  This raw data rate is on par with the SKA, which generates 700 Tbps \citep[refs. therein]{mattmann2014scalable}.  Data rates and volumes of this scale challenge even terrestrial fiber networks and current storage technologies.
\vspace{5pt}

\begin{table}
    \begin{tabular}{|lll|}
    \hline
    \textbf{Parameter} & \textbf{Value} & \textbf{Unit}\\
    \hline
    Number of channels & 6 & \\
    Sampling rate & 33 & Msamples/sec \\
    Bits per channel & 16 & bits/channel \\
    Raw bitrate & 3.168 & Gbps \\
    \hline
    Beamform to 2 channels & 1.056 & Gbps\\
    \hline
    \end{tabular}
    \caption{Raw data generation calculation for one LN}
    \label{tab:datvol}
\end{table}

\begin{figure}
    \includegraphics[width=0.69\textwidth]{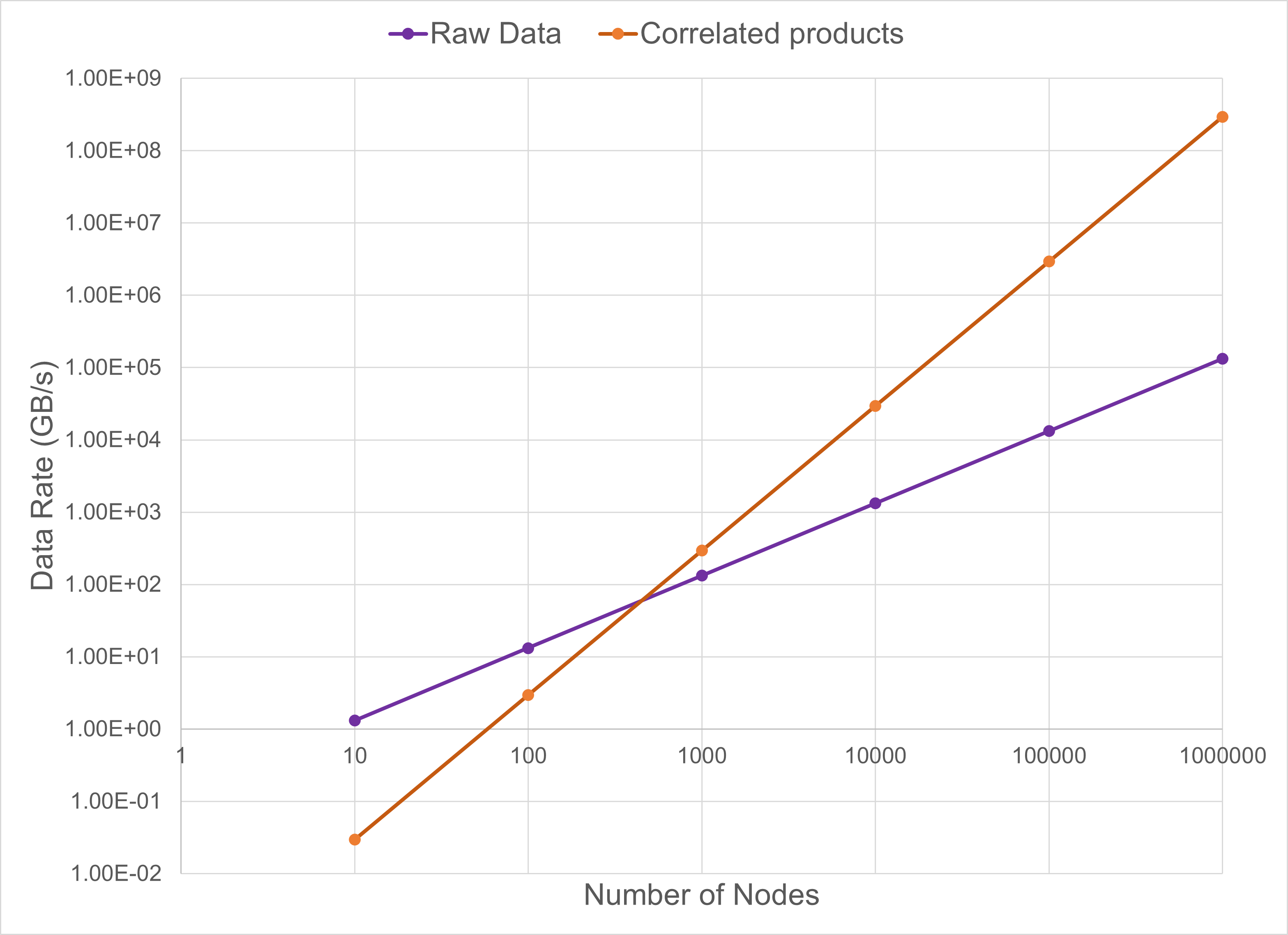}
    \caption{Comparison between raw data volume and correlated data products as a function of number of nodes.}
    \label{fig:raw_corr_compare}
\end{figure}

\subsubsection{Data volume (correlated products)} \label{sec:datvol-corr}
In-space correlation (distributed correlation) makes sense if it reduces the data volume that must be transmitted to the ground.  We compare the raw data volume, discussed above, with the correlated data products.  Figure \ref{fig:raw_corr_compare} shows the production rate of raw data (purple, Table \ref{tab:datvol} multiplied by the number of nodes) compared to the production rate of correlated products (orange), calculated following the approach in \citet{Rajan2013corr}.  The raw products scale with the number of nodes while the correlated products scale as n$^2$.  Correlation is an n$^2$ operation because correlation products (visibilities) are produced for each \textit{baseline}; there are $\sqrt{N(N-1)/2}$ unique baselines in an interferometric array.  Correlation at small n reduces the overall volume of data because of inherent averaging/integration in time and frequency.  At large n, however, the volume of visibilities exceeds the raw data.  Where the crossing point happens will depend on the specific averaging settings during correlation.

\paragraph{GO-LoW's full cross-correlation products will exceed raw data in volume for any reasonable correlation setting.}  Distributed correlation would make the data volume \textit{larger}, which is the opposite of its intended purpose.  GO-LoW therefore has two options:
\begin{enumerate}
    \item Transmit raw data from every node to the ground and correlate there, where the large data volume can be managed more easily, or
    \item Reduce the number of correlations required by subarraying (hybrid architecture).
\end{enumerate}

As noted in \S\ref{sec:corrstudy}, hybrid, station-based architectures are used by ground-based telescopes to solve the same challenge GO-LoW faces.  A hybrid architecture leaves open the option of transmitting the beamformed station data to Earth for correlation or cross-correlating the station data in space in a distributed manner.

\subsection{Direct-to-Earth vs. hybrid trade} \label{sec:dte-vs-hybrid}
The study in the previous section demonstrates that full (n = 100,000) cross-correlation in space is not optimal for GO-LoW.  The next step is to consider whether all raw data from every node can be transmitted directly to Earth for correlation and imaging, or whether the raw data volume of the array must be reduced in order to close the link budget and not impose unnecessary requirements on individual nodes.

In the \textbf{direct-to-Earth architecture}, every node in the constellation must have the capacity to transmit its raw data to the Earth from a 1 AU distance.  In this architecture, all nodes would be identical and all raw data from every node would be preserved.  The amount of data to be transmitted is constant for each node and the total data volume scales linearly with the number of nodes.

In a \textbf{hierarchical architecture}, there are two different types of nodes: Listener Nodes (LN) which only collect data and transmit it locally and Communication and Computation Nodes (CCN) which aggregate data from multiple listeners, process it to reduce the total volume, and then transmit it to Earth.  This architecture is used for several ground-based telescopes (LOFAR, MWA); antennas (equivalent to LN) are arranged into "stations".  All of the antennas in a station are beamformed into a single virtual antenna by a local beamformer (equivalent to CCN) before the beamformed data (reduced from n data streams from each antenna to 1 datastream) is sent to a central correlator.

In order to select one of these architectures, we sought to answer the following questions:
\begin{enumerate}
    \item Is it possible, with reasonable projection of today's technologies, for individual nodes to transmit their raw or lightly processed RF data to Earth?
    \item What communications and computation requirements would distributed correlation impose on individual nodes?
    \item What reduction in required data transmission to Earth, if any, would distributed correlation provide?
\end{enumerate}

\begin{figure}
    \vspace{-15pt}
    \includegraphics[width=0.7\linewidth]{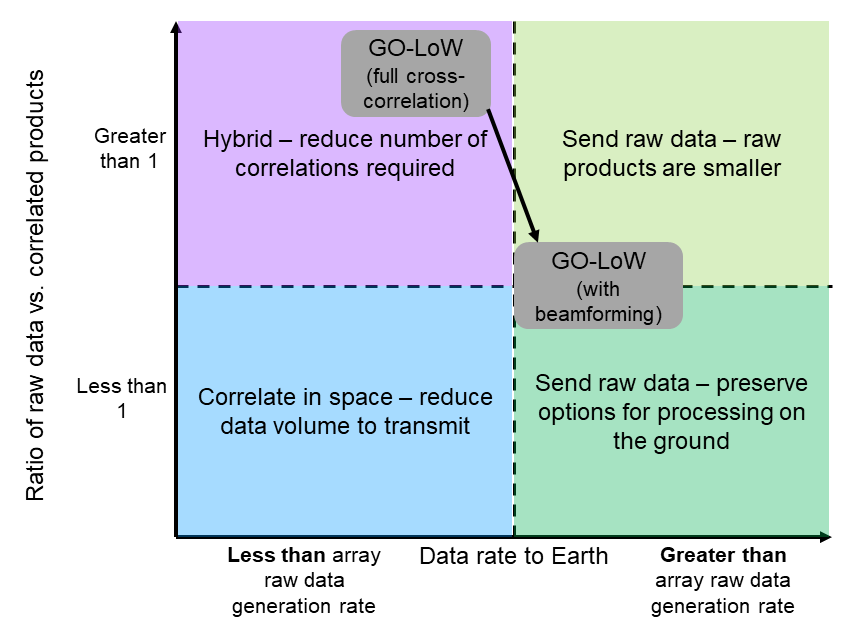}
    \caption{When to use in-space correlation vs. direct transmission of RF data to the ground.  The key parameters are the data rate to Earth as compared to the raw data generation rate of the interferometer (horizontal axis) and the ratio of raw data to correlated products for an observing interval (vertical axis).  When considering full cross correlation between every node in the 100,000 node constellation, GO-LoW is a good candidate for a hybrid architecture that reduces the volume of correlation products (purple quadrant).  When GO-LoW uses CCN to beamform data from many LN into one data stream, it moves into the green quadrants.  Whether correlation products are slightly larger or slightly smaller than raw data products will depend on averaging choices and the exact number of CCN.}
    \label{fig:corr-quad}
\end{figure}

Figure \ref{fig:corr-quad} provides a decision aid once question 1 above is answered.  In the case where the data rate to Earth is very limited and the correlated products are smaller than the raw products, in-space correlation is appropriate (blue quadrant).  We have already shown that GO-LoW is not in this quadrant.  If the size of correlated products exceeds raw products, but the data rate to Earth is less than the rate of raw data generation, a hybrid architecture, which reduces the number of required correlated products, makes sense (purple quadrant).  In the case where the data rate to Earth is larger than the raw products generation rate, it always is best to send the raw data to Earth (light and dark green quadrants).

Adding computation capability and data storage to spacecraft will always be more expensive than making use of the same resources on the ground.  Spacecraft systems need to be optimized for low power consumption, thermal management, and radiation tolerance in a way that ground-based systems do not --- power is essentially unlimited, convection or water cooling makes thermal control a breeze, and radiation tolerance is not required.  Since computation on the ground is always easier, and storage can be redundant and long-term, spacecraft should send the lowest level (most "raw") data products that they can to Earth for archiving and processing.  In summary, \textbf{if you can downlink raw data, you should downlink raw data}.

The following sections describe the link analysis that was used to choose between options 1 and 2 at the end of \S\ref{sec:datvol-corr}.  Section \ref{sec:lasercom_capabilities} describes the requirements for a laser communications (lasercom) link between L4 (equivalently L5) and Earth over a distance of 1 AU.  The results show that a lasercom system on all 100k nodes would make each node unrealistically large, necessitating an economically-infeasible number of launches .  This matches the conclusions in \S\ref{sec:launchpacking} as well.  Direct-to-Earth raw data transfer from each individual node is therefore impractical for GO-LoW; a hybrid architecture, as described in \S\ref{sec:rep-mission-architecture}, is selected.  Section \ref{sec:rf-comm-arch} describes both a backup RF link to Earth via the DSN and the RF link between LNs and CCNs within the constellation.

    \subsection{Laser communications architecture}
        \label{sec:lasercommarchitecture}
            \subsubsection{Current/near-future lasercom capabilities} \label{sec:lasercom_capabilities}
\begin{figure}[hbt!]
    \centering
    \includegraphics[width=\textwidth]{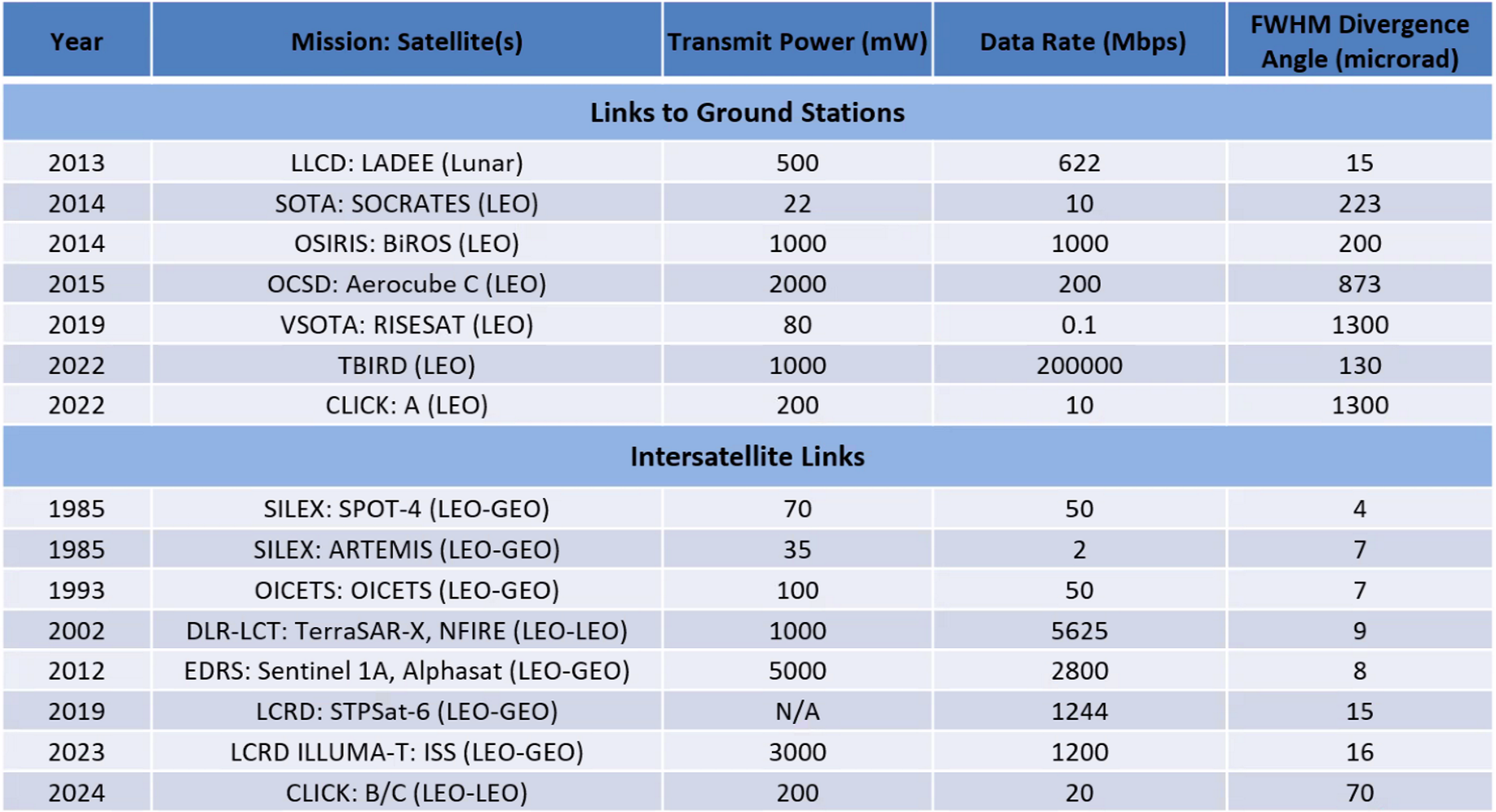}
    \caption{Past and present lasercom demonstrations. Image credit: P. Grenfell, 2023.}
    \label{fig:lasercom_missions}
\end{figure}

Demonstrations of optical communication capabilities in Low Earth Orbit include LLCD \citep{boroson_overview_2014}, CLICK \citep{grenfell_design_2020,tomio_development_2022}, and TBIRD \citep{schieler_onorbit_2023} and MAScOT \citep{gillmer2021demonstration}. Recently launched DSOC will demo deep space lasercom with datarate up to 200 Mbps at distance 1 AU \citep{biswas2017status}. ILLUMA-T \citep{khatri2023system}. See Figure \ref{fig:lasercom_missions} for a summary of key parameters for these missions.

\pagebreak
\subsubsection{Link budget background} \label{sec:laser-link-background}
The laser range equation gives the received power in terms of link parameters. In decibels, it is \citep{grenfell_gnssbased_2020,liang_link_2022,kotake_link_2022}:
\begin{equation}
    P_{rx} = P_{tx} + G_{tx} + G_{rx} + L_{path} + L_{ptg} + L_{tx} + L_{rx} + L_{atm},
\end{equation}
where

\begin{itemize}
\item $P_{tx}$: Transmitter power
\item $G_{tx} = 32 / \theta_0^2$ : Transmitter gain for beam with divergence angle $\theta_{0}$
\item $\theta_0 = 2\lambda/\pi D$ : Beam divergence angle, a measure of the Gaussian beam spread.\footnote{Varying definitions are found in the literature: FWHM, $1/e$, $1/e^2$, all related by constants. FWHM is used here, consistently with \citep{grenfell_gnssbased_2020}.}
\item $G_{rx} = (\pi D / \lambda)^2$ : Diffraction limited gain for receiver of diameter $D$
\item $L_{path} = (\lambda / (4\pi R))^2 $ : Free space loss for link distance $R$
\item $L_{ptg} = \exp(-4\ln(2) \Delta\theta^2 / \theta_0^2)$ : Loss due to pointing error of $\Delta\theta$
\item $L_{tx}$ and $L_{rx}$: Losses due to transmitter and receiver optical components - lenses, filters, couplers
\item $L_{atm}$: Losses due to atmosphere - attenuation, scintillation
\end{itemize}

Energy per bit is equal to received power divided by bit rate:
$$E_b = \frac{P_{rx}}{R}$$
At optical frequencies, shot noise is the dominant noise source (rather than thermal). The noise power spectral density is given by Planck constant and frequency:
$$N_0 = h\nu$$
The ratio, unitless and in dB, is:
$$E_b/N_0 = P_{rx} - 10\log_{10}(h\nu R)$$
From analyzing BER curves, suppose we want $E_b/N_0$ of 8 dB to get BER of $10^{-4}$ using QPSK modulation. PPM can tolerate lower $E_b/N_0$, but it would not be capable of supporting high datarates, so we chose QPSK. The required received power is:
$$P_{req} = P_{bg} + \text{SNR}$$
where $P_{bg}$ is the background brightness. We assume daytime blue sky at 60 deg angle to the sun, at 1550 nm with 3 nm filter, consistently with \citep{sodnik_deepspace_2017}. The SNR term encompases margin needed for the data rate, channel bandwidth, modulation scheme, bit error rate, and link availability. Link margin is the difference between required and actually received power, and it must be positive for the link to close.

\subsubsection{Design of L4 to Earth optical downlink} \label{sec:laser-link-L4}
We baselined our design of the optical downlink from the MAScOT terminal \citep{gillmer2021demonstration} because of its heritage and success in the LLCD and ILLUMA-T demonstrations. However, we found that we need to scale the MAScOT to reach the datarate of 1 Gbps over a link range of 1 AU. Namely, we increase the transmitter power to 50 W, and need a larger aperture with more pointing accuracy.

\begin{figure}
    \centering
    \includegraphics[width=0.75\textwidth]{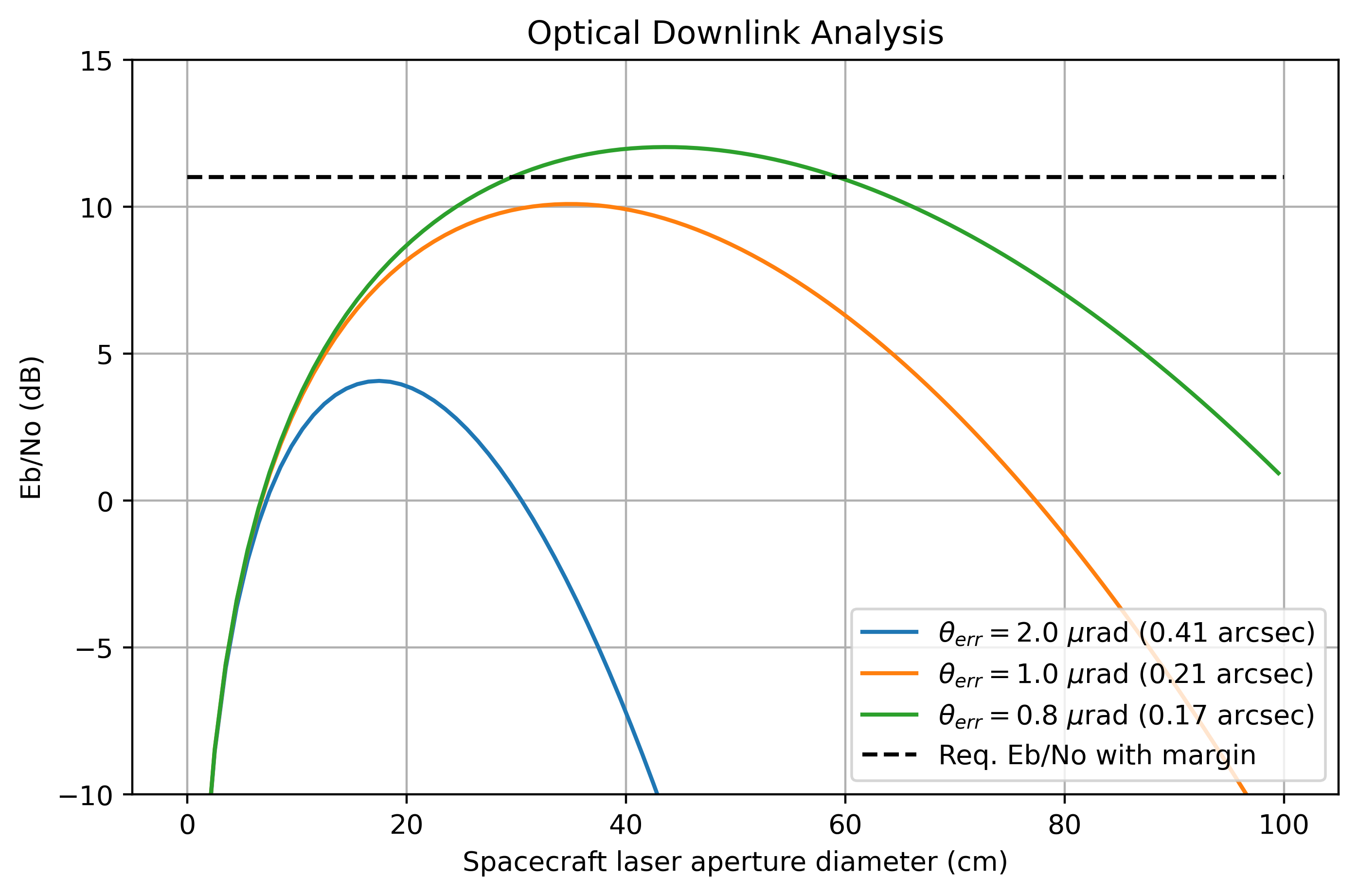}
    \caption{Analysis of available $E_b/N_0$ as a function of optical aperture for several pointing accuracy.}
    \label{fig:aperture_vs_ptg}
\end{figure}

The nonlinear relationship between $E_b/N_0$, aperture diameter, and pointing accuracy is shown in Figure \ref{fig:aperture_vs_ptg}. A small aperture (left side of the curve) performs poorly because of beam divergence losses: a small aperture produces a wide beam with weak gain. Conversely, a large aperture produces a narrow beam that must be pointed extremely accurately. With insufficient pointing accuracy, a large aperture performs poorly because of high pointing loss. For a given pointing accuracy, there is an optimal aperture diameter that corresponds to the minimal combination of divergence losses and pointing losses. 

We can meet our $E_b/N_0$ requirement (including 3 dB margin) by using a 30cm aperture with 0.8 $\mu$rad (0.17 arcsec) pointing accuracy. While the MAScOT was designed for $<4\mu$rad pointing, its performance reported in \citep{gillmer2021demonstration} reached better than $0.5\mu$rad. Furthermore, we expect to see lower disturbances in L4 than in LEO, and expect this to help with pointing stability. So the key change is to scale MAScOT from 10cm to 30cm (Jade said this shouldn't be technically challenging)

\begin{table}
\centering
\begin{tabular}{llll}
Quantity     &  Value   & Units & Notes \\
$P_{tx}$     &    17    & dBW   & 50 W\\
$G_{tx}  $   &   124.7  & dBi   & 30 cm\\
$G_{rx}  $   &   138.2  & dBi   & 4 m\\
$L_{path}$   &  -361.7  & dB    & 1 AU\\
$L_{ptg}$    &    -2.1  & dB    & 0.8 $\mu$rad\\
$L_{tx} $    &    -1.0  & dB    & \\
$L_{rx} $    &    -1.0  & dB    & \\
$L_{atm}$    &    -2.0  & dB    & \\
\hline
$P_{rx}$  &  -87.9  &  dBW   &  \\
$E_b$     & -177.9  &  dBJ   &  \\
$N_o$     & -188.9  &  dBJ   &  \\
$E_b/N_o$ &   11.1  &  dB    &  $10^{-4}$ BER with QPSK\\
\hline
Margin    &   3.07 & dB  & 
\end{tabular}
\caption{Link budget for 1 AU optical downlink}
\label{tab:link_budget}
\end{table}

Given the above assumptions and optimization, Table \ref{tab:link_budget} gives a link budget for L4 to Earth optical downlink. \textbf{The 30 cm aperture and 50 W laser amplifier are realizable on a large spacecraft such as the CCN but prohibitively large to put on each LN. Therefore a hierchical communication architecture, where the long-range downlink is performed at the CCNs, is most practical for GO-LoW.} LNs only need to transmit their data over a short range to their assigned CCN.  The link between LNs and their local CCN is described in \S\ref{sec:rf-comm-arch}.

\subsubsection{Ground station optical terminal}
We model our ground station optical communication terminal based on the SOAR telescope\footnote{\url{https://noirlab.edu/science/index.php/programs/ctio/telescopes/soar-telescope/optical-data-soar-telescope-and-instruments}}. The telescope has a 4.1 m aperture diameter with a 16.625 focal ratio, which provide a 68.16 m focal length. We make modest assumptions about the eyepiece: 10 mm focal length and 52 degree apparent field of view. The telescope's magnification is the ratio of the focal lengths, or 6816 with our assumed eyepiece. The telescope's field of view is the apparent field of view divided by the magnification, which results in a 27.5 arcsec field of view.

The angular diameter of the GO-LoW constellation when viewed from the ground station on Earth is found by
$$\delta = 2\arctan\left(\frac{d}{2D}\right),$$
where $d$ is the actual constellation diameter (assumed to be no greater than 10,000 km) and $D$ is the link range (1 AU). With these assumptions, the resulting angular diameter is 13.8 arcsec. Since it is less than the field of view of the ground station telescope, \textbf{the entire constellation will fit in one telescope pointing.} This greatly simplifies the ground station operation when downlinking because it will only need to track the constellation midpoint, but not slew to view different CCN across the span of the constellation.
            
\pagebreak

\subsection{Radio Frequency (RF) Communications Architecture} \label{sec:rf-comm-arch}

Two RF links were analyzed as part of the GO-LoW Phase I study:

\paragraph*{CCN-to-Earth RF Link}
While lasercom will be the primary method of downlinking science data (\S\ref{sec:lasercommarchitecture}), CCNs will also include a high-gain RF antenna for communicating with the Deep Space Network (DSN). The goal of this analysis was to verify the need for lasercom by demonstrating that RF data rates are not sufficient at a distance of 1 AU, and to appropriately size an antenna to serve as a backup communication system. We anticipate the RF backup to be used primarily for command and control in the event of a) weather conditions that are prohibitively bad for optical frequencies, and b) technical issues with either the in-space lasercom system or Earth-based ground stations.

\paragraph*{LN-to-CCN RF Link}
The Listener Nodes were designed to be the smallest possible form factor in order to accommodate the maximum number of spacecraft per launch vehicle and minimize the total number of launches required. A 3U CubeSat architecture was selected due to its combination of compactness and availability of compatible commercial hardware. Lasercom technology was recently demonstrated on CubeSat platforms \citep{tomio_development_2022,schieler_onorbit_2023}, where it was the main payload for those missions. However, on a Go-LoW LN, a lasercom terminal would require substantial miniaturization to fit alongside the instrument payload and other supporting subsystems.


In contrast, components for RF communication interfaces are mature and miniaturized enough that it is common for CubeSats to host them as subsystems. Therefore, GO-LoW's baseline approach is for all communication between LNs and their controlling CCN to occur over radio frequencies. The distances involved in this link are relatively small (on the order of 100km or smaller) and therefore relatively high data rates can be achieved with modest input powers and antenna sizes.

\paragraph*{CCN antenna type} For the CCNs, both a traditional parabolic dish antenna and a phased array composed of patch antennas were considered. Ultimately, the phased array approach was determined to be preferable because the parabolic dish would need to slew and point at each LN. Given that each CCN will communicate with 100-1000 LNs, it would not be feasible to move the necessary data volume with the parabolic dish. Table \ref{tab:my_label} summarizes the cases that were analyzed alongside key takeaways.

\begin{table}
\setlength{\arrayrulewidth}{0.5mm}
\renewcommand{\arraystretch}{1.5}
    \small
    \centering
    \begin{tabular}{|L{0.160\textwidth} | C{0.060\textwidth} L{0.135\textwidth} L{0.075\textwidth} C{0.060\textwidth} C{0.060\textwidth} L{0.26\textwidth}| } \hline 
         \textbf{Link}               & 
         \textbf{Freq. Band}         &
         \textbf{Spacecraft Antenna} &
         \textbf{Data Rate}          &
         \textbf{Tx Power}           &
         \textbf{Link Close?}        &
         \textbf{Key Takeaways} \\
                            & 
                            &
                            &
         (Mbps)             &
         (W)               &
                            &
                            \\
         \hline\hline
 CCN $\rightarrow$ Earth \newline \scriptsize{(via 34 m DSN)}       & X     & 1 m \newline parabolic                                         & 0.1   &    50 & Yes& Feasible for command, control, and backup, but not science data. 70 m DSN antennas are more desirable but in high demand.\\ \hline 
 CCN $\rightarrow$ Earth \newline \scriptsize{(via 70 m DSN)}      & X     & 1 m \newline parabolic                                         & 0.1   &    12 &Yes&Same as above.\\ \hline
 CCN $\rightarrow$ Earth \newline \scriptsize{(via future 4$\times$ 34 m DSN array)} & Ka    & 1 m \newline parabolic                                         & 0.1   &    12 & Yes& Same as above. Future DSN array will open up high-gain Ka-band as a desirable option.\\ \hline
 CCN $\rightarrow$ Earth \newline \scriptsize{(via future 4$\times$ 34 m DSN array)}  & Ka    & 1 m \newline phased array                                      & 0.1   &    $\sim$39& Yes& Phased array on CCN preferrable to parabolic antenna due to prohibitively long time for downlinking and steering between 100-1000 separate LNs.\\ \hline 
 LN $\rightarrow$ CCN                                  & X     & {\footnotesize\textbf{LN:}} 10 cm phased array \newline \vspace*{0.3cm} {\footnotesize\textbf{CCN:}} 1 m parabolic    & 1000  &  4 \newline (LN)&Yes& Downlink data rates high but achievable. However, parabolic antenna on CCN undesirable (see above).\\ \hline
 LN $\rightarrow$ CCN                                  & Ka    & {\footnotesize \textbf{LN:}}  10 cm phased array \newline \vspace*{0.3cm} {\footnotesize\textbf{CCN:}} 1 m phased array & 1000  &  4 \newline (LN)& Yes& Same as above.\\ \hline
    \end{tabular}
    \caption{Executive summary of RF communication cases analyzed and key takeaways. A positive link margin indicates that communication is feasible at the required data rates. }
    \label{tab:my_label}
\end{table}

\pagebreak
\subsubsection{Methodology}\label{sec:rf_link_method}
All analyses follow the link design methodology outlined in the Space Mission Analysis and Design (SMAD) textbook \cite{SMAD} and are described in detail in the sections that follow.

The governing equation for the link design is reproduced below:
\begin{equation}\label{eq:rf_link_design}
    \frac{E_b}{N_0} = \frac{P_{tx} L_{line} G_{tx} L_{space} L_a G_{rx}} {k T_s R},
\end{equation}
where:
\begin{itemize} \setlength{\itemsep}{-3pt}
    \item $E_b$ = received energy-per-bit
    \item $N_0$ = noise power spectral density
    \item $P_{tx}$ = transmitter power
    \item $L_{line}$ = transmitter-to-antenna loss
    \item $G_{tx}$ = transmit antenna gain
    \item $L_{space}$ = space loss
    \item $L_a$ = transmission path loss (includes atmospheric and rain absorption)
    \item $G_{rx}$ = receive antenna gain
    \item $k$ = Bolzmann's constant
    \item $T_s$ = system noise temperature
    \item $R$ = data rate
\end{itemize}

The energy per bit to noise power spectral density ratio, $E_b/N_0$, is analogous to a signal-to-noise ratio (SNR) for the communications link. Typically an $E_b/N_0$ of 5-10 is desirable to have a low probability of error \cite{SMAD}. A minimum $E_b/N_0$ of 10 was chosen as required for this analysis, which is likely conservative because interferometric data, which relies on time-averaging during correlation, is particularly error-tolerant.

Link budgets were computed by evaluating Equation \ref{eq:rf_link_design} over a variety of parameters (including transmitter powers, antenna gains, data rates), with the goal of appropriately sizing the system. For parabolic antennas, after determining required antenna gains, the following relationship was used to calculate an equivalent dish diameter \cite{balanis_antenna_2016}:
\begin{equation}
    G = \eta_{ae}\left(\frac{\pi d}{\lambda}\right)^2
\end{equation}
where $\eta_{ae}$ is the aperture efficiency, with typical values ranging from 0.50-0.70 \cite[p. 23]{rudge_handbook_1982}, and this analysis assumed 0.70. 

For phased array calculations, a uniform square array with half-wavelength spacing was assumed, and the maximum (broadside) array factor becomes:
\begin{equation}
    G_{AF} = N^2
\end{equation}
and total phased array gain is calculated by:
\begin{equation}\label{eq:G_array}
    G_{array} = G_{AF} G_e
\end{equation}
where $G_{AF}$ is the gain due to the array factor, $N$ is the number of array elements on a side, $G_e$ is the gain of the individual element, and $G_{array}$ is the net (peak) gain of the phased array \cite{balanis_antenna_2016, peebles_radar_1998}. 

\subsubsection{CCN-to-Earth RF link via DSN and in-space parabolic antenna}
GO-LoW will rely on the Deep Space Network (DSN) \cite{nasa_what_2020} for backup radio frequency communications with the constellation. The DSN is an array of large radio antennas around the world operated by NASA's JPL and has traditionally supported spacecraft that venture beyond Earth's orbit. It consists of three facilities, each with multiple antennas, including a 70 m dish and several 34 m ones.

The DSN is capable of communicating using S-, X-, and Ka-band. However, Ka-band is currently limited to 34m antennas, not the higher-gain 70m\footnote{\url{https://deepspace.jpl.nasa.gov/dsndocs/810-005/104/104O.pdf}} \footnote{\url{https://deepspace.jpl.nasa.gov/dsndocs/810-005/101/101G.pdf}}.  There are ongoing plans to retire the 70m DSN antennas and instead utilize an array of the 34m variants that will have equivalent performance (in addition to lower cost and increased flexibility)\footnote{\url{https://www.nasa.gov/image-article/new-generation-of-antennas/}}. It is therefore assumed that future version of the DSN will have high-gain, Ka-band capabilities (and access to the associated higher data rates).

As a starting point, this analysis assumed a high-gain parabolic dish on the CCN, similar to those found on other deep space spacecraft (e.g., Voyager, STEREO, New Horizons). Given the CCN's size, a 1 m diameter was chosen as a feasible target. The gain of the resulting antenna is shown in  (see Section \ref{sec:rf_link_method} for details regarding the calculation).

\begin{figure}[hbt!]
    \centering
    \includegraphics[width=0.5\linewidth]{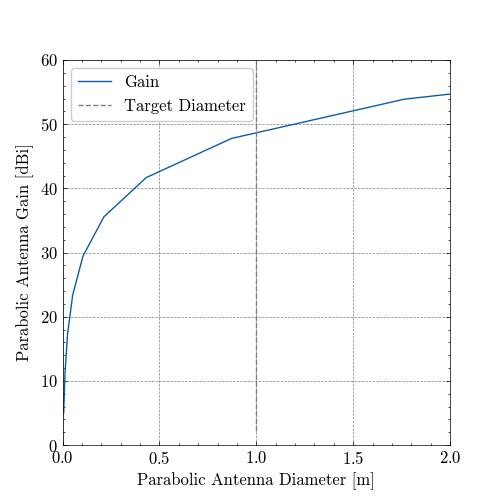}
    \caption{The gain of a notional parabolic antenna on the CCN. 1m is the target size, corresponding to an ESPA-class spacecraft, and a gain of 48.7 dBi is achieved.}
    \label{fig:CCN-parabolic-antenna-gain}
\end{figure}

Next, links were calculated for several spacecraft transmitter powers that were deemed achievable by a future ESPA-class spacecraft (10, 25, 50, 100W). Note that these represent transmitted RF power, not total input power which accounts for transmitter efficiency.

\begin{figure}[hbt!]
    \centering
    \includegraphics[width=1\linewidth]{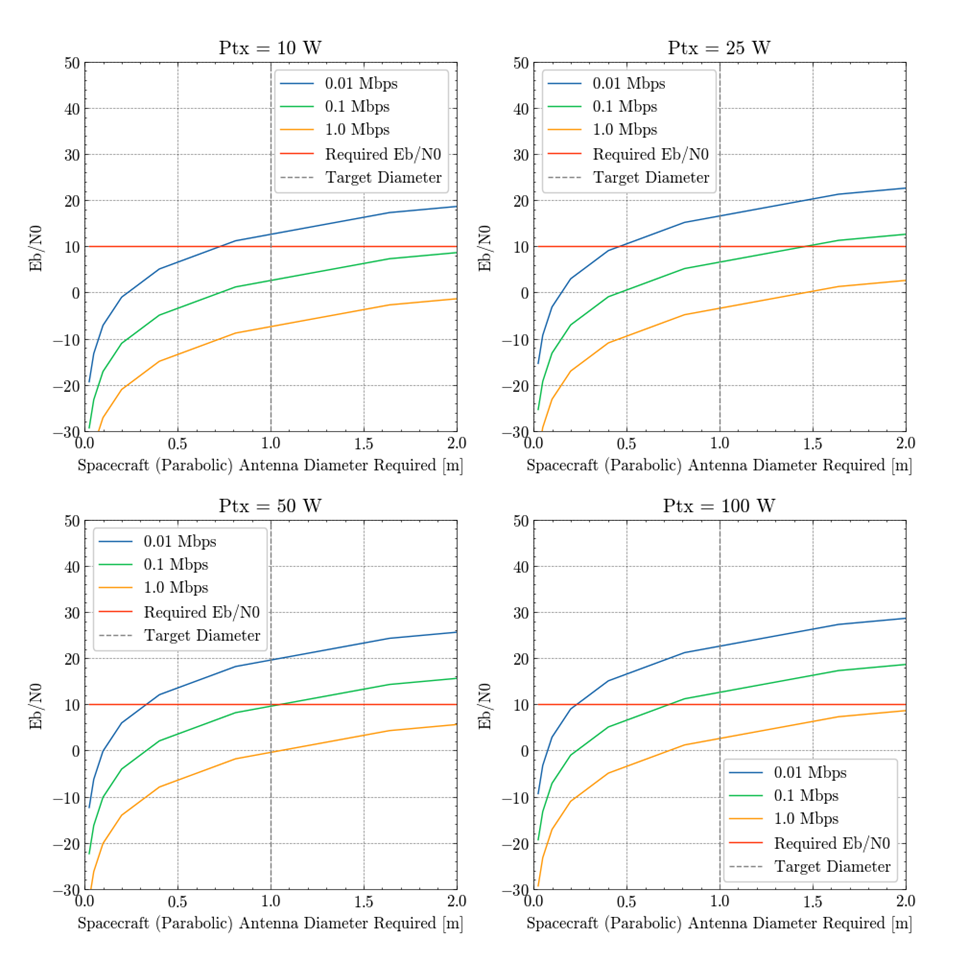}
    \caption{\textbf{Downlink Analysis (Spacecraft-to-Ground) Using 34 m DSN Antenna in X-Band.} The required spacecraft antenna size is shown for various transmitter power level and data rates.}
    \label{fig:34m-DSN-X-downlink}
\end{figure}

\begin{figure}[hbt!]
    \centering
    \includegraphics[width=0.5\linewidth]{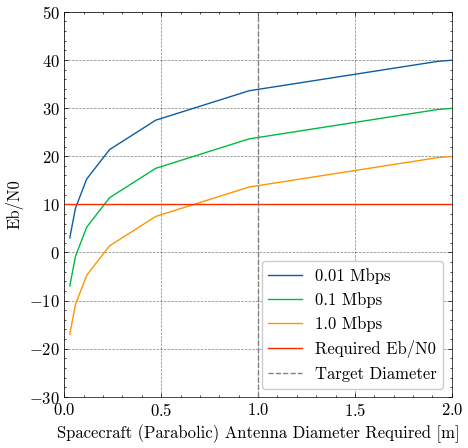}
    \caption{\textbf{Uplink Analysis (Ground-to-Spacecraft) Using 34 m DSN Antenna in X-Band.} The required spacecraft antenna size is shown for various data rates.}
    \label{fig:34m-DSN-X-uplink}
\end{figure}

The initial link analysis was performed at X-band (7145--7190 MHz uplink, 8400-8450 MHz downlink)\footnote{\url{https://deepspace.jpl.nasa.gov/dsndocs/810-005/201/201D.pdf}}, due to current support by the 70 m network and the availability of commercial CubeSat hardware. The results for the 34 m downlink and uplink cases are shown in Figures \ref{fig:34m-DSN-X-downlink} and \ref{fig:34m-DSN-X-uplink} respectively. 

Note that the uplink scenario provides significantly more margin than the downlink one (i.e., $>$1 MBps is feasible with a 1 m dish on the spacecraft). This is as expected, given the powerful transmission capabilities of the DSN. The downlink scenario requires a 50 W (RF power) transmitter on the spacecraft to close the link at 0.1 Mbps with a 1 m in-space antenna. This is feasible, and higher data rates can be achieved by increasing the power further.

For the next analysis, the same link parameters were examined but with the 70 m DSN antenna instead of the 34 m one. While the 70 m variant is preferable due to its higher gain, they are in high-demand and there are not as many of them in the network (only three total); the 34 m are more numerous\footnote{\url{https://www.nasa.gov/directorates/somd/space-communications-navigation-program/what-is-the-deep-space-network/}}. 

\begin{figure}[hbt!]
    \centering
    \centering
    \includegraphics[width=1\linewidth]{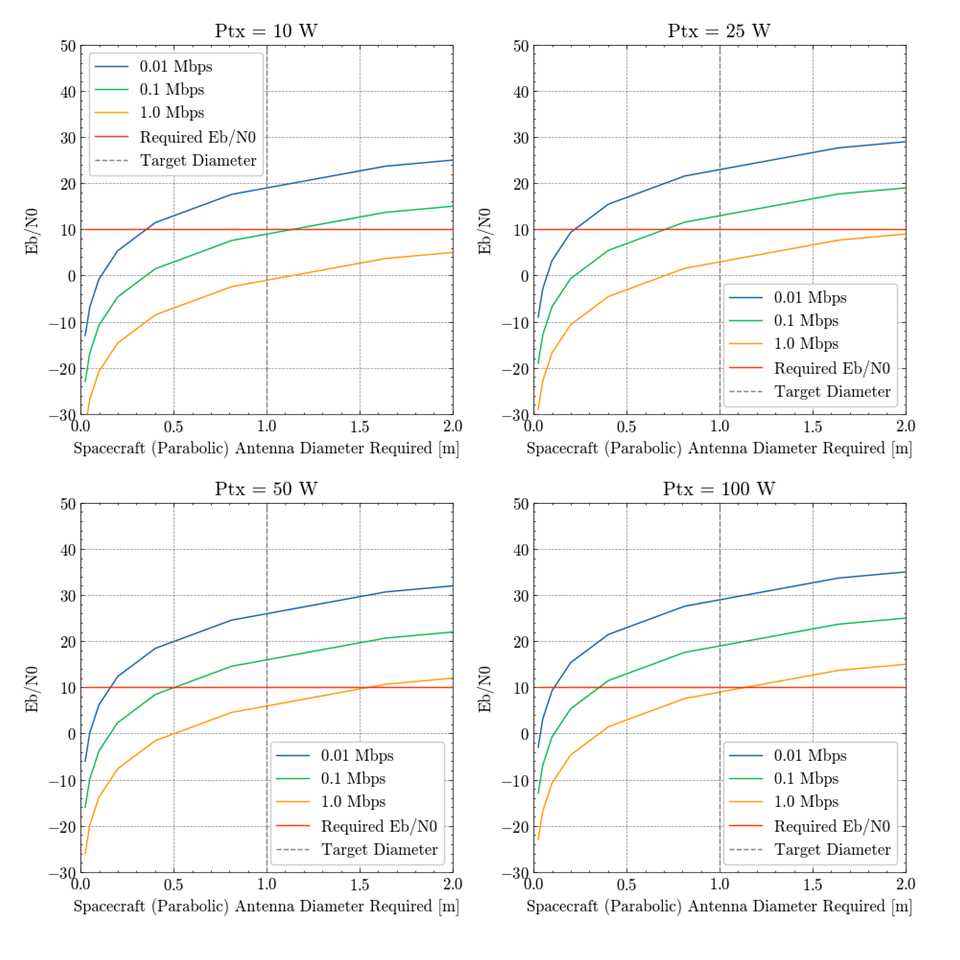}
    \caption{\textbf{Downlink Analysis (Spacecraft-to-Ground) Using 70m DSN Antenna in X-Band.} The required spacecraft antenna size is shown for various transmitter power level and data rates.}
    \label{fig:70m-DSN-X-downlink}
    \end{figure}

\begin{figure}
        \centering
    \includegraphics[width=0.5\linewidth]{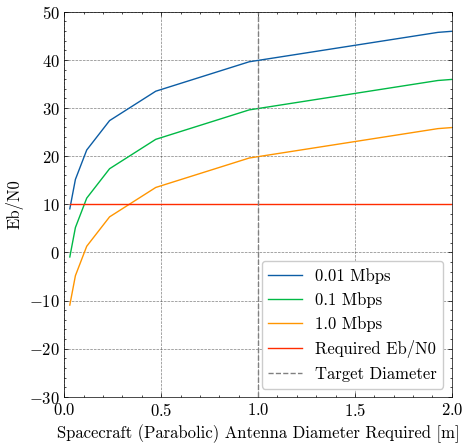}
    \caption{\textbf{Uplink Analysis (Ground-to-Spacecraft) Using 70m DSN Antenna in X-Band.} The required spacecraft antenna size is shown for various data rates.}
    \label{fig:70m-DSN-X-uplink}
\end{figure}

The results for downlink and uplink are shown in Figure \ref{fig:70m-DSN-X-downlink} and Figure \ref{fig:70m-DSN-X-uplink}, respectively. 

Again, the uplink scenario is the easier of the two and data rates above 1 Mbps are achievable with a 1 m spacecraft antenna diameter. The increased gain of the 7 0m antenna is also evident in the downlink scenarios, with 12 W of spacecraft transmitter power required to achieve 0.1 Mbps (compared to 50 W required for the 34 m DSN antenna).

\begin{figure}
    \centering
    \includegraphics[width=1\linewidth]{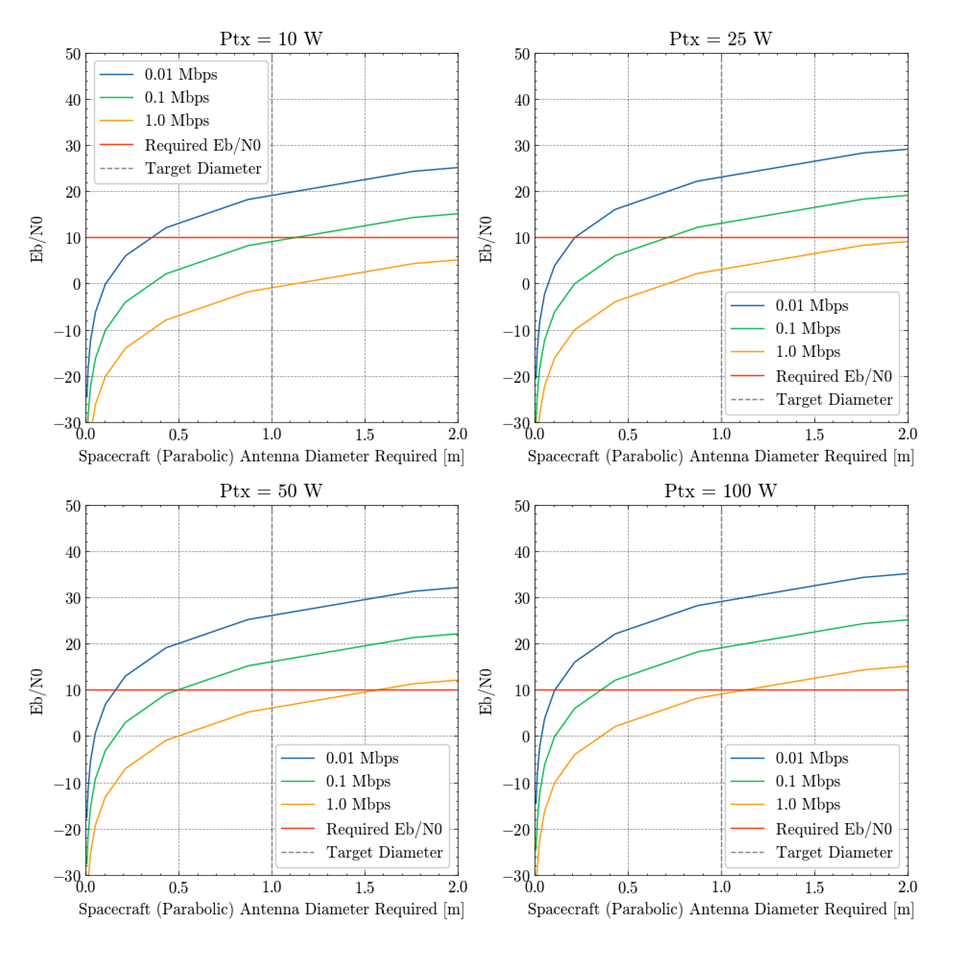}
    \caption{\textbf{Downlink Analysis (Spacecraft-to-Ground) Using Future (Notional) 34 m DSN Array in Ka-Band.} The required spacecraft antenna size is shown for various transmitter power level and data rates.}
    \label{fig:34m-array-DSN-Ka-downlink}
\end{figure}

\begin{figure}
    \centering
    \includegraphics[width=0.5\textwidth]{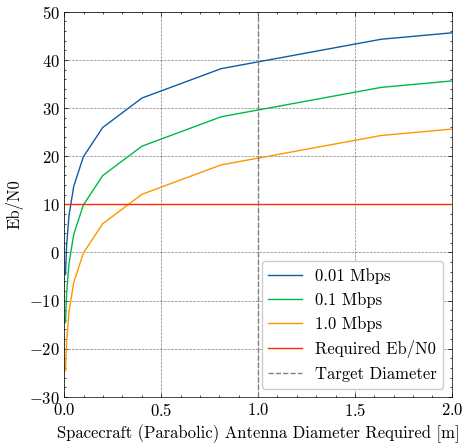}
    \caption{\textbf{Uplink Analysis (Ground-to-Spacecraft) Using Future (Notional) 34 m DSN Array in Ka-Band.} The required spacecraft antenna size is shown for various data rates.}
    \label{fig:34m-array-DSN-Ka-uplink}
    \end{figure}

Next, we analyzed a notional future DSN consisting of an array of 34 m dishes operating in Ka-band (34.2-34.7 GHz uplink, 31.8-32.3 GHz downlink)\footnote{\url{https://deepspace.jpl.nasa.gov/dsndocs/810-005/201/201D.pdf}}. The net gain of the array was set to equal that of a single 70 m\footnote{\url{https://www.nasa.gov/image-article/new-generation-of-antennas/}}. The results for downlink and uplink are shown in Figure \ref{fig:34m-array-DSN-Ka-downlink} and Figure \ref{fig:34m-array-DSN-Ka-uplink}, respectively.

By design, this scenario nearly mirrors that of the single 70 m dish, with the key difference being the communication band used (Ka-band as opposed to X-band). Using Ka-band, which is higher frequency than X-band, provides access to wider bandwidth and associated higher data rates.

\clearpage
\subsubsection{LN-to-CCN RF link, using the CCN's parabolic} \label{sec:LN-CCN-link}
Listener Nodes will need to transmit the science data that they have collected---on the order of 1 Gbps---to their local CCN. This will need to occur over RF, as the 3U LNs are too small to accommodate a lasercom system (in addition to propulsion and the science payload). In order to assess the feasibility of achieving these high data rates, a link analysis was performed using the methodology described in Section \ref{sec:rf_link_method}. First, it was assumed that the CCN's 1 m parabolic antenna would be used for communicating with the LNs (in addition to the DSN). 

\begin{figure}
    \centering
    \includegraphics[width=0.29\textwidth]{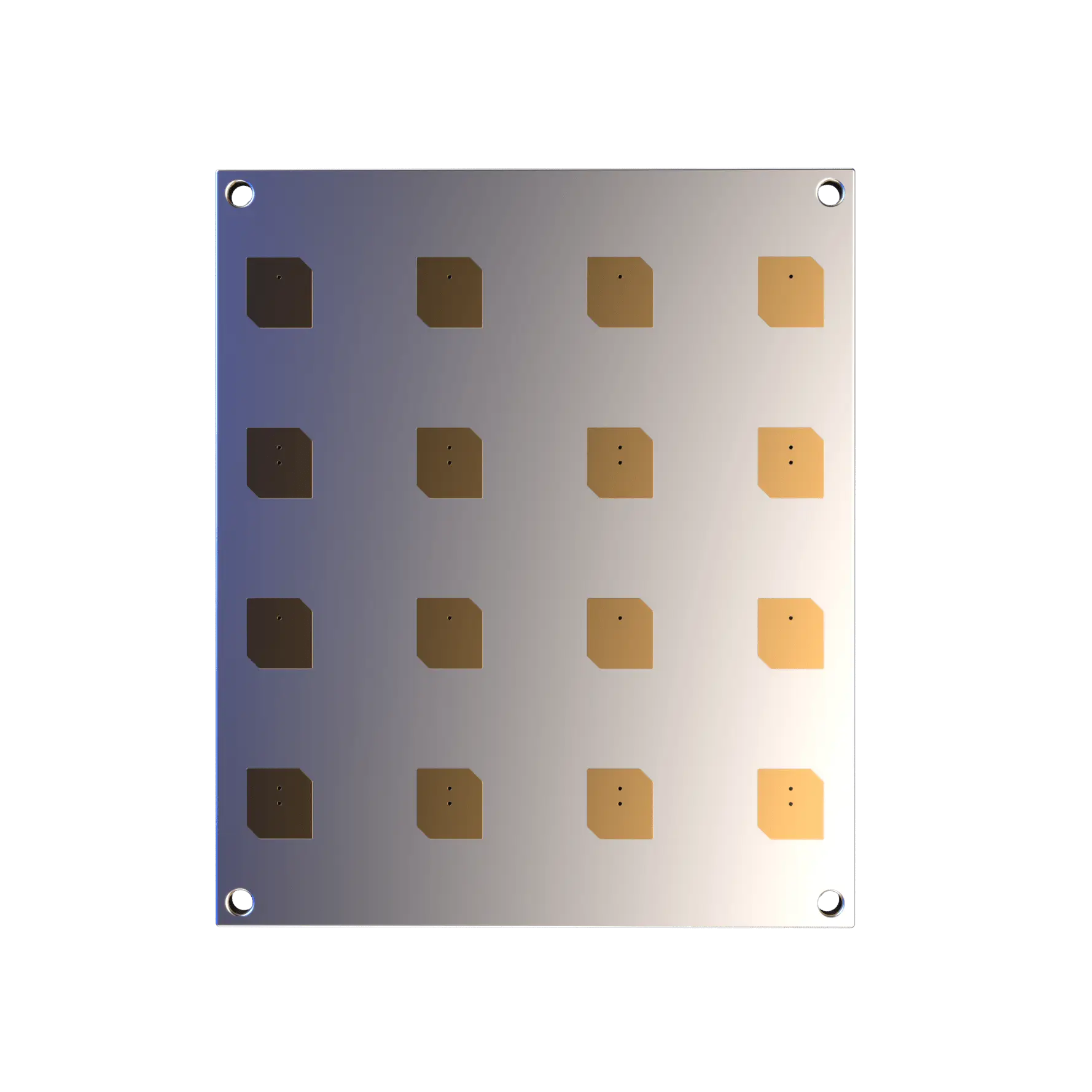}
    \caption{EnduroSat's X-Band 4x4 Patch Phased Array}
    \label{fig:EnduroSat-X-Band-Phased-Array}
\end{figure}

Since accommodating the depth of a parabolic antenna on a 3U spacecraft would be challenging, a flat phased array approach was taken instead. X-band was selected for the initial analysis, due to the CoTS availability of CubeSat form factor antennas. A commercial X-band phased array was selected as a reference point: EnduroSat's X-band 4x4 Patch Array achieves 16+ dBi of gain at up to 4W of RF power in a 3U-compatible size\footnote{\url{https://www.endurosat.com/cubesat-store/cubesat-antennas/x-band-4x4-patch-array/}}.

Tables \ref{tab:LN-to-CCN-X-Inputs} and \ref{tab:LN-to-CCN-X-Results} list the key assumptions and results of the analysis, respectively. Positive margin was achieved for both downlinking of science data and uplinking of commands, which indicates that an X-band RF link is feasible.

\begin{table}[hbt!]
    \centering
    \begin{tabular}{|l|c|} \hline 
         Center Frequency& 8.2125 GHz\\ \hline 
 Link Distance&500 km\\ \hline 
 LN Tx Power&4 W\\ \hline 
 LN Tx/Rx Gain&16 dBi\\ \hline 
 CCN Tx Power&10 W\\ \hline 
 CCN Tx/Rx Gain&37 dBi (1 m antenna)\\ \hline 
 Required Data Rate&1 Gbps\\ \hline
    \end{tabular}
    \caption{Key Assumptions for the LN/CCN X-Band Link Analysis }
    \label{tab:LN-to-CCN-X-Inputs}
\end{table}

\begin{table}[hbt!]
    \centering
    \begin{tabular}{|l|c|c|} \hline 
         & $E_b/N_0$ &Margin with $E_b/N_0\geq10$)\\ \hline 
 LN-to-CCN Data Downlink&10.7 &0.7\\ \hline 
 CCN-to-LN Command Uplink&14.7 &4.7\\ \hline
    \end{tabular}
    \caption{Results of LN/CCN X-Band Link Analysis. The positive margin indicates that achieving the required data rate of 1 Gbps is feasible. }
    \label{tab:LN-to-CCN-X-Results}
\end{table}

\paragraph*{Antenna Pointing}
While the LN-to-CCN link appears feasible from the analysis above, challenges arise when considering the pointing requirements of a parabolic antenna. GO-LoW's architecture is designed around 100-1,000 LNs per CCN (with each group acting as a beamforming cluster). Parabolics are desirable because of their high-gain, but this accompanies a tight beam that results in stringent pointing requirements. The LNs will be distributed in space within a 100km-diameter sphere around the CCN, and the antenna will need to be re-pointed to target each individual node.

A brief, back-of-the-envelope calculation (shown in Table \ref{tab:CCN-Pointing-Analysis}) demonstrates that this approach is not feasible, resulting in over four days of (continuous) time required for a CCN to downlink data from all of the LNs in it's cluster.

\begin{table}
    \centering
    \begin{tabular}{|l|c|} \hline
         Length of Observing Campaign& 1 hour\\
 Data Generation Rate&1 Gbps\\
 Number of LNs&100\\
 Re-Pointing Time&30 sec\\
 Downlink Rate&1 Gbps\\
 Time to Downlink from Single LN&1 hour\\
 Time to Downlink from All LNs&101 hours (4.2 days)\\ \hline
    \end{tabular}
    \caption{Back of the envelope calculations demonstrating that having a parabolic antenna on the CCN is not feasible due to pointing requirements.}
    \label{tab:CCN-Pointing-Analysis}
\end{table}

\paragraph*{Phased Arrays for the CCN}
To address the pointing concerns associated with a parabolic, a phased array was sized for the CCN instead. With phased arrays, the beam can be electronically steered much faster than a mechanically gimbaled parabolic. Additionally, because phased arrays can be composed of patch antennas, the resulting low-profile array could be added to multiple sides of the CCN. These arrays wouldn't be used concurrently due to power and thermal limitations, but would minimize the need to rotate the spacecraft in order to establish links with all of the LNs (and therefore minimize total downlink time).

\begin{figure}[hbt!]
    \centering
    \begin{subfigure}[t]{0.45\textwidth}
    \includegraphics[width=\linewidth]{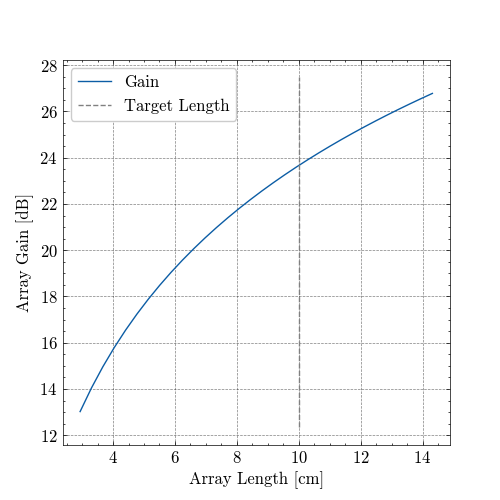}
    \caption{Gain of the Listener Node's 32 GHz (Ka-band) phased array, shown as a function of length. The target maximum array dimension is 10 cm, corresponding to a 1U array, in order be easily accommodated on the 3U LN spacecraft. The resulting gain at 10 cm is about 23.7 dBi}
    \label{fig:LN-Phased-Array-Gain-Ka-32GHz}
    \end{subfigure}
\hfill
\begin{subfigure}[t]{0.45\textwidth}
    \centering
    \includegraphics[width=\linewidth]{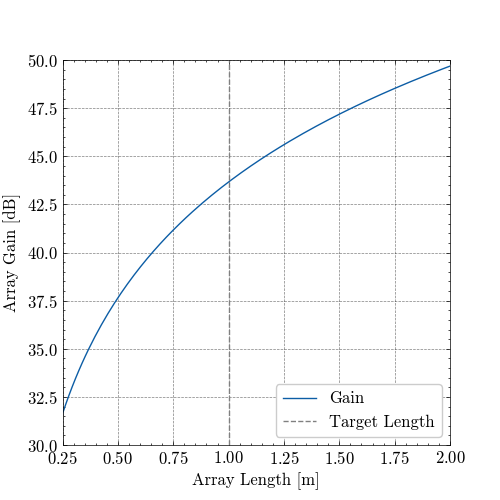}
    \caption{Gain of the CCN's 32 GHz (Ka-band) phased array, shown as a function of length. The target maximum array dimension is 1 m, corresponding to an ESPA-class spacecraft. The resulting gain at 1 m is about 43.7 dBi}
    \label{fig:CCN-Phased-Array-Gain-Ka-32GHz}
    \end{subfigure}
    \caption{}
\end{figure}

\clearpage
A Ka-band patch antenna designed for CubeSat applications was used as a starting reference point for geometry and gain. A gain of 7 dBi was achieved at 28.3 GHz in a total footprint of 7.4 x 9.7 mm \citep{Hammoumi2021}. Equivalent gain was assumed for a notional patch antenna at 32 GHz, and uniform square arrays with half wavelength spacing were assumed for both the CCN and LN arrays. Table \ref{tab:Ka-Phase-Array-Sizing-Assumptions} below lists the key assumption for the phased array sizing and LN-to-CCN link analysis:
\begin{table}
    \centering
    \begin{tabular}{|l|c|}\hline
         Center Frequency& 32 GHz\\
 Link Distance&500 km\\
 LN Tx Power&4 W\\
 Required Data Rate&1 Gbps\\
 Max Array Dimension, LN&10 cm\\
 Max Array Dimension, CCN&1 m\\ \hline
    \end{tabular}
    \caption{Summary of input parameters for the Ka-band phased array analysis.}
    \label{tab:Ka-Phase-Array-Sizing-Assumptions}
\end{table}
The gain of the resulting LN and CCN phased arrays, alongside target sizing, is shown in Figure \ref{fig:LN-Phased-Array-Gain-Ka-32GHz} and Figure \ref{fig:CCN-Phased-Array-Gain-Ka-32GHz}, respectively.

\begin{figure}[hbt!]
    \centering
    \includegraphics[width=1\linewidth]{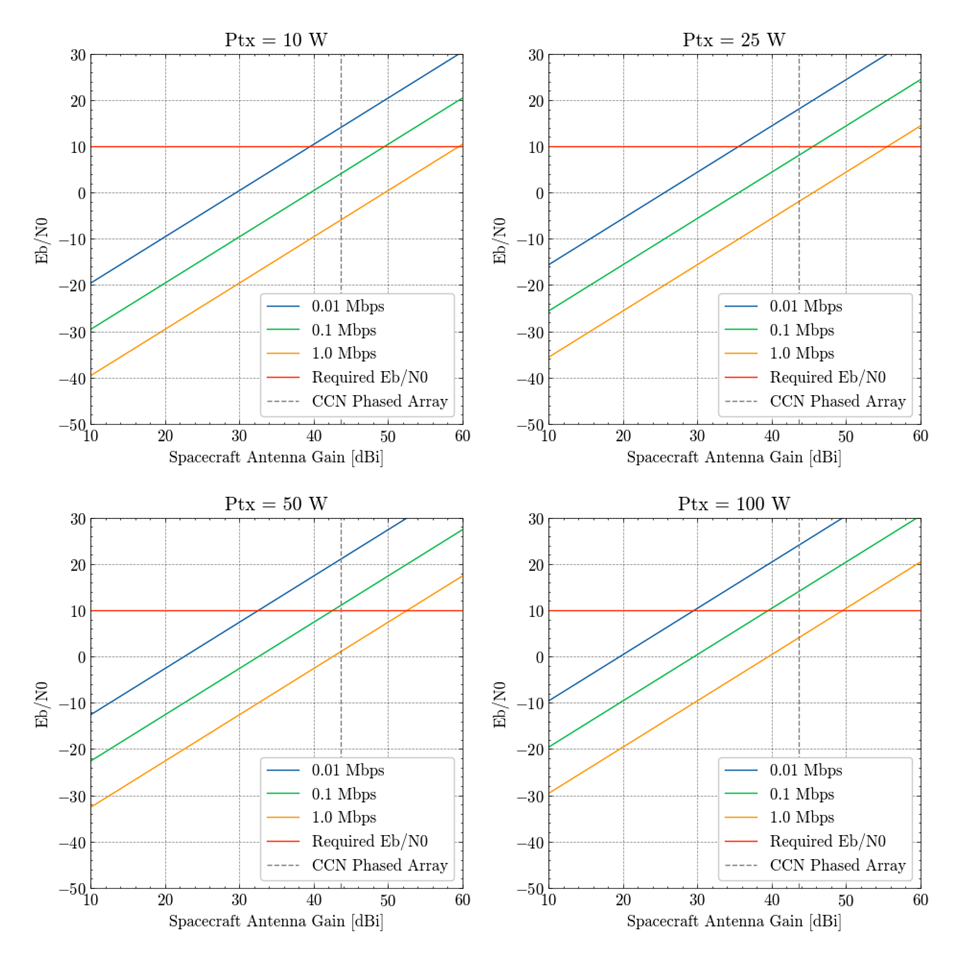}
    \caption{\textbf{Downlink Analysis (Spacecraft-to-Ground) Using Future (Notional) 34m DSN Array in Ka-Band.} The required spacecraft antenna gain is shown for various transmitter power level and data rates. The gain of the CCN phased array (43.7 dBi) is highlighted. The target data rate of 0.1 Mbps is achieved with sufficient margin at a transmitter power of 39 W.}
    \label{fig:34m-array-Ka-phased-array-downlink}
\end{figure}

\begin{figure}[hbt!]
    \centering
    \includegraphics[width=0.5\linewidth]{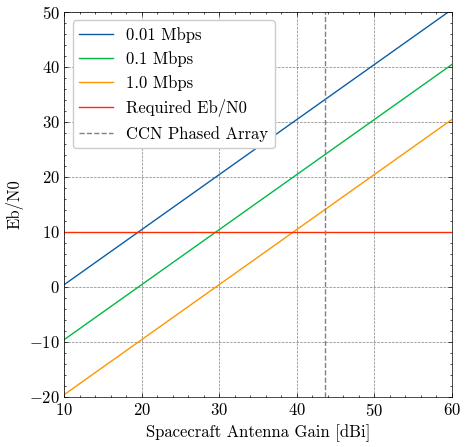}
    \caption{\textbf{Uplink Analysis (Ground-to-Spacecraft) Using Future (Notional) 34 m DSN Array in Ka-Band.} The required spacecraft antenna gain is shown for various data rates. The gain of the CCN phased array (43.7 dBi) is highlighted. Uplink rates of $>$1 Mbps are achievable.}
    \label{fig:34m-array-Ka-phased-array-uplink}
\end{figure}

The calculated gains of both arrays were then used to compute the link margin, with results shown in Table \ref{tab:LN-to-CCN-Ka-Results}. Postive $E_b/N_0$ margin was achieved, indicating that using Ka-band phased arrays on both the LN and CCN, with modest power transmit power, is feasible.
\begin{table}
    \centering
    \begin{tabular}{|l|c|}\hline
     LN Gain with 10x10 cm array & 23.7 dB\\
     CCN Gain with 1x1 m array & 43.7 dB\\
     LN-to-CCN Downlink $E_b/N_0$&13.1\\
     LN-to-CCN Downlink Margin (with $E_b/N_0\geq10$)&3.1\\ \hline
    \end{tabular}
    \caption{Results of the LN/CCN Ka-band phased array sizing and LN/CCN downlink analysis. Note that positive marign was achieved, indicating that the link is feasible at 1 Gbps.}
    \label{tab:LN-to-CCN-Ka-Results}
\end{table}
Finally, given the sizing and performance of the CCN's phased array described above, the RF link between the CCN and Earth was re-examined. The Ka-band DSN downlink and uplink analysis (shown in Figure \ref{fig:34m-array-DSN-Ka-downlink} and Figure \ref{fig:34m-array-DSN-Ka-uplink}) was performed again but with a 43.7 dBi phased array (as compared with a 48.7 dBi parabolic antenna). The results for downlink and uplink are shown in Figure \ref{fig:34m-array-Ka-phased-array-downlink} and Figure \ref{fig:34m-array-Ka-phased-array-uplink}, respectively.

\clearpage          
\section{Mission Architecture}
    \label{sec:mission_architecture}
    
\subsection{Launch \& deployment architecture} \label{sec:launchdeploy}
The GO-LOW concept is centered around a very flexible and easily-scalable launch architecture, an overview of which is shown in Figure \ref{fig:launch-architecture} below. Several super heavy-lift launch vehicles (LVs), such as SpaceX's Starship, are sent to Low Earth Orbit (LEO). Once there, they are refueled (by previously launched Starships, for example), and begin their journey to the fourth (or fifth) Lagrange point in the Earth-Sun system (L4/5). Once at their destination, the LVs deploy their full compliment of LNs and CCNs and GO-LOW's mission begins. 
\begin{figure}[hbt!]
    \centering
    \includegraphics[width=1\linewidth]{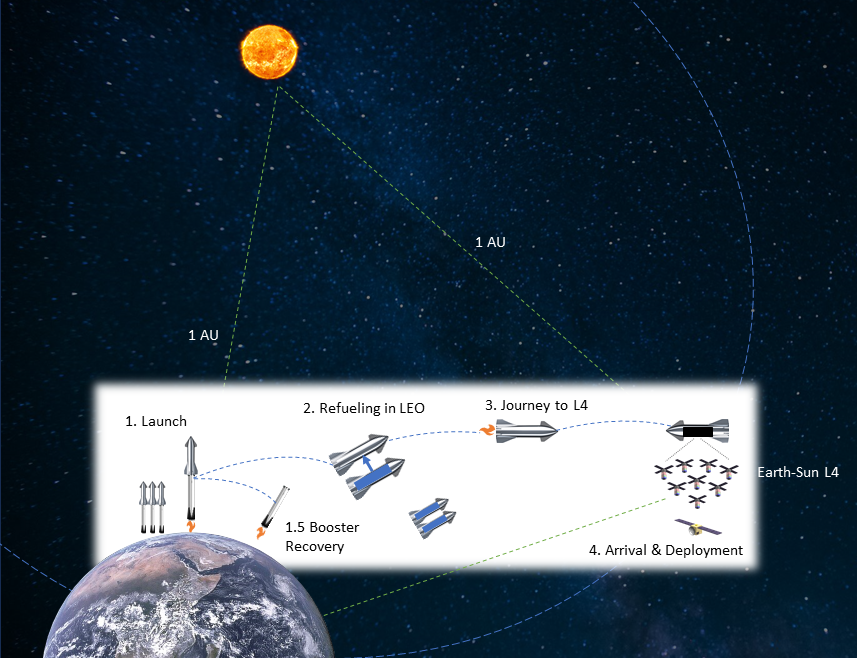}
    \caption{An overview of GO-LOW's flexible and scalable launch architecture. Several heavy-lift launch vehicles (e.g. Starship) are sent to Low Earth Orbit (LEO), where they are refueled before journeying on to Earth-Sun L4. Upon arrival, each launch vehicle deploys its manifest of Listener Nodes and Communication \& Computation Nodes. }
    \label{fig:launch-architecture}
\end{figure}

\begin{figure}[hbt!]
    \centering
    \includegraphics[width=1\linewidth]{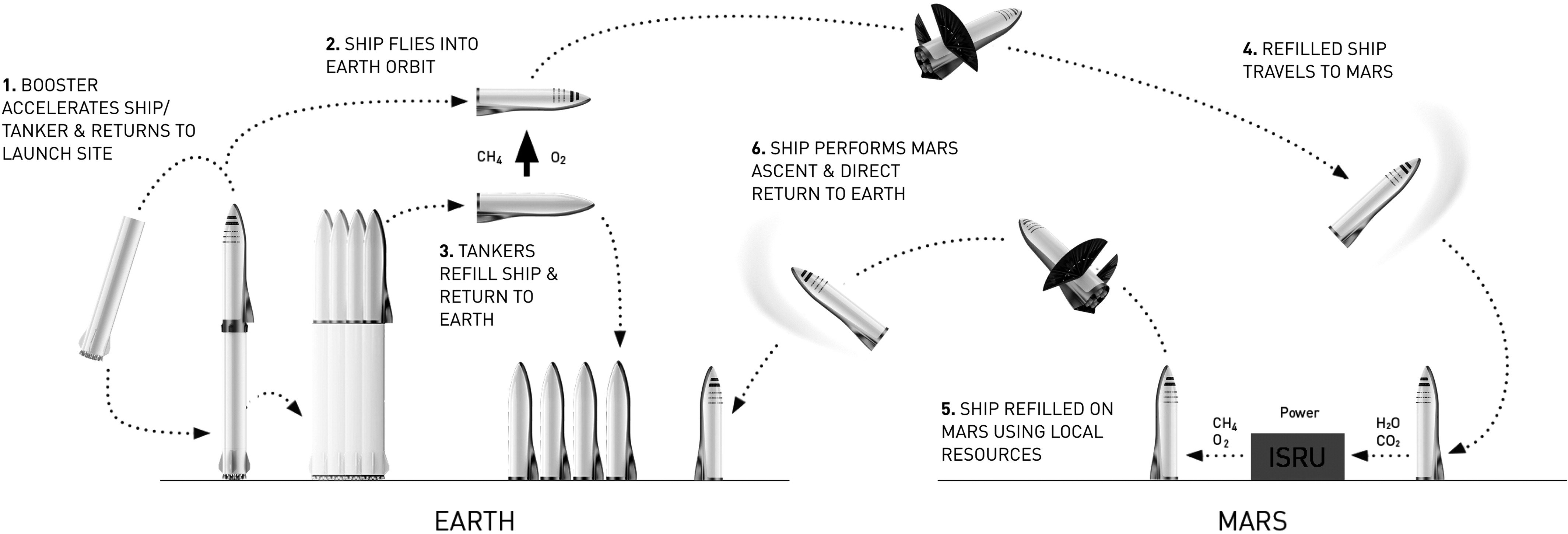}
    \caption{Starship's Mars transportation architecture, featuring refueling on-orbit. Filling Starship's propellant tanks in LEO enables you to launch heavy payloads \textbf{and} journey to a destination requiring high delta-V (e.g. Mars or L4) \citep{Musk2018}}
    \label{fig:Starship-Mars-architecture}
\end{figure}

Refueling in LEO is key to this architecture, because it enables the LVs to launch heavy payloads (e.g. 100 metric tons) \textbf{and} journey somewhere far (like L4) and/or deep down a gravity well (like Mars). The ability to autonomously refuel spacecraft on-orbit has already been demonstrated (back in 2007 by DARPA's Orbital Express on-orbit servicing mission\footnote{\url{https://spacenews.com/after-successful-mission-orbital-express-put-out-pasture/}}) and is currently in active development by the commercial sector\footnote{\url{https://www.orbitfab.com/}}. It is also an integral part of SpaceX's baseline architecture for Starship, which was originally designed to send cargo and humans to Mars (see \ref{fig:Starship-Mars-architecture}). 

\subsection{Launch vehicle packing \& spacecraft geometry analysis} \label{sec:launchpacking}
The primary goals of this analysis were two-fold:
\begin{enumerate}
    \item Determine the form factor of spacecraft that maximizes the number of LNs per launch (while also meeting the needs of GO-LoW's mission).
    \item  Estimate the total number of launches required to deploy the full-scale constellation.
\end{enumerate}

\subsubsection{Super-heavy lift launch vehicles \& support structure} \label{sec:heavylift-supportstruct}
\begin{figure}[hbt!]
    \centering
    \includegraphics[width=0.5\linewidth]{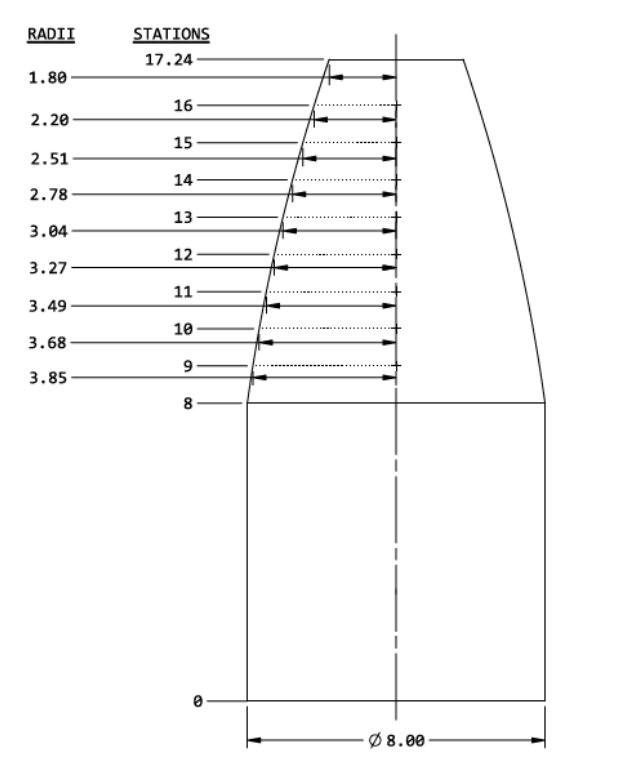}
    \caption[Planned dimensions, in meters, of the payload volume.] {Planned dimensions, in meters, of the payload volume for SpaceX's Starship launch vehicle. Starship will have an 8 m diameter volume available to payloads, whereas current launch vehicles have just 4-5 m\footnotemark.  See also \citet{Elvis2023}.}
    \label{fig:Starship-payload-volume-dimensions}
\end{figure}

In order to achieve GO-LoW's most ambitious science goals (e.g. detecting exoplanet radio emissions), a very large number of sensing nodes are required (100,000+). Launching that many spacecraft with existing LVs would likely be economically infeasible, but next generation super heavy-lift launch vehicles, defined by their ability to deliver 50+ tons to LEO, (e.g. SpaceX Starship, NASA SLS, Chinese Long March) can accommodate \textit{thousands }of small satellites for a dedicated launch. For this analysis, SpaceX's Starship was chosen as a representative LV, due to both its rapid development timeline and the availability of technical information about it. Starship's key specifications are shown in Table \ref{tab:Starship-specs} and the dimensions of the available payload volume is shown in Figure \ref{fig:Starship-payload-volume-dimensions}. Starship will nearly \textbf{double} the diameter of the payload volume of contemporary launch vehicles.

\begin{table}
    \centering
    \begin{tabular}{|lc|}
    \hline
         Payload Volume (inner fairing) Diameter& 8 m\\ \hline
         Payload Volume Height, Total & 17.24 m\\ \hline
         Payload Volume Height, Before Taper & 8 m\\ \hline
         Payload Mass-to-LEO & 100,000 kg\\ \hline
    \end{tabular}
    \caption[Key specifications for SpaceX's planned Starship launch vehicle]{Key specifications for SpaceX's planned Starship launch vehicle\footnotemark.}
    \label{tab:Starship-specs}
\end{table}
\footnotetext{\url{https://www.spacex.com/media/starship_users_guide_v1.pdf}}

\footnotetext{\url{https://www.spacex.com/media/starship_users_guide_v1.pdf}}

\begin{figure}[hbt!]
    \centering
    \includegraphics[width=0.5\linewidth]{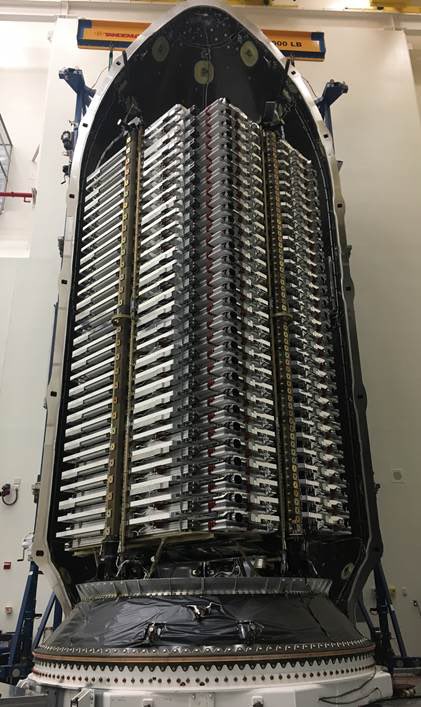}
    \caption[SpaceX Falcon 9 fairing packed with 60 Starlink satellites.] {SpaceX Falcon 9 fairing packed with 60 Starlink satellites. It's evident that the vast majority of the volume is filled with (usable) satellites, as opposed to support and deployment structure (e.g., dead weight)\footnotemark.}
    \label{fig:Falcon-9-with-60-Starlinks}
\end{figure}

    Even with a large payload volume, realizing GO-LoW requires spacecraft to be efficiently packed into the LV in order to make the total number of launches required feasible. The commercial sector (e.g. SpaceX and Starlink) has already made great strides in mass-manufacturing spacecraft and maximizing the space available inside the launch vehicle. Figure \ref{fig:Falcon-9-with-60-Starlinks} shows a Falcon 9 fairing packed with 60 Starlink satellites, and it's clear that the majority of the volume is being used and support structure is minimal.

\footnotetext{\url{https://twitter.com/elonmusk/status/1127388838362378241}}
Assumptions about the payload support structure significantly impact the results of any LV packing analysis. Support structure has two primary roles in a launch vehicle: a) to safely hold and shield satellites during launch and b) to deploy those satellites once the LV reaches its destination. Data from recent Falcon 9 launches dedicated to Starlinks were analyzed in order to come up with a reasonable estimate of the state-of-the-art, see Table \ref{tab:Starlink-support-structure}. The support structure mass fraction, defined as the ratio of support structure mass to total payload mass, varied from 0.2\% to 16.3\%, depending on assumptions about spacecraft mass. Starlink satellites are frequently being updated, and because there is some uncertainty as to what the mass of the "v1.5" generation is\footnote{\url{https://spaceflightnow.com/2023/01/26/falcon-9-starlink-5-2-coverage/}} \footnote{\url{https://spacenews.com/big-constellations-no-longer-necessarily-mean-small-satellites/}} \footnote{\url{https://www.teslarati.com/spacex-elon-musk-next-gen-starlink-satellite-details/}} a range was modeled (260--310 kg).  

For this Phase I analysis, 15\% was selected as a reasonably achievable support structure mass fraction (this may even be conservative given likely gains in satellite manufacturing efficiencies over coming decades).

\begin{table}
\begin{tabular}{lllll}
               & F9 v1.2 (Block 5) & \multicolumn{2}{l}{F9 v1.2 (Block 5)}   &  \\
               & L1 11-11-2019 & \multicolumn{2}{l}{Group 5-2, 1-26-2023}   &  \\
Reference Mission               & Starlink v1.0  & \multicolumn{2}{l}{Starlink v1.5}   &  \\

\hline & \\[-1.5ex]

LV-Mass-to-LEO                  & 17,400 mt & \multicolumn{3}{c}{17,400 mt} \\
Satellites/LV                         & 60        & \multicolumn{3}{c}{56}        \\
Starlink Mass                   & 260 kg    & between 260 kg      &  and    & 310 kg   \\
Support Structure Mass Fraction & 10.3\%   & between 16.3\%     &   and   & 0.2\%    
\end{tabular}
        \caption[]{Analysis of two Starlink launches on dedicated Falcon 9 rockets. Support structure mass is estimated using knowledge of spacecraft mass, number of spacecraft launched, and Falcon 9's mass-to-LEO. The support structure mass fraction ranged from 0.2--10.3\%, depending on the underlying assumptions\footnotemark.}
        \label{tab:Starlink-support-structure}
\end{table}

\footnotetext{\url{ https://spaceflightnow.com/2023/01/26/falcon-9-starlink-5-2-coverage/}, \url{https://spaceflightnow.com/2019/11/11/successful-launch-continues-deployment-of-spacexs-starlink-network/}, \url{https://space.skyrocket.de/doc_sdat/starlink-v1-0.htm}}
    
\subsubsection{Listener node geometry} \label{sec:LN-geom}
\begin{figure}[hbt!]
    \centering
    \includegraphics[width=1\linewidth]{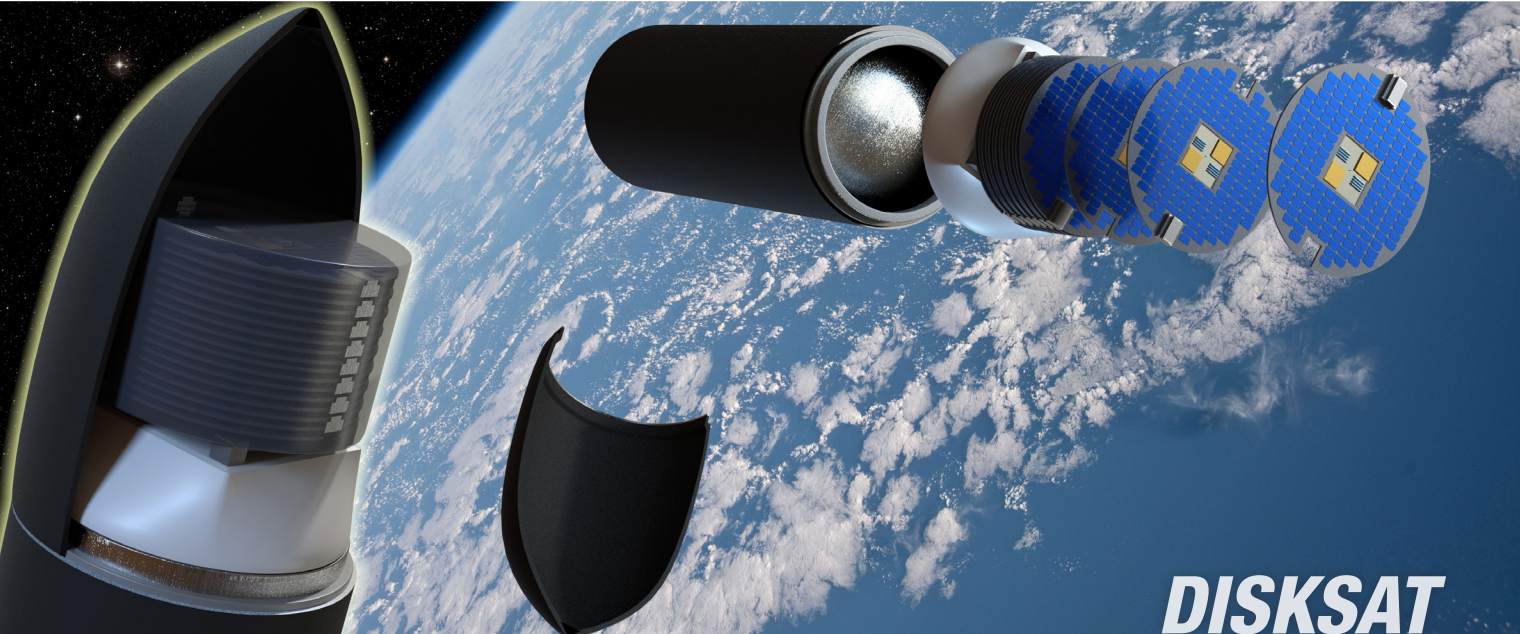}
    \caption[]{The Aerospace Corporation's DiskSat concept. DiskSats are nominally 1m in diameter and 2.5cm thick. NASA's Space Technology Mission Directorate (STMD) is funding a demonstration mission that is anticipated to launch in 2024\footnotemark.}
    \label{fig:DiskSat-overview}
\end{figure}

\begin{figure}[htb!]
    \centering
    \begin{subfigure}[b]{0.5\textwidth}
    \includegraphics[width=\linewidth]{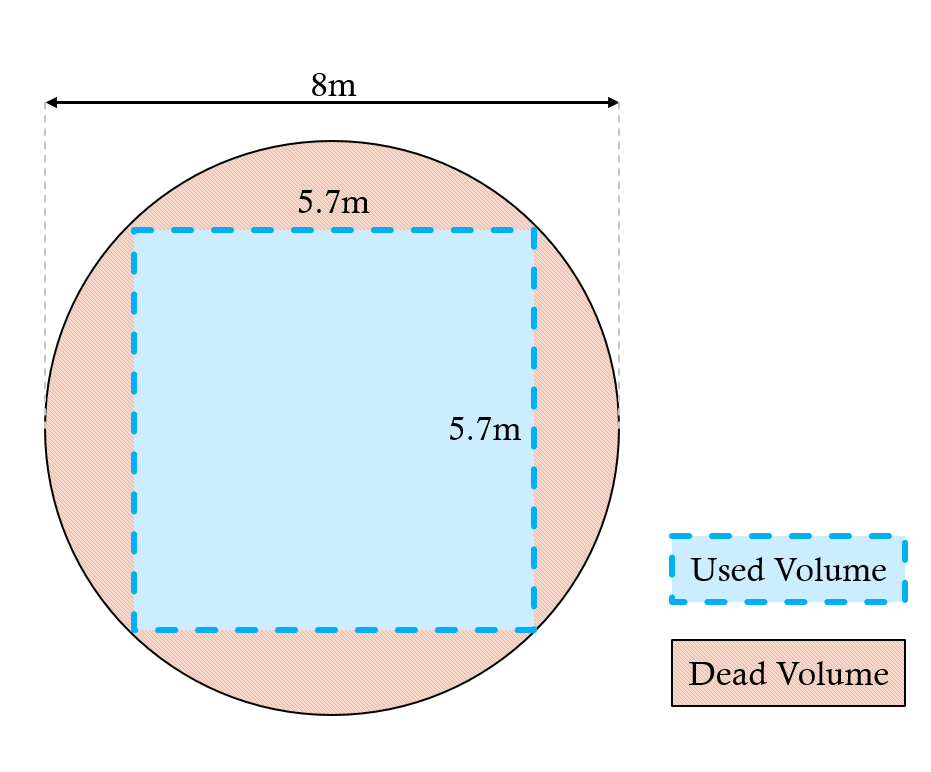}
    \caption{Dimensions of notional square support structure within the 8m Starship payload volume. While this results in significant dead volume (i.e. the space outside of the blue square remains unused) our analysis shows that this is an appropriate simplification to make. The 3U, 6U, and ESPA-class satellite cases are all mass-limited---Starship's 100mt-to-LEO capacity is reached before the assumed square volume is filled. This simplification is not appropriate for the DiskSat case, however, which fills Starship's volume before it reaches the mass limit.}
    \label{fig:support-structure-square-volume}
    \end{subfigure}
    \hfill
\begin{subfigure}[b]{0.4\textwidth}
    \centering
    \includegraphics[width=\linewidth]{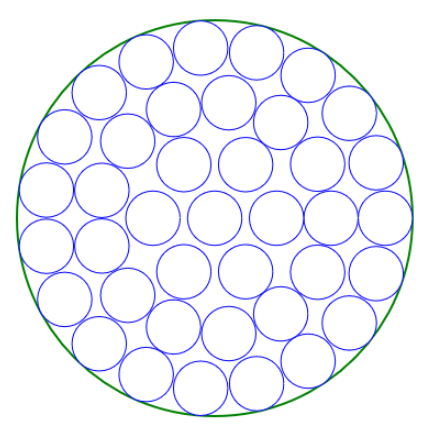}
    \caption{39 DiskSats (with clearance added) packed into an 8m diameter Starship fairing\footnotemark.}
    \label{fig:DiskSat-fairing-packing}
    \end{subfigure}
    \caption{}
\end{figure}

Two primary LN geometries were considered for this analysis: rectangular (e.g. CubeSats) and thin circular disks (e.g. Aerospace Corporation DiskSat, see Figure \ref{fig:DiskSat-overview}). DiskSats were designed to be an alternative to the CubeSat standard and to be efficiently stacked inside of a launch vehicle. With a nominal 1m diameter and 2.5cm thickness, DiskSats provide a larger-than-CubeSat surface area that could be used to provide LNs with more power and larger radiators and RF phased arrays \citep{welle2022disksat}. The team particularly considered DiskSats for the case of a single-spacecraft architecture for GO-LoW (e.g. larger LNs and no CCNs).
\footnotetext{\url{https://aerospace.org/sites/default/files/2022-08/DiskSat_0822.pdf}, \url{https://www.nasa.gov/mission/disksat/}}
For the case of a rectangular satellite (e.g. 3U, 6U), a square columnar support structure was assumed, see Figure \ref{fig:support-structure-square-volume}. While this results in a large section of wasted volume (red), the analysis will go on to demonstrate that this is an appropriate simplification. In fact, packing rectangular satellites into this square volume results in a mass-limited scenario: the LV's mass limit is reached before the entire volume is filled. This, however, is not the case for the DiskSats.

\footnotetext{\url{https://www.engineeringtoolbox.com/smaller-circles-in-larger-circle-d_1849.html}}
DiskSats were assumed to more efficiently fill the LV fairing, see Figure \ref{fig:DiskSat-fairing-packing}. In the tapered portion of the LV (see Figure \ref{fig:Starship-payload-volume-dimensions}),  the support structure for both the rectangular and DiskSat cases was assumed to follow the profile of the fairing.

A summary of the analysis results is shown in Table \ref{tab:Spacecraft-packing-analysis}. Two options for LN spacecraft were studied: 3U CubeSat and 1m-diameter DiskSat. An ESPA-class satellite, modeled as a 1x1x1m spacecraft, was also studied as a baseline for the CCN. The total possible number of satellite per Starship was calculated using  assumptions about satellite mass, the support structure mass fraction, and Starship's geometry. A clearance  was assumed around each satellite, to account for the support structure volume.

\pagebreak

It's clear from this analysis that DiskSats are not a viable option for GO-LoW's LNs---they would require over 25 launches to deploy the full constellation, which is likely prohibitive. 3U CubeSats, however, require an order of magnitude less launches and are still large enough to fit all critical subsystems. The total number of CCNs depends on how many LN-per-beamforming cluster the constellation has, and two cases were modeled: 100 or 1,000 total CCNs. For the case of 3U LNs and ESPA-class CCNs, the \textbf{total number of Starship launches required to deploy GO-LoW at its full size ranges from 5.7--10.6}.

\begin{table}[]
\begin{tabular}{@{}lcccl@{}}
\toprule
                       & \multicolumn{1}{l}{\textbf{3U}} & \multicolumn{1}{l}{\textbf{DiskSat}} & \multicolumn{1}{l}{\textbf{ESPA-}} & \textbf{ESPA-}                 \\
\textbf{}              & \multicolumn{1}{l}{\textbf{}}   & \multicolumn{1}{l}{\textbf{}}        & \multicolumn{1}{l}{\textbf{Class}} & \textbf{Class}                 \\ \midrule
                       & \multicolumn{1}{l}{}            & \multicolumn{1}{l}{}                 & (100 total)                        & \multicolumn{1}{c}{(1k total)} \\
Dimensions             & 10x10x30cm                      & 1m Dia., 2.5cm thick                 & \multicolumn{2}{c}{1x1x1m}                                               \\ \midrule
Satellite Mass         & 4 kg                            & 15 kg                                & \multicolumn{2}{c}{500 kg}                                               \\ \midrule
Support Structure      & 15\%                            & 15\%                                 & \multicolumn{2}{c}{15\%}                                                 \\
Mass Fraction          & \multicolumn{1}{l}{}            & \multicolumn{1}{l}{}                 & \multicolumn{1}{l}{}                    &                                \\ \midrule
Nodes Req.             & 100,000                         & 100,000                              & 100                                     & \multicolumn{1}{c}{1,000}      \\
in Constellation       & \multicolumn{1}{l}{}            & \multicolumn{1}{l}{}                 & \multicolumn{1}{l}{}                    &                                \\ \midrule
Max. Satellites per LV & 21,250                          & 3,984                                & \multicolumn{2}{c}{170}                                                  \\
                       & (mass-limited)                  & (volume-limited)                     & \multicolumn{2}{c}{(mass-limited)}                                       \\ \midrule
LN Launches Req.       & 4.7                             & 25.1                                 & \multicolumn{2}{c}{N/A}                                                  \\
for Full Constellation & \multicolumn{1}{l}{}            & \multicolumn{1}{l}{}                 & \multicolumn{1}{l}{}                    &                                \\ \midrule
CCN Launches Req.      & N/A                             & N/A                                  & 1                                       & \multicolumn{1}{c}{5.9}        \\
for Full Constellation & \multicolumn{1}{l}{}            & \multicolumn{1}{l}{}                 & \multicolumn{1}{l}{}                    &       \\                        
\bottomrule
\end{tabular}

\vspace{8pt}
    \caption{Results of spacecraft packing analysis, with highlighted assumptions. Several different form factors were considered for LNs, including 3U CubeSats and Aerospace Corp's DiskSats. An ESPA-class spacecraft was considered for the CCNs. After making informed assumptions about support structure mass fractions, the total possible number of satellites per launch vehicle was calculated.}
    \label{tab:Spacecraft-packing-analysis}
\end{table}

\subsection{Multi-level autonomy} \label{sec:autonomy}

While several different classification systems for hierarchies of autonomy have been proposed across the aerospace, automotive, and robotics fields, the European Cooperation for Space Standardization (ECSS) definition \citep{ECSS} is most applicable to GO-LoW. According to \citet{Tipaldi2018}, ``Current European space missions typically use level E2, while some advanced earth observation and science missions implement level E3” and ``Achieving level E4 means designing a spacecraft able to make decisions based on a set of defined rules and goals, while autonomously replanning activities in case of off-nominal conditions” … GO-LoW requires E4.

As a whole, GO-LoW exhibits Level E4 mission autonomy (execution of goal-oriented mission operations on-board, and goal-oriented mission re-planning), Level D1 data management autonomy, and Level F2 failure management autonomy. To our knowledge, full E4 has yet to be implemented, while both E2 and E3 have been.

Autonomy is a crucial component of GO-LoW, but not all constellation tasks need to happen in-situ, without human support. Table \ref{tab:autonomy} shows a breakdown of primary constellation tasks and their associated level of autonomy, applying the ECSS definition at a task- (instead of mission-) level. 

\begin{table}
\centering
\begin{tabular}{|L{0.3\textwidth}|L{0.2\textwidth}|L{0.4\textwidth}|} \hline
\textbf{Task}  & \textbf{Autonomy Tier}  & \textbf{Comments}  \\
\hline
Position and attitude determination and control  & E3 (event-based autonomous ops)  & Impractical to do fully from the ground because of limited comms infrastructure/time and scale of constellation.  \\
\hline
Collision avoidance  & E4 (goal-oriented)  &   \\
\hline
Fault determination and recovery  & F2  & Basic fault determination can be done autonomously (i.e., enter fault state). Basic recovery (i.e., try a reset) can also be done autonomously. Difficult to manage all faults manually from the ground with this number of spacecraft.  \\
\hline
Correlation and final science product generation  & E1 (ground-controlled)  & Computation is abundant and relatively low cost on the ground. Beamforming into ``stations'' will be performed in space, cross correlation to form visibilites and then images will be done on the ground.  \\
\hline
Science target selection  & E1 (ground-controlled)  & While the constellation may get a schedule of science observations to perform, the human science team generates the parameters for the observations.  \\
\hline
Determination of constellation geometry (i.e. baseline selection)  & E3/4  & GO-LoW can determine baselines between elements, ground operators determine whether/how those baselines must be adjusted to suit the science goals.  \\
\hline
Re-arranging beamforming clusters  & E4  & Depends on the timescale of orbital intermixing. One of the key outcomes of the Phase II will be to determine this.  \\
\hline
Aggregation of data by CCNs  & E4  &   \\
\hline

\end{tabular}
\caption{GO-LoW mission activities, shown with the required level of autonomy.}
\label{tab:autonomy}
\end{table}

\subsection{Constellation location} \label{sec:const-loc}
\begin{wrapfigure}{r}{0.5\textwidth}
\vspace{-15pt}
    \includegraphics[width=0.45\textwidth]{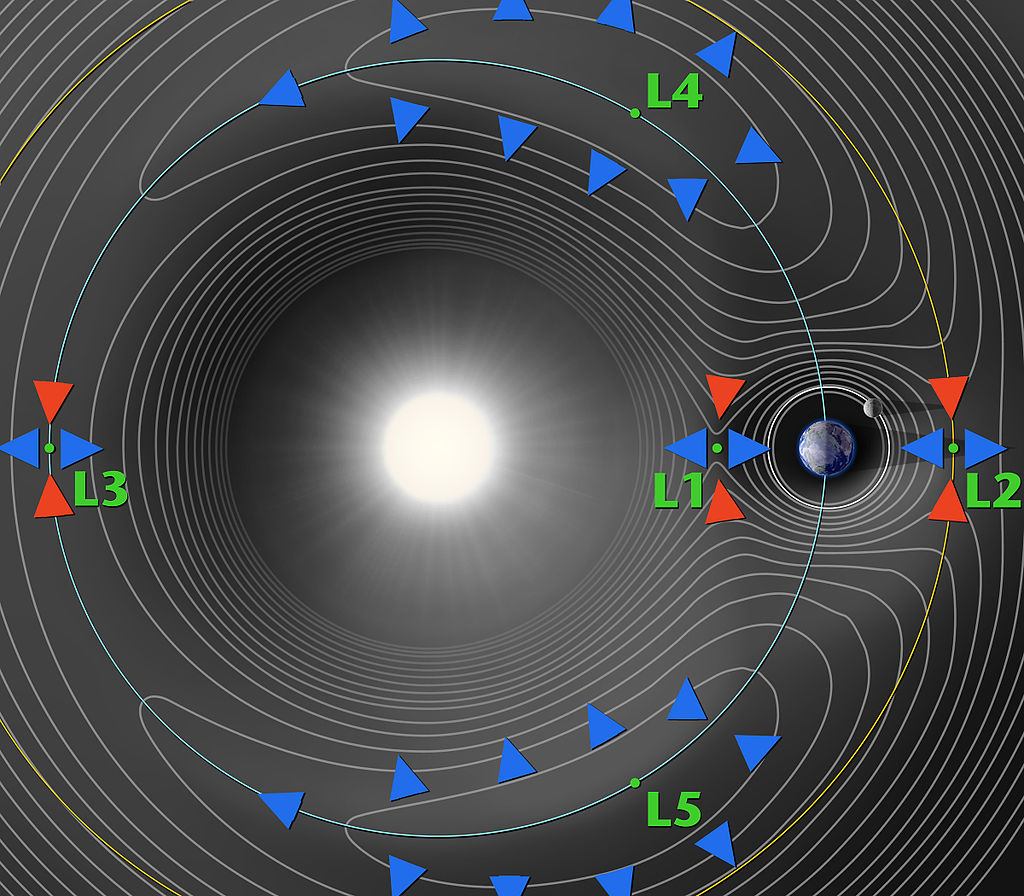}
    \caption{Earth-Sun Lagrange points shown via potential curves in a rotating frame.  L4 and L5 are stable points as opposed to the quasi-stable saddle points L1, L2, and L3.  Placing GO-LoW at L4 or L5 provides separation from the Earth, a stable location for reduced stationkeeping requirements, and low risk of space debris impacting Earth or satellites in Earth orbit.  Image credit: NASA.}
    \label{fig:lagrange}
\end{wrapfigure}

Radio signals from exoplanets and their stars are of such low intensity ($<$1 mJy, \citet{Vidotto2019}) that natural and human-made sources of radio frequency interference (RFI) from the Earth easily overwhelm them. GO-LoW therefore requires a location that is sufficiently far from Earth and yet is feasible for communications and frequent launches. The Earth-Sun Lagrange points L4 and L5 (Figure \ref{fig:lagrange}) are natural choices because of their 1 AU distance from Earth and the reasonable delta-v required to get there (on order of 1 km/s beyond Earth-escape, \citep{Gopalswamy2011, walker2017miniaturised}). Additionally, their stable gravitational potential requires less propulsion for station-keeping. placing a mega-constellation at L4 or L5 also greatly reduces the risk of space debris interfering with other spacecraft or human spaceflight in Earth or lunar orbit. 

L5 is a particularly attractive location for GO-LoW's representative mission because there are several proposed heliophysics missions to this location as well: providing an opportunity for sharing infrastructure and launches. Earth-Sun L5 is ideal for heliophysics observations \citep{Lavraud2016, Liewer2014}; at 60$^{\circ}$ away from Earth, a heliophysics observatory at L5 could view coronal mass ejections (CMEs) as they transit from the Sun toward the Earth \citep{Gopalswamy2011}.  The L4 and L5 points are also of interest to planetary science because they may contain Trojan asteroids and dust trapped since the formation of the solar system.  Only one Earth Trojan is known, but observations from Earth-based telescopes are difficult due to the position of L4/5 relative to the sun \citep{John2015,davis2019disposal}.  The multiple launches needed to build up GO-LoW at L5 could also transport spacecraft focused on heliophysics or planetary science, opening new opportunities for remote and in-situ measurements in these fields. 

\subsection{Comparison to other concepts} \label{sec:compare_other_missions}
GO-LoW can be compared to two different categories of mission: Lunar surface low frequency interferometric arrays and other flagship astronomical telescopes. 

\subsubsection{Comparison to lunar arrays} \label{sec:compare_lunar}
The Lunar farside has been considered for radio observations since RAE-2 showed a significant drop in noise while passing behind the Moon \citep{alexander_scientific_1975}. Low frequency telescopes fixed to the Lunar surface, whether single dish \citep{Bandyopadhyay2020} or interferometers \citep{Burns2019, Polidan2022}, share science goals with GO-LoW. GO-LoW represents an alternative technical path towards exploring the low frequency sky with challenges that are distinct from landed telescopes. 
Landed telescopes cannot be easily reconfigured or upgraded due to the challenges of landing and Lunar surface operations. Landed arrays also face some of the same challenges as Earth-based telescopes – they cannot see all of the sky all the time, they may be affected by radio frequency interference (RFI) from, for example, lunar orbiters, rovers, or human bases, and the local surface dielectric properties under the array will vary, possibly making calibration challenging. Lunar landed telescopes also depend on the Lunar launch and landing infrastructure that is still in its infancy; GO-LoW depends on heavy lift launch capability and in-orbit refueling that is also still developing.
The GO-LoW constellation architecture can be replenished and upgraded more easily than landed telescopes by simply delivering new spacecraft as individual elements fail or become obsolete. Antenna positions can be rearranged to optimize spacing between nodes for a variety of observation programs which is essential for a Great Observatory with many scientific stakeholders. 

\subsubsection{Comparison to free-flying array concepts}
Several concepts for free-flying low frequency interferometric arrays have been developed.  Examples include ALFA \citep{Jones2000}, SIRA \citep{MacDowall2005}, PARIS \citep{Oberoi2003}, OLFAR \citep[references therin]{Bentum2020}, SOLARA \citep{Knapp2013}, XSOLANTRA \citep{Banazadeh2013} NOIRE \citep{Cecconi2018}.  See \citet[ch.~6]{Knapp2018} for further overview of past free-flying interferometer concepts.  All of these past concepts have similar, if more limited, science goals to GO-LoW, but all are smaller in number of nodes.  Some of these concepts take advantage of CubeSats/smallsats, while others use more traditional spacecraft architectures.  Without exception, these concepts use crossed dipoles or tripole antennas, while GO-LoW takes advantage of the more capable vector sensor antenna (Section \ref{sec:antenna_options}).

GO-LoW builds on these past concepts and adds mega-constellation mass production for scale and redundancy.  The hybrid mission architecture splits the difference between raw voltage data downlink and full in-space correlation, using the station concept employed by ground-based arrays (LOFAR \citep{van2013lofar}, MWA \citep{Lonsdale2009}, SKA \citep{Dewdney2009}) to reduce data volumes in space.

\subsubsection{Comparison to flagship observatories at other wavelengths} \label{sec:compare_flagship}
GO-LoW is a paradigm shift away from expensive, single-point-of-failure spacecraft that must be extensively tested to a massively redundant and replenishable constellation. GO-LoW leverages mass production, autonomy, and a sustained upgrade model to make a revolutionary new type of Great Observatory.

\subsection{Benefits Beyond GO-LoW} \label{sec:broaderimpacts}
The massive multi-node architecture that we will model in Phase II, if awarded, is applicable far beyond radio astronomy. Other potential applications include heliospheric sensing constellations with nodes scattered throughout the heliosphere to sample the solar wind at thousands of points representing many spatial scales, or an atmospheric sensing net dropped into the turbulent atmosphere of Venus or one of the gas giants. The graduated autonomy we plan to develop could also be employed for a network of bots swimming in the oceans beneath the surface of an icy moon. The technology we develop could be used for remote sensing in challenging terrestrial environments as well.

As described in \S\ref{sec:resiliency}, GO-LoW challenges the current mission status quo in several ways. First, GO-LoW relies on massive numerical redundancy for reliability rather than the extensive testing and design redundancy that characterize large single spacecraft missions (\emph{e.g.}, JWST \citep{Matthews2016}, Parker Solar Probe \citep{Smith2020}). This feature takes advantage of assembly line and quality control paradigms used in manufacturing rather than bespoke engineering. The mass production of GO-LoW units will benefit the local economy where they are produced, with the possibility of sustained production for replacement units. Second, GO-LoW is intended to grow and change over time as new units are added and old ones fail. It will be a Ship of Theseus in telescope form; unlike single spacecraft, GO-LoW can gain capability through the injection of new technology in replacement units (e.g., faster processors, improved propulsion systems). The technology that makes up GO-LoW will not be fixed to specific hardware that is outdated before it flies.

\clearpage
\section{Technology Roadmap}
    \label{sec:techmap}
    
Eight key enabling technologies were identified during the Phase I study (see below). In particular, the study focused on defining detailed requirements for the lasercom, launch vehicles, small satellite technology, and science payload (VSA and radio) elements. 

\subsection{Free-space optical communications (aka lasercom)} 
Radio interferometry generates very high data volumes (1 Gbps per node, in the case of GO-LoW) which are not readily compatible with radio-based transmission systems. Lasercom can handle the required volumes and rates and the technology is already being developed for space applications beyond LEO (e.g., Pysche DSOC [34], Artemis O2O[35]). 

\subsection{Super heavy lift launch vehicles (LVs) and low cost to LEO} 
The full GO-LoW constellation, at 100,000 nodes, requires sending a lot of mass to LEO before venturing on even further. In order to keep the total number of launches economically and practically feasible, next generation launch vehicles are required. Luckily, the launch vehicle industry has already driven cost-per-kg to LEO down to historic levels (mostly with the advent of SpaceX’s Falcon 9 and Heavy) and new super heavy lift LVs are on the horizon (e.g. SpaceX Starship, NASA SLS, Chinese Long March). GO-LoW’s launch architecture is based around Starship, as its development is progressing in-line with GO-LoW’s timeline and its being designed with the key capability of in-orbit refueling.

\subsection{In-orbit (e.g. LEO) refueling }
Refueling in LEO is key to GO-LoW’s launch architecture, because it enables LVs to launch heavy payloads (e.g. 100 metric tons) \textit{and }journey somewhere far (like L4/5) and/or deep down a gravity well (like Mars). If an LV was required to reach L4/5 in “one shot”, without refueling, its payload capacity would be drastically reduced and the total number of launches required for GO-LoW would likely be infeasible. The ability to autonomously refuel spacecraft on-orbit has already been demonstrated (back in 2007 by DARPA's Orbital Express on-orbit servicing mission) and is currently in active development by the commercial sector by companies like Orbit Fab. It is also an integral part of SpaceX's baseline architecture for Starship, which was originally designed to send cargo and humans to Mars. 

\subsection{Small satellite technologies}
GO-LoW’s Listener Nodes (LNs) are 3Us and rely on packing a lot of functionality into a very small volume (10x10x30cm). The key underlying technologies required include: robust spacecraft platforms, compact propulsion systems, and radiation tolerant electronics. Luckily, the SmallSat commercial sector has been very active and several companies have flown successful 3U missions (e.g. Blue Canyon Technologies, NanoAvionics). Additionally, CubeSats are moving beyond LEO and demonstrating the radiation tolerance required to go further (e.g. Blue Canyon’s BCT-12 which recently flew to GEO). And with the success of deep space CubeSats like NASA’s CAPSTONE (12U) and MarCO (6Us), not to mention the investment in the Artemis I CubeSats (6Us), it’s anticipated that the capabilities of CubeSats will only continue to increase. While the delta-v requirements for station-keeping at Earth-Sun L4/5 are minimal, a propulsion system is still required due to the unavailability of options such as magnetorquers beyond LEO. Propulsion technology has been rapidly developing and miniaturizing, and there are several $\leq$1U systems currently available (or in active development) spanning chemical, cold gas, and electric propulsion. And with the recent attention to the LEO debris problem and the FCC’s 5-year rule for deorbiting satellites, there is an increased incentive for the development of extremely compact and capable propulsion systems.

\subsection{Vector sensor antenna (VSA)}
There are two key challenges to address: 1) maximizing performance by matching the effective height of all elements while 2) refining the deployable design for maximum reliability. Progress on the deployable design has been made through the AERO-VISTA mission (ref) and will be further developed as the follow-on SWIPE mission concept is developed.

\subsection{Science Receiver System / Data Acquisition System (DAQ)}
Like the VSA, the receiver system has already undergone substantial development. Further work will include optimization to reduce receiver noise to ensure externally noise limited performance. Reducing size and power can be achieved through a more integrated design and system-on-a-chip optimization.

\subsection{In-space networking architecture and protocols}
GO-LoW’s large size necessitates a robust networking architecture to facilitate local communication between LNs and CCNs over radio frequencies. With the rapid growth of constellations and intersatellite links, significant resources are being invested in developing robust networking architecture that can support space-to-space and space-to-Earth communications (e.g. SDA’s Transport Layer ). As the size of both commercial (e.g. Starlink, OneWeb, Kuiper) and science (e.g. SunRISE, CINEMA) constellations continue to grow, so will the investment in communication and networking technologies. Additional, key technologies for deep space like Delay/Disruption Tolerant Networking (DTN) are already in active development. 

\subsection{Autonomous spacecraft}
A key Phase II task will be determining the appropriate level of autonomy for LNs and CCNs. We will model this system first at a modest scale (\~100 nodes) and then scaled up to the full 100k+ constellation in order to assess how automation scales with constellation size. This modeling effort will allow us to further refine the levels of autonomy and identify any holes or bottlenecks in our constellation management scheme.

\clearpage
\section{Conclusions, Open Questions, \& Future Work}
    \label{sec:conclusion}
    In this Phase I study we demonstrated that GO-LoW is promising and feasible concept for a wide variety of science cases and pushes the state-of-the-art forward for a number of broadly-applicable space technologies, including science constellations, small spacecraft subsystems, data transport and processing, and mission autonomy. Building on this success, the next phase of GO-LoW's development should focus on three primary areas:

1.  Simulate the \textbf{cutting-edge autonomous behavior} required to execute GO-LoW's mission. This could involve a multi-agent simulation of the full-scale constellation during an observing campaign (with individual spacecraft modeled as intelligent, interacting agents).

2. Modeling the science performance of the constellation, producing key findings: angular resolution and sensitivity during constellation growth and in various configurations, calibration performance and expected noise/uncertainty, and \textbf{simulated maps of the radio sky} produced by each growth stage of the constellation.

3. Identifying viable paths toward \textbf{technology demonstration missions} that test and validate the key elements required to build GO-LoW in the future: autonomous flight operations, science data collection, data processing and high-speed transport, robustness to individual element failures.

\subsection{Space Mission Autonomy} \label{sec:autonomy_future}
Realizing GO-LoW's ambitious mission requires an unprecedented degree of spacecraft- and constellation-level autonomy. The next step to advance the concept is to build on the strong modeling foundations established in this Phase I study and focus on defining the autonomous operations architecture required to achieve a large constellation telescope outside of Earth’s orbit. \textbf{This Phase I study showed that it is the ‘system of systems’ that makes GO-LoW innovative and unique, rather than any specific technology, component, or subsystem}. Future work should therefore focus on demonstrating the complex, and autonomous, interactions between GO-LoW’s many spacecraft and proving that the concept can efficiently scale. 

The best way to do this would be to developing a multi-agent simulation of the GO-LoW constellation that demonstrates all of the components necessary for success:

1. \textbf{Mission planning and execution; }turning high-level science goals into specific constellation tasks within well-defined constraints.

2. Local \textbf{intelligent sensing }(not Earth-reliant)\textbf{; }monitoring spacecraft states and the status of the constellation as a whole, including construction of a dynamic map of constellation geometry with adjustments made as needed to suit the science objective. 

3. \textbf{Fault management}; responding to anomalies locally (e.g., taking anomalous LNs out of the constellation or executing intelligent reset strategies), as well as state identification and escalation to human-in-the-loop investigations as needed.

4. \textbf{Distributed decision making}; collaboratively building a model of constellation geometry and dynamically changing subgroupings of LNs (beamforming clusters), and their controlling CCN, as orbital trajectories evolve.

\subsection{Timing and Calibration} \label{sec:timecal_future}
This report (\S\ref{sec:ranging}) demonstrated basic feasibility of ranging, timing distribution, and calibration by analogy to ground-based and near-space systems.  The next step will be detailed error budget development and simulation of data with realistic timing/ranging errors.  Once this imperfect data is generated, we will test calibration procedures to demonstrate that we can recover the input model science data.

Developing a detailed calibration strategy, including the use of artificial coded beacons, is another area in need of further study.  Simulated beacon signals from local CCNs and CCNs in Earth orbit can be added into science data forward modeling in order to assess near- and far-field calibration strategies.

\subsection{Orbital Modeling} \label{sec:orbit}
As part of this study, we confirmed that Earth-Sun L4 or L5 would be an appropriate location for GO-LoW (\S\ref{sec:const-loc}) with appropriate orbits.  Specifying orbits around Lagrange points is a subtle art, however, and beyond the scope of the Phase I study.  The next step for GO-LoW is to engage an orbital dynamics expert and simulate constellation orbit injection, maintenance, and adjustments to realize different baseline configurations.  Key questions will include the rate at which spacecraft move relative to one another in different orbital scenarios.  This will drive the required position update rate and possibly the delta-V budget if frequent small adjustments are needed.  We will also examine how CCNs form beamforming clusters as the constellation evolves.

\subsection{Technology Demonstration Mission}\label{sec:techdemo}
GO-LoW is ambitious and challenges existing mission paradigms.  A technology demonstration mission would provide an opportunity to showcase the autonomy, graduated growth, failure tolerance, and orbit management that full-scale GO-LoW will require.  As described in \S\ref{sec:const-growth} and Figure \ref{fig:ConstProg}, even a small constellation of 10-100 nodes would provide groundbreaking heliophysics, planetary science, and astrophysics science results.  A constellation of this size would enable continuous, detailed monitoring of solar radio bursts and aurorae from Earth, Jupiter, Saturn, Uranus, and Neptune – an unprecedented simultaneous dataset. A 100 node constellation would also provide sufficient sensitivity and angular resolution for a first-of-its-kind low frequency sky map. This sky map would provide insight into the properties and distribution of plasma in the interstellar medium of the galaxy. A detailed map of the galactic “foreground” is also enabling for 21-cm cosmology, both for the global Dark Ages signal and future Dark Ages tomography. Beyond these exciting science investigations, a 100 node constellation would be a tractable testbed for the autonomous approaches developed in \S\ref{sec:autonomy} as well as a first step toward mass production of GO-LoW spacecraft.  \textbf{Unlike traditional tech demo missions, a GO-LoW tech demo would raise the TRL of the \textit{concept} of a constellation telescope} as well as specific techologies and subsystems laid out in the technology roadmap (\S\ref{sec:techmap}).

\clearpage
\section{Acknowledgments}
    \label{sec:ack}
    
A portion of this work was supported under Air Force Contract No. FA8702-15-D-0001. Any opinions, findings, conclusions or recommendations expressed in this material are those of the author(s) and do not necessarily reflect the views of the U.S. Air Force.

\section{Team} 
    \label{sec:team}
    
\begin{wrapfigure}{l}{0.2\textwidth}
\vspace{-15pt}
    \includegraphics[width=0.195\textwidth]{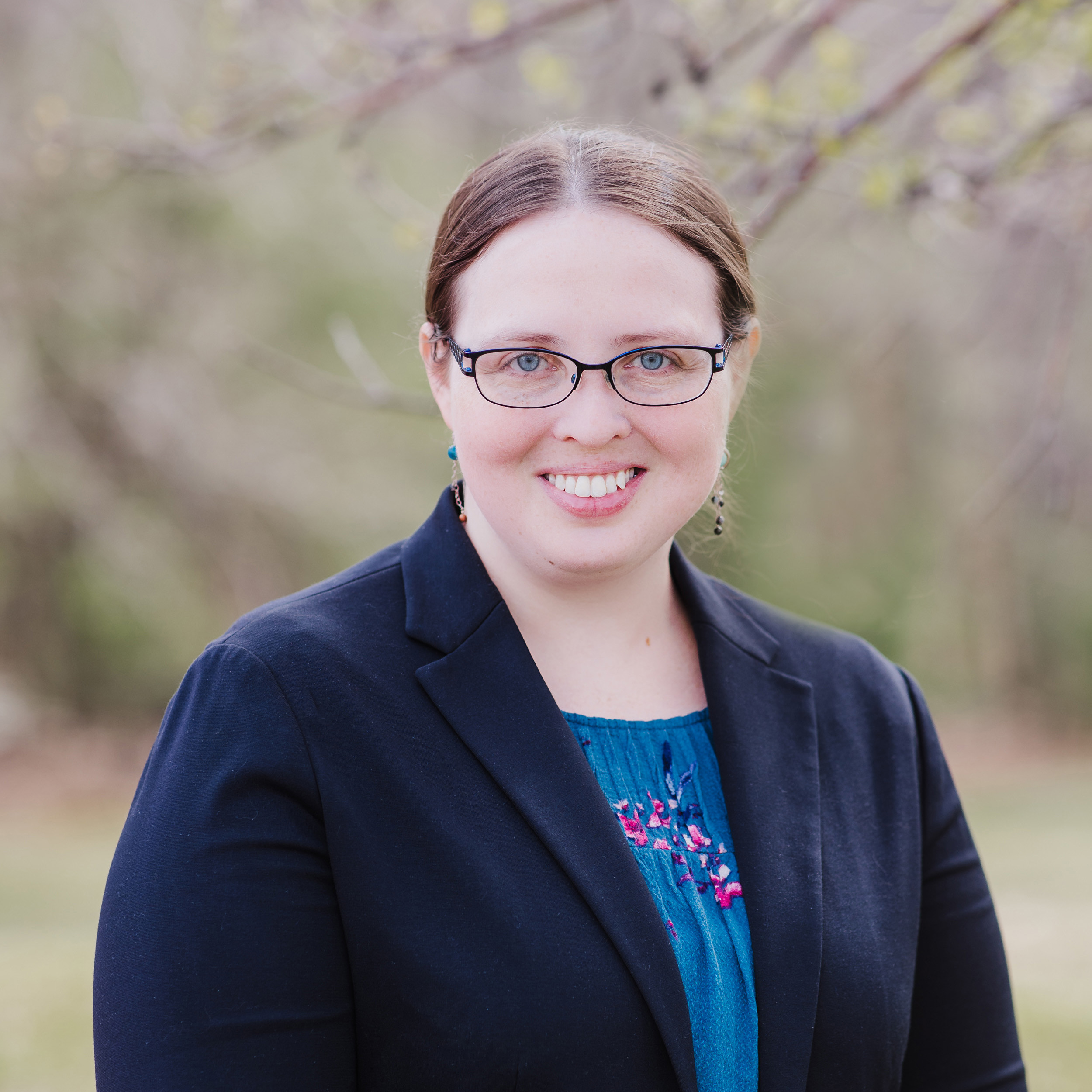}
\end{wrapfigure}
\paragraph{Mary Knapp}
 is a Research Scientist at MIT Haystack Observatory.  She obtained a BS in Aeronautics and Astronautics from MIT in 2011 and a PhD in Planetary Science from MIT in 2018.  Dr. Knapp has extensive experience with ground-based interferometric data collection and reduction.  Her PhD work focused on searching for exoplanets with Jovian-strength magnetic fields and magnetic star-planet interaction using the Jansky Very Large Array (JVLA) and LOFAR \citep{Knapp2018, deGasperin2020}.  Dr. Knapp also has substantial experience in design and operation of CubeSats for science applications, including as Project Scientist for the ASTERIA mission \citep{Smith2018, Knapp2020}.  Currently, Dr. Knapp is the deputy PI/deputy PM for the AERO-VISTA mission \citep{Erickson2018, Lind2019}, which will demonstrate the use of vector sensor interferometry in space in order to map the Earth’s radio auroral emission.

 \begin{wrapfigure}{l}{0.2\textwidth}
 \vspace{-15pt}
    \includegraphics[width=0.195\textwidth]{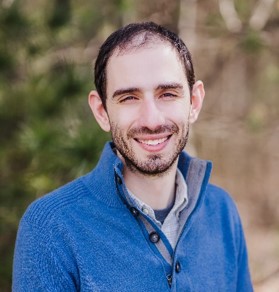}
\end{wrapfigure}
\paragraph{Lenny Paritsky}
is the Space Technology Lead Engineer at MIT Haystack Observatory. Mr. Paritsky has previously held systems engineering, analysis, and management roles at a number of organizations in the space industry. He has extensive experience developing technologies for small spacecraft (\emph{e.g.}, \citet{Tsay2016, Paritsky2013, Wrobel2013, Tsay2016b}), designing mission architectures, and creating models of complex physical systems. Mr. Paritsky has flown technologies on NASA’s PTD-1 (HYDROS, water-based propulsion), two of the Artemis-1 CubeSats (BIT-3, ion engine), and two spacecraft aboard SpaceX’s Transporter-2 (TILE electrospray).

\begin{wrapfigure}{l}{0.2\textwidth}
\vspace{-15pt}
    \includegraphics[width=0.195\textwidth]{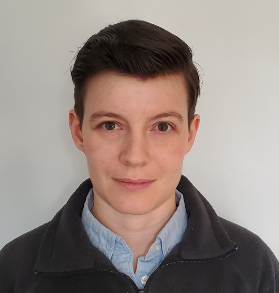}
\end{wrapfigure}
\paragraph{Ekaterina Kononov}
currently holds a joint appointment as a Ph.D. candidate in the Department of Aeronautics and Astronautics at MIT (expected completion May 2024), and an associate technical staff member at MIT Lincoln Laboratory. Her Ph.D. work included creating tools for forward modeling all-sky astronomical data, simulating the output of electrically-small antenna arrays (including VSAs), and developing algorithms for computational imaging with these arrays. Kononov has been with Lincoln Laboratory for 9 years, working on radar technology, including leading the integration of a real-time multi-agent hardware-in-the-loop electromagnetic engagement simulator prior to starting the Ph.D. program.
\pagebreak
\begin{wrapfigure}{l}{0.2\textwidth}
    \includegraphics[width=0.195\textwidth]{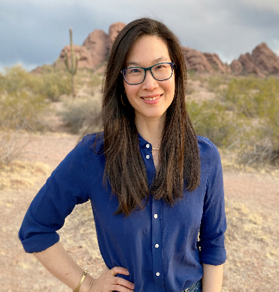}
\end{wrapfigure}
\paragraph{Melodie M. Kao}
is an expert on magnetospheric radio emissions from brown dwarfs and other substellar objects as well as star-planet interactions. Her work includes establishing the existence of auroral phenomena on brown dwarfs \citep{Kao2016}, the first and only direct measurement of a magnetic field using radio emissions from extrasolar planetary mass objects \citep{Kao2016, Kao2018}, and the only proven statistical framework for assessing substellar magnetospheric radio emission occurrence rates \citep{Kao2023}. Dr. Kao’s body of work lays the statistical and interpretative foundations for future detections of radio emissions from exoplanets.

\clearpage
\bibliography{z_references.bib}



\end{document}